\documentclass[reqno]{amsart}

\usepackage{array,multirow}
\usepackage{amsmath}
\usepackage{amssymb}
\usepackage{amsfonts}
\usepackage{graphicx}
\usepackage{bm}
\usepackage{enumerate}

\newtheorem{theorem}{Theorem}
\newtheorem{lemma}{Lemma}
\newtheorem{proposition}{Proposition}
\newtheorem{corollary}{Corollary}
\newtheorem{definition}{Definition}

\numberwithin{equation}{section}

\newcounter{mnote}
\newcommand{\Sec}{\S}

\newcommand{\ID}{I\!\!D}          
\newcommand{\IF}{I\!\!F}         
          
\newcommand{\IK}{I\!\!K}          
\newcommand{\IL}{I\!\!L}
\newcommand{\IN}{I\!\!N}
          
\newcommand{\codim}{\mbox{\rm codim\,}}
\newcommand{\Span}{\mbox{{\rm Span}}}

\newcommand{\bperp}{\bot\mspace{-8.5mu}\flat} 

\newcommand{\closedw}{closed${}_{w}\,$}
\newcommand{\lscw}{lsc${}_{w}\,$}

\newcommand{\Ri}{{}^{\mbox{\tiny \rm 3}}\!R} 
\newcommand{\hRi}{{}^{\mbox{\tiny \rm 3}}\!\hat{R}} 

\newcommand{\Proof}{\noindent{\bf Proof.~}}

\newcommand{\Outline}{\noindent{\bf Outline of the Proof.~}}
\newcommand{\Remark}{\noindent{\bf Remark.~}}

\newcommand{\LRI}{\quad\Leftrightarrow\quad}

\newcommand{\RI}{\quad\Rightarrow\quad}
\newcommand{\wto}{\rightharpoonup} 
\newcommand{\di}{\displaystyle}
\newcommand{\leqs}{\leqslant}      
     
\newcommand{\geqs}{\geqslant}      
\newcommand{\ngeqs}{\ngeqslant}      
\newcommand{\Tr}{{\sf tr}}

\newcommand{\tiD}{\mbox{{\tiny $D$}}}

\newcommand{\tiF}{\mbox{{\tiny $F$}}}
\newcommand{\tiG}{\mbox{{\tiny $G$}}}
\newcommand{\tiH}{\mbox{{\tiny $H$}}}

\newcommand{\tiJ}{\mbox{{\tiny $J$}}}

\newcommand{\tiL}{\mbox{{\tiny $L$}}}

\newcommand{\tiN}{\mbox{{\tiny $N$}}}
\newcommand{\tiR}{\mbox{{\tiny $R$}}}
\newcommand{\tiX}{\mbox{{\tiny $X$}}}

\newcommand{\tiZ}{\mbox{{\tiny $Z$}}}

\newcommand{\tiID}{\mbox{{\tiny $\ID$}}}

\newcommand{\tiIF}{\mbox{{\tiny $\IF$}}}

\newcommand{\tiIL}{\mbox{{\tiny $\IL$}}}
\newcommand{\tiIN}{\mbox{{\tiny $\IN$}}}

\newcommand{\tiwedge}{\mbox{{\tiny $\wedge$}}}
\newcommand{\tivee}{\mbox{{\tiny $\vee$}}}

\newcommand{\N}{{\mathbb N}}       
\newcommand{\R}{{\mathbb R}}       

\newcommand{\cI}{{\mathcal I}}

\newcommand{\cL}{{\mathcal L}}
\newcommand{\cM}{{\mathcal M}}

\newcommand{\ttK}{{\tt K}}

\newcommand{\ttk}{{\tt k}}

\newcommand{\bomega}{\boldsymbol{\omega}}

\def\mathbi#1{\textbf{\em #1}}

\newcommand{\biA}{\mathbi{A}}

\newcommand{\biC}{\mathbi{C}}

\newcommand{\biL}{\mathbi{L}}

\newcommand{\biW}{\mathbi{W\,}}

\newcommand{\bib}{\mathbi{b \!\!}}

\newcommand{\bif}{\mathbi{f\,}}

\newcommand{\bij}{\mathbi{j\,}}

\newcommand{\bin}{\mathbi{n}}

\newcommand{\biu}{\mathbi{u}}
\newcommand{\biv}{\mathbi{v}}
\newcommand{\biw}{\mathbi{w}}

\newcommand{\tbA}{\textbf{A}}
\newcommand{\tbL}{\textbf{L}}
\newcommand{\tbW}{\textbf{W\,}}

\newcommand{\tbb}{\textbf{b}}

\newcommand{\tbf}{\textbf{f\,}}

\newcommand{\tbj}{\textbf{j}}
\newcommand{\tbu}{\textbf{u}}

\newcommand{\tbw}{\textbf{w}}

\newcommand{\hh}{\hat h}
\newcommand{\hj}{\hat \jmath}
\newcommand{\hk}{\hat k}

\newcommand{\hs}{\hat s}

\newcommand{\hD}{\hat D}

\newcommand{\hrho}{\hat \rho}
\newcommand{\htau}{\hat \tau}
\newcommand{\hDelta}{\hat \Delta}

\newcommand{\un}{\underline}       

\begin{document}

\title[Rough Non-CMC Solutions to the Einstein Constraints]
      {Rough solutions of the Einstein constraint equations
       with nonconstant mean curvature}

\author[M. Holst]{M. Holst}
\email{mholst@math.ucsd.edu}
\thanks{MH was supported in part by NSF Awards~0715145, 0411723, 
and 0511766, and DOE Awards DE-FG02-05ER25707 and DE-FG02-04ER25620.}

\author[J. Kommemi]{J. Kommemi}
\email{jkommemi@ucsd.edu}
\thanks{JK was supported in part by UCSD Academic Enrichment Fellowship 
and a UCSD/CalIT2 Summer Research Fellowship.}

\author[G. Nagy]{G. Nagy}
\email{gnagy@math.ucsd.edu}
\thanks{GN was supported in part by NSF Awards~0715145 and 0411723.}

\address{Department of Mathematics\\
         University of California San Diego\\ 
         La Jolla CA 92093}

\date{\today}


\begin{abstract}
We consider the conformal decomposition of Einstein's constraint
equations introduced by Lichnerowicz and York, on a compact manifold
with boundary.  We first develop some technical results for the
momentum constraint operator under weak assumptions on the problem
data, including generalized Korn inequalities on manifolds with
boundary not currently in the literature.  We then consider the
Hamiltonian constraint, and using order relations on appropriate
Banach spaces we derive weak solution generalizations of known sub-
and super-solutions (barriers).  We also establish some related {\em
a~priori} $L^{\infty}$-bounds on any $W^{1,2}$-solution.  The barriers
are combined with variational methods to establish existence of
solutions to the Hamiltonian constraint in $L^{\infty} \cap W^{1,2}$.
The result is established under weak assumptions on the problem data,
and for scalar curvature $R$ having any sign; non-negative $R$
requires additional positivity assumptions either on the matter energy
density or on the trace-free divergence-free part of the extrinsic
curvature. Although the formulation is different, the result can be
viewed as extending the regularity of the recent result of Maxwell on
``rough'' CMC solutions in $W^{k,2}$ for $k>3/2$ down to $L^{\infty}
\cap W^{1,2}$.  The results for the individual constraints are then
combined to establish existence of non-CMC solutions in $W^{1,p}$,
$p>3$ for the three-metric and in $L^{q}$, $q=6p/(3+p)$ for the
extrinsic curvature. The result is obtained using fixed-point
iteration and compactness arguments directly, rather than by building
a contraction map.  The non-CMC result can be viewed as a type of
extension of the regularity of the 1996 non-CMC result of Isenberg and
Moncrief down to $W^{1,p}$ for $p>3$, and extending their result to
$R$ having any sign.  Similarly, the result can also be viewed as type
of extension of the recent work of Maxwell on rough solutions from the
CMC case to the non-CMC case.  Although our presentation is for
3-manifolds, the results also hold in higher dimensions with minor
adjustments.  The results should also extend to other cases such as
closed and (fully or partially) open manifolds without substantial
difficulty.
\end{abstract}

\maketitle

\clearpage

{\tiny
\tableofcontents
}

\section{Introduction}
\label{S:I}

In this article, we give an analysis of the coupled Hamiltonian and
momentum constraints in the Einstein equations on 3-dimensional
compact manifolds with boundary. We consider the equations with matter
sources satisfying an energy condition implied by the dominant energy
condition in the 4-dimensional spacetime; the unknowns are a
Riemannian three-metric and a two-index symmetric tensor. The
equations form an under-determined system; therefore, we focus
entirely on a standard reformulation used in both mathematical and
numerical general relativity, called the conformal method, introduced
by Lichnerowicz and York~\cite{aL44,jY71,jY72}. The conformal method
assumes that the unknown metric is known up to a scalar field called a
conformal factor, and also assumes that the trace and a term
proportional to the trace-free divergence-free part of the two-index
symmetric tensor is known, leaving as unknown a term proportional to
the traceless symmetrized derivative of a vector. Therefore, the new
unknowns are a scalar and a vector field, transforming the original
under-determined system for a metric and a symmetric tensor into a
(potentially) well-posed elliptic system for a scalar and a vector
field.  See~\cite{rBjI04} for a recent review article.
We point out just some of the quite substantial number of previous 
related works, including:
the original work on the Lichnerowicz equation
\cite{aL44};
the development of the conformal method
\cite{jY71,jY72,jY73,jY74};
the initial solution theory for the Hamiltonian constraint
\cite{nOMjY73,nOMjY74,nOMjY74a};
the thin sandwich alternative to the conformal method
\cite{rBgF93,cMkTjW70};
the complete classification of CMC initial data
\cite{jI95}
and the few known non-CMC results
\cite{jIvM96,jIjP97,yCBjIjY00};
various technical results on transverse-traceless tensors
and the conformal Killing operator
\cite{rB96,rBnOM96};
the more recent development of the conformal thin sandwich formulation
\cite{jY99};
initial data for black holes
\cite{rB00,jBjY80};
initial data for Kerr-like black holes
\cite{sD99,sD00b};
and the gluing approach to generating initial data
\cite{jC00}.

The conformal method gives rise to a coupled nonlinear elliptic system
for the unknown scalar and vector fields; the trace of the symmetric
tensor plays an important role: in the case that the trace is constant
(referred to as the {\em constant mean curvature or CMC case}), the
two equations decouple, giving rise to the term ``semi-decoupling
decomposition'' which is sometimes used to describe the conformal
method~\cite{rBjI04}.  In this case, a linear equation for the unknown
vector can be solved first, and then a semi-linear equation for the
scalar field can be solved, where a coefficient in the nonlinearity
depends quadratically on derivatives of the vector unknown.  Almost
all of the previous work on developing a solution theory for the
constraints has focussed on the conformal decomposition in the CMC
case, primarily in the case of compact manifolds without
boundary~\cite{rBjI04}.  A notable exception is the non-CMC existence
and uniqueness result in H\"{o}lder-classes for a particular physical
scenario, which was established in~\cite{jIvM96}.

In this article, we extend the solution theory for the individual
and coupled Hamiltonian and momentum constraints on compact manifolds 
with boundary in three ways:
\begin{enumerate}[{\it(i)}]
\item Some technical results, including generalized Korn inequalities,
      are established for the conformal Killing operator on compact
      Riemannian manifolds with boundary, under several different
      boundary condition assumptions.  The results, which are not
      currently in the literature, allow us to establish
      well-posedness of the momentum constraint equation in
      $W^{1,2}(\cM)$ on a manifold $\cM$ with boundary, using either
      variational or Riesz-Schauder methods.  The assumptions we make
      on the data using either method are weak enough that standard
      techniques to establish additional regularity are not available.
\item Existence (and in some cases, uniqueness) results are 
      established for weak (or rough) CMC solutions to Hamiltonian
      constraint, for weaker solution spaces than appeared previously
      in~\cite{jIvM96,dM05}.  In particular, using variational methods
      we establish existence of weak solutions to the Hamiltonian
      constraint in $L^{\infty}(\cM) \cap W^{1,2}(\cM)$ on compact
      manifolds with boundary, under assumptions on the data that do
      not allow for the use of standard techniques to establish
      additional regularity. The variational methods we employ make
      use of (generalized) barriers for the Hamiltonian constraint
      equation; in \Sec\ref{S:HC-SB} we summarize the barriers we use
      for different values of the Ricci scalar of the background
      metric.  We also establish some related {\em a~priori}
      $L^{\infty}$-bounds on any $W^{1,2}$-solution to the Hamiltonian
      constraint in \Sec\ref{S:HC-apriori}. Although such results are
      standard for semi-linear scalar problems with monotone
      nonlinearities (see for example~\cite{jJ85}), our results hold for 
      a class of non-monotone nonlinearities that includes the Hamiltonian
      constraint nonlinearity and appear to be new.
\item Existence results are established for non-CMC solutions to the 
      coupled system of constraints on compact manifolds with
      boundary, in the setting of weaker (rougher) solutions spaces
      and for more general physical scenarios than appeared previously
      in~\cite{jIvM96}.  In particular, we establish existence of
      solutions to the coupled Hamiltonian and momentum constraints,
      in $W^{1,p}(\cM)$ for $p > 3$ for the conformal factor and in
      $W^{1,q}(\cM)$ for $q=6p/(3+p)$ for the momentum vector, on
      compact manifolds with boundary, with no restrictions on the
      sign of the scalar curvature $R$. For the case of non-negative
      $R$, either the matter energy density or the trace-free
      divergence-free part of the extrinsic curvature must be globally
      positive. The technical condition on the trace of the extrinsic
      curvature (called the ``near-CMC'' condition in~\cite{rBjI04})
      used to produce the coupled system result in~\cite{jIvM96} is
      still present here, although it now involves weaker norms (see
      \Sec\ref{S:CS}).  In addition, this condition is only used here
      to construct a global super-solution to the Hamiltonian
      constraint, and is not used a second distinct time as part of
      the fixed-point argument as was needed in \cite{jIvM96}.
\end{enumerate}
The results above imply that the weakest differentiable solutions of
the Einstein constraint equations we have found correspond to CMC
hypersurfaces with physical spatial metric $h_{ab}$ and extrinsic
curvature $k_{ab}$ satisfying
\begin{equation}
h_{ab} \in L^{\infty}(\cM)\cap W^{1,2}(\cM),\qquad k_{ab} 
\in L^2(\cM).
\end{equation}
The curvature of such data can be computed in a distributional sense,
following~\cite{rGjT87}.

There are at least four distinct, but related, motivations for
establishing the extensions outlined above.
First, as outlined in~\cite{rBjI04}, new results for the non-CMC case, 
beyond the case analyzed in~\cite{jIvM96}, are of great interest in both
mathematical and numerical relativity.
Second, there is currently substantial research activity in rough solutions 
to the Einstein evolution equations, which rest on rough/weak solution 
results for the initial data~\cite{sKiR01}.
Third, the role of boundary conditions and bounded domains in 
the solution and approximation theory is of importance particularly in
numerical relativity; most existing results are for closed (compact
without boundary), open, or only partially bounded domains.
Finally, the approximation theory for Petrov-Galerkin-type methods 
(including finite element, wavelet, spectral, and other methods) for the 
constraints and similar systems previously developed in~\cite{mH01a} 
establishes convergence of numerical solutions in very general physical 
situations, but rests on assumptions about the solution 
theory; the results in the present paper help to complete this approximation 
theory framework. 

An outline of the paper describing the results is as follows.

In \Sec\ref{S:NC}, we give a brief outline of the notation used
throughout the paper.  In \Sec\ref{S:CE}, we quickly overview the
conformal decomposition, describe the classical strong formulation of
the resulting coupled elliptic system, and then define weak
formulations of the constraint equations that will allow us to develop
solution theories for the constraints in the spaces with the weakest
possible regularity.  Our formulation allows for a mix of Dirichlet
and Robin boundary conditions for modeling e.g. black hole and other
physically important scenarios.

In \Sec\ref{S:MC}, we develop some basic technical results for the
momentum constraint equation on compact manifolds with boundary that
we will need later for analysis of the Hamiltonian constraint and for
analysis of the coupled system.  We first develop the weak formulation
of the momentum constraint for a given scalar conformal factor in
\Sec\ref{S:SMC}, and then develop some preliminary results related to
the Korn inequality in \Sec\ref{S:GKI}.  In particular, we establish
generalized Korn inequalities for the conformal Killing operator on
compact manifolds with boundary under several boundary condition
scenarios; these results do not appear to be in the literature.  In
\Sec\ref{S:MC-ExVM}, we use the preliminary results from
\Sec\ref{S:GKI}, together with a variational argument, to establish
existence and uniqueness of weak solutions to the momentum constraint
in $W^{1,2}(\cM)$ when the Dirichlet part of the boundary is
non-empty, with assumptions on the data that do not allow for
additional regularity in the sense described earlier.  In
\Sec\ref{S:MC-WP} we give a second (non-variational) argument for
existence and uniqueness again in $W^{1,2}(\cM)$ using a
Riesz-Schauder (Fredholm alternative) argument following. This second
argument is more general than the variational argument in the sense
that the Dirichlet part of the boundary might be empty. While the
techniques we use in the analysis of the momentum constraint are
standard, this collection of results of the momentum constraint
operator on compact domains with boundary apparently are not in the
literature.  Regularity of solutions to the momentum constraint is
discussed briefly in \Sec\ref{S:MC-RS}.

In \Sec\ref{S:HC}, we give a corresponding analysis of the Hamiltonian
constraint on compact manifolds with boundary. We first develop the
weak formulation of the Hamiltonian constraint for a given momentum
vector variable in \Sec\ref{S:WF-HC}, and establish some preliminary
results on local and global barriers (constant sub- and
super-solutions) for weak solutions in \Sec\ref{S:WF-LB}. The term
local means that the barrier depends on the given momentum vector
variable solution of the momentum constraint equation, while global
means that the barrier is not local. These barriers are non-trivial
extensions of those in~\cite{jI95,jIvM96} to nonlinearities with
coefficients in $W^{-1,2}(\cM)$.  We also establish some related {\em
a~priori} $L^{\infty}$-bounds on any $W^{1,2}$-solution to the
Hamiltonian constraint; although such results are standard for
semi-linear problems with monotone nonlinearities, our results hold
for a class of non-monotone nonlinearities that includes the
Hamiltonian constraint nonlinearity and appear to be new.  The weak
solution barriers are critical to extending the solution theory for
the Hamiltonian constraint to the weakest possible setting of
$W^{1,2}(\cM)$ in \Sec\ref{S:HC-ExVM}, and are also key to extending
the solution theory for the coupled system to weaker spaces and to new
physical scenarios in \Sec\ref{S:CS}. In \Sec\ref{S:HC-ExVM}, we use
the barriers from \Sec\ref{S:WF-LB}, together with a variational
argument, to establish existence and uniqueness of solutions to the
Hamiltonian constraint in the weakest possible setting of
$L^{\infty}(\cM) \cap W^{1,2}(\cM)$.  Due to the lack of
G\^ateaux-differentiability of the nonlinearity in $W^{1,2}(\cM)$, the
connection between the energy used for the variational argument and
the Hamiltonian constraint as its Euler condition is non-trivial, and
is established through several Lemmas.  We note that our arguments
allow for scalar curvature to have any sign, and the assumptions on
the data are such that additional regularity is not possible in the
sense described earlier. The results for non-negative $R$ require an
assumption of global positivity either on the trace-free and
divergence-free part of the extrinsic curvature or on the matter
energy density. Although our problem formulation is somewhat
different, the result can be viewed as extending the regularity of the
recent result of Maxwell~\cite{dM05} on ``rough'' CMC solutions in
$W^{k,2}(\cM)$ for $k>3/2$ down to $L^{\infty}(\cM) \cap
W^{1,2}(\cM)$.  In \Sec\ref{S:WF-Existence-LB} we give a second
(non-variational) argument for existence and uniqueness, using the
barriers (sub-/super-solution) approach as in most of the earlier
work~\cite{jI95,jIvM96,dM05}. Unlike the case of the momentum
constraint in \Sec\ref{S:MC-WP}, where the non-variational technique
allows for the development of a solution theory in same weak setting
of $W^{1,2}(\cM)$ as does the variational method, the Hamiltonian
constraint barriers approach requires additional regularity beyond
what the variational approach in \Sec\ref{S:HC-ExVM} requires.
Regularity of solutions is discussed briefly in \Sec\ref{S:HC-Reg}.

Finally, in \Sec\ref{S:CS} we use the results for the individual
constraints derived earlier to establish a new non-CMC result for the
coupled system. In \Sec\ref{S:CS-Ex}, we establish existence of
non-CMC solutions to the coupled constraints through fixed-point
iteration and compactness arguments directly, rather than by using the
Contraction Mapping Theorem as was done in the original work of
Isenberg and Moncrief in~\cite{jIvM96}.  The ``near-CMC'' assumption
on the trace of the extrinsic curvature required in~\cite{jIvM96} is
still present here, although it now involves weaker norms (see
\Sec\ref{S:CS}).  In addition, the condition is only used here to
construct a global super-solution to the Hamiltonian constraint, and
is not used a second distinct time as part of the fixed-point argument
as was necessary in~\cite{jIvM96}.  If a global super-solution can be
constructed without the near-CMC assumption, then this new coupled
system result would still hold for ``far-from-CMC'' scenarios. The
result requires more regularity than that needed for the result
established in \Sec\ref{S:MC} and \Sec\ref{S:HC} for the individual
constraints, with solutions in $W^{1,p}(\cM)$ for $p > 3$ for the
conformal factor and in $W^{1,q}(\cM)$ for $q=6p/(3+p)$ for the
momentum vector, but still extends the existing theory for the system
in two distinct ways.  First, although our problem formulation is
somewhat different (bounded domains with matter), the result can be
viewed as a type of extension of the 1996 non-CMC result of Isenberg
and Moncrief in \cite{jIvM96} from the scalar curvature $R=-1$ case to
$R$ having any sign (with non-negative $R$ requiring the assumption
that either the matter energy density or the trace-free
divergence-free part of the extrinsic curvature be globally positive),
and to weaker solution spaces.  Second, again although the problem
formulation is different, the result could be viewed as a type of
extension of the recent rough CMC solution work of Maxwell
in~\cite{dM05} to the non-CMC case.  Although our presentation is for
3-manifolds, the results hold in higher spatial dimensions with minor
adjustments, and the techniques we employ should extend to other cases
such as closed and (fully or partially) open manifolds through the use
of tools such as weighted Sobolev spaces.

We summarize our results in \Sec\ref{S:summary}.

\subsection{Notation and conventions}
\label{S:NC}

Let $(\cM,h_{ab})$ be a a Riemannian manifold, where $\cM$ is a
3-dimensional, smooth, compact manifold with non-empty boundary
$\partial\cM$, and $h_{ab}$ is a $C^2$ metric on $\overline\cM$, that
is, a symmetric, positive definite, covariant, two-index tensor on
$\cM$ with all components in a smooth coordinate system having two
continuous derivatives. Latin indices denote abstract indices as they
are defined in~\cite{Wald84}, \Sec 2.4. The metric defines an inner
product on $T_x\cM$, the vector space tangent to $\cM$ at the point
$x\in\cM$. Denote by $h^{ab}$ the inverse of the metric tensor
$h_{ab}$, that is, $h_{ac}h^{bc} =\delta_a{}^b$, where $\delta_a{}^b:
T_x\cM\to T_x\cM$ is the identity map. We use the convention that
repeated indices, one upper-index and one sub-index, denote
contraction. Let $\nabla_a$ be the Levi-Civita connection associated
with the metric $h_{ab}$, that is, the unique torsion-free connection
satisfying $\nabla_ah_{bc}=0$. Let $R_{abc}{}^d$ be the Riemann tensor
of the connection $\nabla_a$, where the sign convention used in this
article is $(\nabla_a\nabla_b -\nabla_b\nabla_a)v_c := R_{abc}{}^d
v_d$. Denote by $R_{ab} := R_{acb}{}^c$ the Ricci tensor and by $R
:=R_{ab}h^{ab}$ the Ricci curvature scalar of this connection. (Only
in \Sec\ref{S:SDD} we modify the notation for the connection
$\nabla_a$.)

Indices on tensors will be raised and lowered with $h^{ab}$ and
$h_{ab}$, respectively. For example, given the tensor $s^{ab}{}_c$ we
denote $s_{abc}=h_{aa_1}h_{bb_1}\,s^{a_1b_1}{}_{c}$, and $s^{abc}
=h^{cc_1}\,s^{ab}{}_{c_1}$; notice that the order of the indices is
important in the case that the tensor $s_{abc}$ or $s^{abc}$ is not
symmetric. We say that a tensor is an {\bf $n$-index} tensor iff it
can be transformed into a tensor $s_{a_1\cdots a_n}$ by lowering
appropriate indices. We denote by $C^{\infty}(\cM,n)$ the set of all
smooth $n$-index tensor fields on $\cM$. Given an arbitrary tensor
$s^{a_1\cdots a_n}{}_{b_1\cdots b_m}$, which is an $n+m$-index
tensor, we define its magnitude at any point $x\in\cM$ as the
real-valued function given by 
\begin{equation}
\label{tensor-magnitude}
|s| := (s^{a_1\cdots b_m}s_{a_1\cdots b_m})^{1/2}.
\end{equation}
Integration on $\cM$ is performed with the volume element $dx$
associated to the metric $h_{ab}$.  A norm of an arbitrary smooth
tensor field $s^{a_1\cdots a_n}{}_{b_1\cdots b_m}$ on $\cM$ can be
defined for any $1\leqs p < \infty$ and for $p=\infty$ respectively 
using~(\ref{tensor-magnitude}) as follows,
\begin{equation}
\label{N-Lp-norm}
\|s\|_p := \Bigl[\int_{\cM} |s|^p\,dx\Bigr]^{1/p},\qquad
\|s\|_{\infty}:= \mbox{ess~} \sup_{x\in\cM}|s|.
\end{equation}
We introduce the {\bf Lebesgue spaces} $L^p(\cM,n)$, for $1\leqs
p\leqs\infty$, of $n$-index tensor valued fields as the completion of
$C^{\infty}(\cM,n)$ under the norm in Eq.~(\ref{N-Lp-norm}), and this
norm is called the {\bf $L^p$-norm}. The Lebesgue spaces $L^p(\cM,n)$
are Banach spaces; they are separable when $1\leqs p < \infty$ and
reflexive when $1 < p < \infty$.  For the case $p=2$ the spaces
$L^2(\cM,n)$ form a Hilbert space with the inner product and norm
given by
\begin{equation}
\label{N-L2-inner}
(s,r):= \int_{\cM} s_{a_1\cdots a_n}r^{a_1\cdots a_n}\, dx,\qquad
\|s\|:=\sqrt{(s,s)}=\|s\|_2.
\end{equation}
Covariant derivatives of tensor fields are denoted as
\[
\nabla^{m}s := 
\nabla_{b_1,\cdots,b_m} s^{a_1\cdots a_n}
:=\nabla_{b_1}\cdots\nabla_{b_m}s^{a_1\cdots a_n},
\]
where the super-script $m$ indicates the total number of derivatives,
which plays the role of the number $|\alpha|$ for a multi-index
$\alpha$, in the multi-index notation used in the PDE literature.
Using again
Eq.~(\ref{tensor-magnitude}) introduce the real-valued function
\[
|\nabla^{m}s| := 
\bigr[(\nabla_{b_1}\cdots\nabla_{b_m}s^{a_1\cdots a_n})
(\nabla^{b_1}\cdots\nabla^{b_m}s_{a_1\cdots a_n})\bigr]^{1/2},
\]
as the starting point for defining $L^p$-type norms involving derivatives.
One such norm in the vector space $C^{\infty}(\cM,n)$ is given for any
non-negative integer $k$ and a real number $p$ with $1\leqs p <
\infty$, and separately for $p=\infty$, as follows
\begin{equation}
\label{N-Wkp-norm}
\|s\|_{k,p} := \Bigl[\sum_{m=0}^k \,\|\nabla^{m}s\|_p^p \Bigl]^{1/p},
\qquad \qquad
\|s\|_{k,\infty} := \max_{0\leqs m\leqs k}\|\nabla^{m}s\|_{\infty}.
\end{equation}
We introduce the {\bf Sobolev spaces} $W^{k,p}(\cM,n)$ of $n$-index
tensor valued fields as the completion of $C^{\infty}(\cM,n)$ under
the norm in Eq.~(\ref{N-Wkp-norm}), and this norm is called the {\bf
$W^{k,p}$-norm}. The Sobolev spaces $W^{k,p}(\cM,n)$ are Banach
spaces; being based on $L^p(\cM,n)$, they are separable when $1\leqs p
< \infty$ and reflexive when $1 < p < \infty$.  For the case $p=2$ the
spaces $W^{k,2}(\cM,n)$ form a Hilbert space with the inner product
and norm given by
\begin{equation}
\label{N-W2-inner}
(s,r)_k:= \sum_{m=0}^k \,(\nabla^{m}s,\nabla^{m}r),
\qquad \qquad \|s\|_{k,2} = \sqrt{(s,s)_k},
\end{equation}
where we have introduced the notation
\[
(\nabla^{m}s,\nabla^{m} r):= 
\int_{\cM} (\nabla_{b_1}\cdots \nabla_{b_m}s_{a_1\cdots a_n})
(\nabla^{b_1}\cdots\nabla^{b_m}r^{a_1\cdots a_n})\, dx.
\]
Therefore we have that $L^p(\cM,n)=W^{0,p}(\cM,n)$ and $\|s\|_p
=\|s\|_{0,p}$. These definitions follow~\cite{Hebey96} for the case of
scalar fields and~\cite{Palais65} for the case of arbitrary tensor
fields. These definitions can also be extended, using appropriate
partitions of the unity and Fourier transforms, from non-negative
integers $k$ to real numbers $s$. For example
see~\cite{Taylor96a}.

In this article we are mainly concerned with spaces of scalar-valued
fields and vector-valued fields on $\cM$, so we introduce the
following notation for these special cases,
\begin{align*}
C^{\infty} &:= C^{\infty}(\cM,0),& 
\biC^{\infty} &:= C^{\infty}(\cM,1),\\
L^p &:= L^p(\cM,0),& \biL^p &:= L^p(\cM,1), \\
W^{k,p} &:= W^{k,p}(\cM,0),& \biW^{k,p} &:= W^{k,p}(\cM,1).
\end{align*}
However, we will not suppress the manifold from the notation of these
spaces when this information is important in a given situation. In a
similar way, we use both notations $\biw$ and $w^a$ to denote a vector
field.  We consider in this article that the boundary $\partial\cM$ of
$\cM$ can be divided in the following two, possible different, ways as
follows,
\begin{gather}
\label{b-DN}
\partial\cM =\partial\cM_{\tiD} \cup \partial \cM_{\tiN},\quad
\overline{\partial\cM}_{\tiD} \cap 
\overline{\partial \cM}_{\tiN} =\emptyset,\\
\label{b-IDN}
\partial\cM =\partial\cM_{\tiID} \cup \partial \cM_{\tiIN},\quad
\overline{\partial\cM}_{\tiID} \cap
\overline{\partial \cM}_{\tiIN} =\emptyset.
\end{gather}
Introduce the trace operators
\[
\Tr_{\tiD}: W^{s,p} \to W^{s-\frac{1}{p},p}(\partial\cM_{\tiD},0),
\qquad
\Tr_{\tiID}:\biW^{s,p} \to W^{s-\frac{1}{p},p}(\partial \cM_{\tiID},1),
\]
for $s > 1/p$, which are the continuous extensions to Sobolev spaces
of the operators defined on smooth fields given by $\Tr_{\tiD}\phi
:=\phi|_{\partial\cM_{\tiD}}$, and $\Tr_{\tiID}\biw
:=\biw|_{\partial\cM_{\tiID}}$, see for example~\cite{Schwarz95}.
Both spaces $W^{s-\frac{1}{p},p}(\partial\cM_{\tiD},0)$ and
$W^{s-\frac{1}{p},p}(\partial\cM_{\tiID},1)$ are Banach spaces and we
denote their norms as $\|\phi\|_{s-\frac{1}{p},p,\tiD}$ and
$\|\biw\|_{s-\frac{1}{p},p,\tiID}$, respectively. In the particular
case $p=2$ these spaces become Hilbert spaces and we denote their
inner product as $(\phi,\un\phi)_{s-\frac{1}{2},\tiD}$ and
$(\biw,\un\biw)_{s-\frac{1}{2},\tiID}$. We will be mainly concerned
with the case $s=1$, $p=2$, and in this case we denote their inner
product and norms as follows
\[
(\hat\phi,\un{\hat\phi})_{\tiD},\quad
\|\hat\phi\|_{\tiD}=(\hat\phi,\hat\phi)_{\tiD}^{1/2},\quad
(\hat\biw,\un{\hat\biw})_{\tiID},\quad
\|\hat\biw\|_{\tiID}=(\hat\biw,\hat\biw)_{\tiID}^{1/2},
\]
for all $\hat\phi$, $\un{\hat\phi}\in
W^{\frac{1}{2},2}(\partial\cM_{\tiD},0)$ and all $\hat\biw$,
$\un{\hat\biw}\in W^{\frac{1}{2},2}(\partial\cM_{\tiID},1)$.

In an analogous way, introduce the trace operators $\Tr_{\tiN}$ and
$\Tr_{\tiIN}$ and the spaces $W^{s-\frac{1}{p},p}(\cM_{\tiN},0)$ and
$W^{s-\frac{1}{p},p}(\cM_{\tiIN},1)$. We use the notation
\begin{align*}
W_{0}^{1,p} &:= \{ \phi \in W^{1,p} : \Tr_{\tiD}\phi =0,~
\Tr_{\tiN}\phi=0  \},&
W_{\tiD}^{1,p} &:= \{ \phi \in W^{1,p} : \Tr_{\tiD}\phi =0  \},\\
\biW_{0}^{1,p} &:= \{\biw\in W^{1,p} : \Tr_{\tiID}\biw =0,~
\Tr_{\tiIN}\biw =0\},&
\biW_{\tiID}^{1,p} &:= \{\biw\in W^{1,p} : \Tr_{\tiID}\biw =0 \},
\end{align*}
for the function spaces. The dual spaces of some Sobolev spaces will
be denoted as follows:
\begin{gather*}
W^{-k,p} := \bigl[ W^{k,p'} \bigr]^{*},\quad
W^{-k,p}_0 := \bigl[ W^{k,p'}_0 \bigr]^{*},\quad
W^{-k,p}_{\tiD} := \bigl[ W^{k,p'}_{\tiD} \bigr]^{*},\\
\biW^{-k,p} := \bigl[ \biW^{k,p'} \bigr]^{*},\quad
\biW^{-k,p}_0 := \bigl[ \biW^{k,p'}_0 \bigr]^{*},\quad
\biW^{-k,p}_{\tiID} := \bigl[ \biW^{k,p'}_{\tiID} \bigr]^{*},
\end{gather*}
where we denote by $p'$ the conjugate of $p$ in the sense $\frac{1}{p}
+\frac{1}{p'}=1$.  These are Banach spaces with the norm
\[
\|s^*\|_{-k,p} := \sup_{0\neq r\in W^{k,p'}(\cM,n)} 
\frac{|s^{*}(r)|}{~\|r\|_{k,p'}},
\]
where $s^{*}\in W^{-k,p}(\cM,n)$, and the asterisk is introduced to
emphasize that $s^{*}$ is a linear and bounded map $s^{*}:
W^{k,p'}(\cM,n)\to\R$. The product of an element in $u\in
L^{\infty}(\cM,0)$ by an element in $s^{*}\in W^{-1,p}(\cM,n)$, with
$1<p<\infty$ will be denoted by $(su)^{*}$. Such element is a
well-defined functional in $W^{-1,p}(\cM,n)$. The proof of this
statement is given in the Appendix using appropriate Gelfand triple
structures.

We will need order structure in some Sobolev spaces. See the Appendix
for a review on ordered Banach spaces, where we explain the particular
notation from this field we use in this article, and where we define
the main order cones needed in this article: $L^{\infty}_{+}$,
$L^p_{+}$, and $W^{k,p}_{+}$. We use both notations $(u-v)\in X_{+}$
and $u\geqs v$ to state that the element $u$ in a Banach space $X$ is
bigger than or equal to another element $v$ in that space. The former
notation specifies which Banach space the elements belong to and also
which order cone is used in that particular Banach space; while both
pieces of information are not explicitly displayed in the latter
notation. We use the notation $u\geqs v$ when there is no ambiguity,
otherwise we use the notation $(u-v)\in X_{+}$. We also write
$-(u-v)\in X_{+}$ to denote $u\leqs v$.

Given Banach spaces $X$, $Y$ and an operator $A:D_{A}\subset X \to
R_{A}\subset Y$, we denote by $D_{A}$ and $R_{A}$ the domain and
range of $A$, respectively, while $N_{A}$ denotes the null space of
$A$.

We recall the {\bf generalized H\"older inequality}
(see~\cite{Gasinski06}, page 904), which is used in several places in
this article, and says that given $n$-index tensor fields $s_i\in
L^{p_i}(\cM,n)$ with $i=1,\cdots,k$ for $k\in\N$, the pointwise
product is well-defined a.e. in $\cM$, the tensor field $s_1\cdots
s_k\in L^p(\cM,nk)$, where $p=\sum_{i=1}^k 1/p_i$, and the following
estimate holds,
\[
\|s_1\cdots s_k\|_p \leqs \|s_1\|_{p_1} \cdots \|s_k\|_{p_k}.
\]

One last comment on the notation: Given a Banach space $X$, the
elements $x$, $\un x\in X$ denote different elements. This notation
usually appears in equations written in weak form, where the element
$x$ denotes the trial function and the element $\un x$ denotes the
test function.

\section{The constraint equations}
\label{S:CE}

We give a quick overview of the conformal decomposition, describe the
classical strong formulation of the resulting coupled elliptic system,
and then define weak formulations of the constraint equations that
will allow us to develop solution theories for the constraints in the
spaces with the weakest possible regularity. Our formulation allows
for a mix of Dirichlet and Robin boundary conditions for modeling
e.g. black hole and other physically important scenarios.

\subsection{The conformal decomposition method}
\label{S:SDD}

Let $(M,g_{\mu\nu})$ be a smooth 4-di\-men\-sional spacetime, that is,
$M$ is a 4-dimensional, smooth manifold, and $g_{\mu\nu}$ is a smooth,
Lorentzian metric on $M$ with signature $(-,+,+,+)$. Let
$\nabla_{\mu}$ be the Levi-Civita connection associated with the
metric $g_{\mu\nu}$, that is, the unique torsion-free connection
satisfying $\nabla_{\sigma}g_{\mu\nu}=0$.  The Einstein equation is
\[
G_{\mu\nu} = \kappa T_{\mu\nu},
\]
where $G_{\mu\nu} = R_{\mu\nu} - R\,g_{\mu\nu}/2$ is the Einstein
tensor, $T_{\mu\nu}$ is the stress-energy tensor, and $\kappa = 8\pi
G/c^4$, with $G$ the gravitation constant and $c$ the speed of
light. The Ricci tensor is $R_{\mu\nu} = R_{\mu\sigma\nu}{}^{\sigma}$
and $R= R_{\mu\nu}g^{\mu\nu}$ is the Ricci scalar, where $g^{\mu\nu}$
is the inverse of $g_{\mu\nu}$, that is $g_{\mu\sigma} g^{\sigma\nu}
=\delta_{\mu}{}^{\nu}$. The Ricci tensor is defined as a contraction
of the Riemann tensor $R_{\mu\nu\sigma}{}^{\rho} w_{\rho}
=\big(\nabla_{\mu}\nabla_{\nu} -\nabla_{\nu}\nabla_{\mu}\bigr)
w_{\sigma}$, where $w_{\mu}$ is any 1-form on $M$. The stress-energy
tensor $T_{\mu\nu}$ is assumed to be symmetric, to satisfy the
integrability condition $\nabla_{\mu}T^{\mu\nu} = 0$, and the {\bf
dominant energy condition}, that is, the vector $-T^{\mu\nu}v_{\nu}$
is timelike and future-directed, where $v^{\mu}$ is any timelike and
future-directed vector field (see \cite{Wald84}, page 219). In this
section Greek indices $\mu$, $\nu$, $\sigma$, $\rho$ denote abstract
spacetime indices, that is, tensorial character on the 4-dimensional
manifold $M$. They are raised and lowered with $g^{\mu\nu}$ and
$g_{\mu\nu}$, respectively. Later on Latin indices $a$, $b$, $c$, $d$
will denote tensorial character on a 3-dimensional manifold.

The map $t:M\to \R$ is a {\bf time function} iff the function $t$ is
differentiable and the vector field $-\nabla^{\mu}t$ is a timelike,
future-directed vector field on $M$. Introduce the hypersurface $\cM
:=\{ x\in M : t(x)=0\}$, and denote by $n_{\mu}$ the unit 1-form
orthogonal to vector fields tangent to $\cM$. By definition of $\cM$
the 1-form $n_{\mu}$ has the form $n_{\mu} = -\alpha\,\nabla_{\mu}t$,
where $\alpha$ is a positive function such that $n_{\mu} n_{\nu}\,
g^{\mu\nu} = -1$, which is called the {\bf lapse function}. Since the
lapse function is positive, the vector field $n^{\mu}$ is
future-directed. Let $\hh_{\mu\nu}$ and $\hk_{\mu\nu}$ be the first
and second fundamental forms of the hypersurface $\cM$, that is,
\[
\hh_{\mu\nu} := g_{\mu\nu} + n_{\mu} n_{\nu},\qquad
\hk_{\mu\nu} := -\hh_{\mu}{}^{\sigma} \nabla_{\sigma} n_{\nu}.
\]
The Einstein constraint equations on $\cM$ are given by
\[
\bigl( G_{\mu\nu} -\kappa T_{\mu\nu}\bigr) \, n^{\nu} =0.
\]
It is a straightforward albeit long computation to express these
equations involving tensors on $M$ as an equation involving tensors on
$\cM$. The result is the following equations,
\begin{align}
\label{CE-def-H}
\hRi + \hk^2 - \hk_{ab}\hk^{ab} - 2\kappa \hrho &=0,\\
\label{CE-def-M}
\hD^a\hk - \hD_b\hk^{ab} + \kappa \hj^a &= 0,
\end{align}
where tensors $\hh_{ab}$, $\hk_{ab}$, $\hj^a$ and $\hrho$ on a
3-dimensional manifold are the pull-back on $\cM$ of the tensors
$\hh_{\mu\nu}$, $\hk_{\mu\nu}$, $\hj^{\mu}$ and $\hrho$ on the
4-dimensional manifold $M$. We have introduced the energy density
$\hrho := n_{\mu} n_{\mu} T^{\mu\nu}$ and the momentum current density
$\hj^{\mu} := -\hh^{\mu}{}_{\nu} n_{\sigma} T^{\nu\sigma}$. We have
denoted by $\hD_{a}$ the Levi-Civita connection associated to
$\hh_{ab}$, so $(\cM,\hh_{ab})$ is a 3-dimensional Riemannian
manifold, with $\hh_{ab}$ having signature $(+,+,+)$, and we use the
notation $\hh^{ab}$ for the inverse of the metric $\hh_{ab}$. Indices
have been raised and lowered with $\hh^{ab}$ and $\hh_{ab}$,
respectively. We have also denoted by $\hRi$ the Ricci scalar of
curvature of the metric $\hh_{ab}$. Finally, recall that the
constraint Eqs.~(\ref{CE-def-H})-(\ref{CE-def-M}) are indeed equations
on $\hh_{ab}$ and $\hk_{ab}$ due to the matter fields satisfying the
energy condition $-\hrho^2 +\hj_a\hj^a < 0$, which is implied by the
dominant energy condition on the stress-energy tensor $T^{\mu\nu}$ in
spacetime.

Let $\phi$ be a positive scalar field on $\cM$, and decompose the
extrinsic curvature tensor $\hk_{ab} = \hs_{ab} + \hh_{ab} \htau/3$,
where $\htau := \hk_{ab}\hh^{ab}$ is the trace and then $\hs_{ab}$ is
the traceless part of the extrinsic curvature tensor. Then, introduce
now the following conformal rescaling:
\begin{gather}
\label{CE-def1}
\hh_{ab} =: \phi^4 \, h_{ab},\qquad
\hs^{ab} =: \phi^{-10} \,s^{ab},\qquad
\htau =: \tau,\\
\label{CE-def-mf}
\hj^a =: \phi^{-10}\; j^a,\qquad
\hrho =: \phi^{-8}\, \rho.
\end{gather}
We have introduced the Riemannian metric $h_{ab}$ on the 3-dimensional
manifold $\cM$, which determines the Levi-Civita connection $D_a$, and
so we have that $D_a h_{bc}=0$. We have also introduced the symmetric,
traceless tensor $s_{ab}$, and the non-physical matter sources $j^a$
and $\rho$. The different powers of the conformal rescaling above are
carefully chosen so that the constraint
Eqs.~(\ref{CE-def-H})-(\ref{CE-def-M}) transform into the following
equations
\begin{gather}
\label{CE-cr1H}
-8 \Delta \phi + \Ri \phi + \frac{2}{3}\, \tau^2 \phi^5 
- s_{ab}s^{ab} \,\phi^{-7} -2\kappa \rho\,\phi^{-3} =0,\\
\label{CE-cr1M}
-D_bs^{ab} + \frac{2}{3} \phi^6 D^a \tau +\kappa j^a =0,
\end{gather}
where in equation above, and from now on, indices of unhatted fields
are raised and lowered with $h^{ab}$ and $h_{ab}$ respectively. We
have also introduced the {\bf Laplace-Beltrami} operator with respect
to the metric $h_{ab}$, acting on smooth scalar fields; it is defined
as follows
\[
\Delta \phi:= h^{ab}D_aD_b\phi.
\]
Eqs.~(\ref{CE-cr1H})-(\ref{CE-cr1M}) can be obtained by a long, but
otherwise straightforward computation. In order to perform this
calculation it is useful to recall that both $\hD_a$ and $D_a$ are
connections on the manifold $\cM$, and so they differ on a tensor
field $C_{ab}{}^c$, which can be computed explicitly in terms of
$\phi$, and has the form
\[
C_{ab}{}^c = 4 \delta_{(a}{}^cD_{b)} \ln(\phi) 
- 2 h_{ab}h^{cd}D_d \ln(\phi).
\]
We remark that the power four on the rescaling of the metric
$\hh_{ab}$ and $\cM$ being 3-dimensional imply that $\hRi =\phi^{-5}
(\Ri\phi - 8\hDelta\phi)$, and for any other power in the rescaling,
terms proportional to $h^{ab}(D_a\phi)(D_b\phi)/\phi^2$ appear in the
transformation. Similar reasons force the power negative ten on the
rescaling of the tensor $\hs^{ab}$ and $\hj^a$, so terms proportional
to $(D_a\phi)/\phi$ cancel out in Eq.~(\ref{CE-cr1M}). Finally, the
ratio between the conformal rescaling powers of $\hrho$ and $\hj^a$
is chosen such that the inequality $-\rho^2 + h_{ab} j^aj^b <0$
implies the inequality $-\hrho^2 + \hh_{ab}\hj^a\hj^b <0$.

There is one more step to convert the original constraint
Eq.~(\ref{CE-def-H})-(\ref{CE-def-M}) into a determined elliptic
system of equations. This step is the following: Decompose the
symmetric, traceless tensor $s_{ab}$ into a divergence-free part
$\sigma_{ab}$, and the symmetrized and traceless gradient of a vector,
that is, $s^{ab} =: \sigma^{ab} + (\cL w)^{ab}$, where
$D_a\sigma^{ab}=0$ and we have introduced the {\bf conformal Killing}
operator $\cL$ acting on smooth vector fields and defined as follows
\begin{equation}
\label{CF-def-CK}
(\cL w)^{ab} := D^a w^b + D^b w^a - \frac{2}{3}\,(D_c w^c) \,h^{ab}.
\end{equation}
Therefore, the constraint Eqs.~(\ref{CE-def-H})-(\ref{CE-def-M}) are
transformed by the conformal rescaling into the following equations
\begin{gather}
\label{CE-cr2H}
-8 \Delta \phi + \Ri \phi 
+ \frac{2}{3}\, \tau^2 \phi^5 -\bigl(\sigma_{ab} 
+(\cL w)_{ab}\bigr) \bigl(\sigma^{ab}+(\cL w)^{ab}\bigr)
\,\phi^{-7} -2\kappa \rho\,\phi^{-3} =0,\\
\label{CE-cr2M}
-D_b(\cL w)^{ab} + \frac{2}{3} \phi^6 D^a \tau +\kappa j^a =0.
\end{gather}
In the next section we interpret these equations above as partial
differential equations for the scalar field $\phi$ and the vector
field $w^a$, while the rest of the fields are considered as given
fields. Given a solution $\phi$ and $w^a$ of
Eqs.~(\ref{CE-cr2H})-(\ref{CE-cr2M}), the physical metric $\hh_{ab}$
and extrinsic curvature $\hk^{ab}$ of the hypersurface $\cM$ are given
by
\[
\hh_{ab} = \phi^4 \, h_{ab},\qquad
\hk^{ab} = \phi^{-10}\bigl[\sigma^{ab}
+ (\cL w)^{ab}\bigr] + \frac{1}{3}\, \phi^{-4} \,\tau \, h^{ab},
\]
while the matter fields are given by Eq~(\ref{CE-def-mf}).

\subsection{Classical formulation}
\label{S:CF}

Beginning in this section, we will change the notation slightly from
the classical notation used to introduce the conformal method in
\Sec\ref{S:SDD}. In particular, the Levi-Civita connection of the
metric $h_{ab}$ on the 3-dimensional manifold $\cM$ will denoted by
$\nabla_a$ rather than $D_a$, and the Ricci scalar of $h_{ab}$ will be
denoted by $R$ instead of $\Ri$.  This change will simplify the
presentation in the remainder of the paper.

Let $(\cM, h)$ be a 3-dimensional Riemannian manifold, where $\cM$ is
a smooth, compact manifold with smooth boundary $\partial\cM$, and
$h\in C^{\infty}(\overline\cM ,2)$ is a positive definite metric. Let
$L: C^{\infty}\to C^{\infty}$ and $\IL :\biC^{\infty}\to\biC^{\infty}$
be the {\bf Laplace-Beltrami} and {\bf momentum} operators,
respectively, with actions on a scalar field $\phi\in C^{\infty}$ and
a vector field $\biw\in\biC^{\infty}$ given by
\begin{align}
\label{CF-def-L}
L\phi &:= -\Delta \phi,\\
\label{CF-def-IL}
(\IL \biw)^a &:= -\nabla_b (\cL \biw)^{ab},
\end{align}
where $\Delta\phi :=\nabla_a\nabla^a\phi$, and $\cL$ denotes the
conformal Killing operator defined in Eq.~(\ref{CF-def-CK}). We will
also use the index-free notation $\IL\biw$ and $\cL\biw$.  Assume that
$\partial\cM$ is divided according to Eqs.~(\ref{b-DN})-(\ref{b-IDN}).
Then, consider boundary conditions for the scalar equation of
Dirichlet type on $\partial\cM_{\tiD}$ and of Robin type on
$\partial\cM_{\tiN}$. Analogously, consider boundary conditions for
the vector equation of Dirichlet type on $\partial\cM_{\tiID}$ and of
Robin type on $\partial\cM_{\tiIN}$. The conditions
\[
\overline{\partial\cM}_{\tiD}\cap\overline{\partial\cM}_{\tiN}
= \emptyset, \quad
\overline{\partial\cM}_{\tiID}\cap\overline{\partial\cM}_{\tiIN}
=\emptyset,
\]
are needed to simplify the proofs regarding the regularity at the
intersection points of the two types of boundaries of solutions of
elliptic equations with Dirichlet-Robin boundary conditions. In what
follows we include the cases given by $\partial\cM_{\tiD} =\emptyset$
or $\partial\cM_{\tiN} =\emptyset$, and $\partial\cM_{\tiID}
=\emptyset$ or $\partial\cM_{\tiIN} =\emptyset$.

The freely specifiable functions of the problem are a scalar function
$\tau$, interpreted as the trace of the physical extrinsic curvature;
a symmetric, traceless, and divergence-free, contravariant, two-index
tensor $\sigma$; the non-physical energy density $\rho$ and the
non-physical momentum current density vector $\bij$ subject to the
requirement $-\rho^2 + h_{ab}j^aj^b < 0$. The term non-physical refers
here to a conformal rescaled field, while physical refers to a
conformally non-rescaled field. The requirement on $\rho$ and $\bij$
mentioned above and the particular conformal rescaling used in the
semi-decoupling decomposition imply that the same inequality is
satisfied by the physical energy and momentum current densities.
Introduce the nonlinear operators $F:
C^{\infty}\times\biC^{\infty}\to C^{\infty}$ and $\IF
:C^{\infty}\to\biC^{\infty}$ given by
\begin{align}
\label{CF-def-F}
F(\phi,\biw) &:= a_{\tau} \phi^5 + a_{\tiR} \phi - a_{\rho} \phi^{-3}
- a_{w}\phi^{-7},\\
\label{CF-def-IF}
\IF (\phi) &:= \bib_{\tau} \, \phi^6 + \bib_{j},
\end{align}
where the coefficient functions are defined as follows
\begin{gather}
\label{CF-def-coeff1}
a_{\tau} := \frac{\tau^2}{12},\qquad
a_{\tiR} = \frac{R}{8},\qquad
a_{\rho} := \frac{\kappa}{4} \rho,\\
\label{CF-def-coeff2}
a_{w} := \frac{1}{8} \,(\sigma +\cL \biw)_{ab}(\sigma + \cL \biw)^{ab},
\qquad b_{\tau}^a := \frac{2}{3} \,\nabla^a \tau,\qquad
b_{j}^a := \kappa j^a.
\end{gather}
Notice that the scalar coefficients $a_{\tau}$, $a_{w}$, and
$a_{\rho}$ are non-negative, while there is no sign restriction on
$a_{\tiR}$.

The {\bf classical Dirichlet-Robin boundary value formulation} for the
semi-de\-cou\-pling Einstein constraint equations is the following:
Given the freely specifiable smooth fields $\tau$, $\sigma$,
$\rho$, and $\bij$ in $\cM$ and the smooth Dirichlet boundary data
$\hat\phi_{\tiD}$ on $\partial\cM_{\tiD}$ and $\hat \biw_{\tiID}$ on
$\partial\cM_{\tiID}$, and smooth Robin boundary data
$\hat\phi_{\tiN}$, $K$, scalar fields on $\partial\cM_{\tiN}$ and
$\hat\biw_{\tiIN}$, $\IK$, a vector and a two-index tensor fields on
$\partial\cM_{\tiIN}$, find a scalar field $\phi$ and a vector field
$\biw$ in $\cM$ solution of the system
\begin{align}
\label{CF-LYs}
L \phi + F(\phi,\biw) &= 0 \mbox{~in~} \cM,&
&\left\{\begin{aligned}
\phi &= \hat \phi_{\tiD}
\mbox{~on~} \partial \cM_{\tiD},\\
\bin\cdot\nabla\phi + K\phi &= \hat \phi_{\tiN} 
\mbox{~on~} \partial \cM_{\tiN},
\end{aligned}\right. \\
\label{CF-LYm}
\IL \biw  + \IF(\phi) &=0 \mbox{~in~} \cM,&
&\left\{\begin{aligned}
\biw &= \hat \biw_{\tiID}
\mbox{~on~} \partial\cM_{\tiID},\\
\bin \cdot \nabla \biw + \IK \biw &= \hat \biw_{\tiIN}
\mbox{~on~} \partial\cM_{\tiIN},
\end{aligned}\right.
\end{align}
where $\IK\biw$ denotes the vector field $(\IK\biw)^a =\IK^{ab}w_b$,
and $\bin\cdot\nabla\biw$ denotes the vector $n^b\nabla_bw^a$.

The classical formulation has been done on spaces of smooth fields,
which are not complete spaces under any known norm defined on them.
This is inconvenient for finding solutions to PDE, because these
solutions are usually found as limits of appropriate
approximations. If a normed vector space is not complete, then a
Cauchy sequence may not converge. In the next section we introduce the
weak formulation of the equations above, where we rewrite the
classical formulation in appropriate normed vector spaces which are
also complete.

\subsection{Weak formulation}
\label{S:WF}

We present the weak formulation associated with the classical
formulation with Eqs.~(\ref{CF-LYs})-(\ref{CF-LYm}). We introduce one
of the weakest forms of the constraint equations, that is, we assume
the weakest regularity of the equation coefficients such that the
equation itself is well-defined. We will be able to obtain existence
and uniqueness results for the momentum constraint using either
variational methods in \Sec\ref{S:MC-ExVM} or Riesz-Schauder methods
in \Sec\ref{S:MC-WP}. We will also be able to obtain existence (and
when possible, uniqueness) for the Hamiltonian constraint in
\Sec\ref{S:HC-ExVM} using variational methods in this weakest setting.
However, the barrier-based existence and uniqueness results for the
Hamiltonian constraint equation in \Sec\ref{S:WF-Existence-LB}, and
the compactness argument in \Sec\ref{S:CS-Ex} giving existence for the
coupled system of constraints, require higher regularity on the
equation coefficients, but we still obtain some non-CMC results for
the coupled system in weaker settings and in more general physical
situations than have been previously obtained.  These additional
assumptions are clearly stated in those sections.

Let $(\cM, h)$ be a 3-dimensional Riemannian manifold, where $\cM$ is
a smooth, compact manifold with Lipschitz boundary $\partial\cM$, and
$h\in C^2(\overline\cM ,2)$ is a positive definite metric. Introduce
the bilinear forms
\begin{align}
\label{WF-def-aL}
&a_{\tiL} : W^{1,2} \times W^{1,2} \to \R,&
a_{\tiL}(\phi,\un\phi) 
&:= (\nabla\phi,\nabla\un\phi) 
+ (K\, \Tr_{\tiN} \phi,\Tr_{\tiN}\un\phi)_{\tiN},\\
\label{WF-def-aIL}
&a_{\tiIL} : \biW^{1,2} \times \biW^{1,2} \to \R,&
a_{\tiIL}(\biw,\un\biw) 
&:= (\cL\biw,\cL\un\biw)
+ (\IK \,\Tr_{\tiIN} \biw, \Tr_{\tiIN}\un\biw )_{\tiIN},
\end{align}
where the Robin scalar field $K\in L^{\infty}(\partial\cM_{\tiN},0)$
and the two-index, symmetric tensor field $\IK\in
L^{\infty}(\partial\cM_{\tiIN},2)$ satisfy the bounds
\begin{align}
\label{WF-cttk}
\hat \ttk\, \|\Tr_{\tiN}\phi\|^2_{\tiN} &\leqs 
(K \Tr_{\tiN}\phi, \Tr_{\tiN}\phi)_{\tiN},
\qquad \forall \phi\in W^{1,2},\\
\label{WF-cttK}
\hat \ttK\, \|\Tr_{\tiIN}\biw\|^2_{\tiIN} &\leqs 
(\IK \Tr_{\tiIN}\biw, \Tr_{\tiIN}\biw)_{\tiIN},
\qquad \forall \biw\in \biW^{1,2},
\end{align}
with $\hat\ttk$ and $\hat\ttK$ being non-negative constants. In this
Section set the number $p=12/5$, and then fix the source functions
\begin{equation}
\label{WF-coeff}
\tau \in L^{p},\quad
\rho^{*} \in W^{-1,p}_{\tiD+},\quad
\sigma \in L^{p}(\cM,2),\quad
\bij^{*} \in \biW^{-1,p}_{\tiID},
\end{equation}
where $\sigma$ is symmetric, traceless and divergence-free in weak
sense, that is, it satisfies $(\sigma,\cL\un\bomega)=0$ for all
$\un\bomega\in\biW^{1,2}_0$. The asterisk on the matter fields is to
emphasize that they are elements of spaces of linear functionals. That
is, $\rho^{*} : W^{1,p'}_{\tiD}\to\R$ is linear and bounded, and an
analogous definition holds for $\bij^{*}$. In the Appendix it
is shown that the spaces $W^{1,q'}_{\tiD}\subset L^2\equiv
[L^2]^{*}\subset W^{-1,q}_{\tiD}$, with $1< q < \infty$, form a
Gelfand triple, so given any element $\rho^{*}\in W^{-1,p}_{\tiD+}$
there exists a sequence $\{\rho_n\}\subset L^2$ such that
\[
\rho^{*}(\un\varphi) := \lim_{n\to \infty}(\rho_n,\un\varphi),\qquad
\forall \, \un\varphi \in W^{1,p'}_{\tiD}.
\]
An analogous statement holds for $\bij^{*}$. We say that the matter
fields $\rho^{*}$ and $\bij^{*}$ satisfy the {\bf energy condition in
weak sense} iff there exist sequences $\{\rho_n\}\subset L^2$ and
$\{\bij_n\}\subset\biL^2$ such that
\begin{equation}
\label{WF-weakEC}
[\rho_n^2 - \bij_n\cdot\bij_n] \in L^1_{+}
\qquad \forall\, n\in \N.
\end{equation}
(We have required $\rho^{*}\in W^{-1,p}_{\tiD+}$ in
Eq.~(\ref{WF-coeff}) instead of $\rho^{*}\in W^{-1,p/2}_{\tiD+}$
because $\bij^{*}$ must belong to $\biW^{-1,p}_{\tiID}$ and
Eq.~(\ref{WF-weakEC}) must hold.) Given any function $\tau\in L^{p}$,
then it is known that $(\nabla\tau)^{*}\in\biW^{-1,p}_0$, where the
asterisk, we repeat for the last time, is added only to reinforce the
idea that $(\nabla\tau)^{*}$ is a linear functional on elements in
$W^{1,p'}_0$, and it is not meant to indicate the adjoint operator of
$\nabla$. We assume here that
$(\nabla\tau)^{*}\in\biW^{-1,p}_{\tiID}$, which is indeed an extra
assumption due to $\biW^{-1,p}_{\tiID}\subset\biW^{-1,p}_0$; this
assumption is needed because the functional $(b_{\tau}^a)^{*}:= (2/3)
(\nabla^a\tau)^{*}$ is the source in the momentum constraint equation,
which requires this particular type of boundary conditions. The
assumptions above on $\tau$ and $\sigma$ imply that for every
$\biw\in\biW^{1,p}$ the functions $a_{\tau}$ and $a_{w}$ belong to
$L^{p/2}$. The assumption on the background metric implies that
$a_{\tiR}$ is a continuous function on $\overline\cM$. These three
functions define the elements $a_{\tau}^{*}$, $a_{\tiR}^{*}$ and
$a_{w}^{*}$ in the space $W^{-1,2}_{\tiD}$ as follows
\begin{equation}
\label{WF-dual-coeff}
a_{\tau}^{*}(\un\varphi):=(a_{\tau},\un\varphi),\quad
a_{\tiR}^{*}(\un\varphi):= (a_{\tiR},\un\varphi),\quad
a_{w}^{*}(\un\varphi):=(a_{w},\un\varphi),\quad
\forall \, \un\varphi \in W^{1,2}_{\tiD}.
\end{equation}
The proof that these functionals are well-defined is based on H\"older
inequality, for example, consider the functional $a_{\tau}^{*}$, then
\[
|(a_{\tau},\un\varphi)| 
\leqs \|a_{\tau}\|_{\frac{6}{5}}\, \|\un\varphi\|_{6}
\leqs c_s\, \|a_{\tau}\|_{\frac{6}{5}} \, \|\un\varphi\|_{1,2}.
\]
These functionals above belong to a particular class of elements in
$W^{-1,2}_{\tiD}$, while the functionals
$a_{\rho}^{*}:=(\kappa/4)\rho^{*}$ and $\bib_{j}^{*} :=\kappa\bij^{*}$
are not restricted to such a particular form, and they can be any
element in $W^{-1,p}_{\tiD+}$ and $\biW^{-1,p}_{\tiID}$,
respectively compatible with the energy condition. Given any two
functions $\phi_1$, $\phi_2 \in L^{\infty}$ with $\phi_1\leqs\phi_2$,
define the interval
\[
[\phi_1,\phi_2] 
:= \{ \phi \in L^{\infty} : \phi_1 \leqs \phi \leqs \phi_2\} 
\subset L^{\infty},
\]
which is a closed, bounded set in $L^{\infty}$. Assume 
$\phi_1>0$, and then introduce the nonlinear operators
\[
f_{\tiF}: [\phi_1,\phi_2]\subset L^2 \times \biW^{1,p} 
\to W^{-1,2}_{\tiD},\qquad
\bif_{\tiIF} : [\phi_1,\phi_2]\subset L^2\to\biW^{-1,p}_{\tiID},
\]
\begin{align}
\label{WF-x0}
f_{\tiF}(\phi,\biw) &:=  (a_{\tau} \phi^5)^{*} + (a_{\tiR} \phi)^{*}
- (a_{\rho}\phi^{-3})^{*} - (a_{w}\phi^{-7})^{*},\\
\label{WF-x1}
\bif_{\tiIF}(\phi) &:=  (\bib_{\tau}\phi^6)^{*} + \bib_{j}^{*},
\end{align}
where the product of an element $\phi\in L^{\infty}$ by an element in
$a^{*}\in W^{-1,q}$, with $1<q<\infty$, denoted by $(a\phi)^{*}$, is a
well-defined element in $W^{-1,q}$. The proof of this statement is
given in the Appendix using appropriate Gelfand triple structures. The
functionals $f_{\tiF}$ and $\bif_{\tiIF}$ are the generalizations of
the functionals $F$ and $\IF$ defined in
Eq.~(\ref{CF-def-F})-(\ref{CF-def-IF}). We remark that the operators
defined in Eqs.~(\ref{WF-x0})-(\ref{WF-x1}) are continuous but {\em
not G\^ateaux differentiable}. They have G\^ateaux derivatives only
along directions in $L^{\infty}$, not on the whole space $L^2$.  This
fact introduces some technical complexity with the use of variational
methods for the individual Hamiltonian and momentum constraints (see
\Sec\ref{S:MC-ExVM} and \Sec\ref{S:HC-ExVM}).  Recall that the trace
operators
\[
\Tr_{\tiD}: W^{1,2} \to W^{\frac{1}{2},2}(\partial\cM_{\tiD},0),
\qquad
\Tr_{\tiID}: \biW^{1,p} \to 
W^{\frac{1}{p'},p}(\partial\cM_{\tiID},1),
\]
satisfy the following property: given any element $\hat\phi_{\tiD} \in
W^{\frac{1}{2},2} (\partial\cM_{\tiD},0)$, there exists an element
$\phi_{\tiD} \in W^{1,2}$ such that $\Tr_{\tiD}\phi_{\tiD}
=\hat\phi_{\tiD}$; analogously, given a boundary data element
$\hat\biw_{\tiID}\in W^{\frac{1}{p'},p}(\partial\cM_{\tiID},1)$,
there exists an element $\biw_{\tiID}\in\biW^{1,p}$ such that
$\Tr_{\tiID}\biw_{\tiID} =\hat\biw_{\tiID}$. The elements
$\phi_{\tiD}$ and $\biw_{\tiID}$ are called here extensions of
$\hat\phi_{\tiD}$ and $\hat\biw_{\tiID}$, respectively. They are not
uniquely determined by the boundary data.

The {\bf weak Dirichlet-Robin boundary value formulation} which is
associated with Eqs.~(\ref{CF-LYs})-(\ref{CF-LYm}) is the following:
Fix Dirichlet boundary data
\[
0 < \mbox{ess}\inf_{\partial\cM_{\tiD}}\hat\phi_{\tiD} 
\leqs\hat\phi_{\tiD} \in  L^{\infty}(\partial\cM_{\tiD},0)\cap 
W^{\frac{1}{2},2}(\partial\cM_{\tiD},0),
\quad
\hat\biw_{\tiID} \in W^{\frac{1}{p'},p}(\partial\cM_{\tiID},1),
\]
with extensions
$\inf_{\partial\cM_{\tiD}}\hat\phi_{\tiD}\leqs\phi_{\tiD}\in W^{1,2}$
and $\biw_{\tiID}\in\biW^{1,p}$, respectively.  Choose the extension
function $\phi_{\tiD}$ as a harmonic extension of the Dirichlet
boundary data using the Laplace-Beltrami operator on $\cM$:
\[
\Delta \phi_{\tiD} = 0 \mbox{~in~}\cM,\quad
\Tr_{\tiD}\phi_{\tiD} =  \hat\phi_{\tiD} > 0
\mbox{~on~} \partial\cM_{\tiD},\quad
\Tr_{\tiN}\phi_{\tiD} = \inf_{\partial\cM_{\tiD}}\hat\phi_{\tiD} > 0
\mbox{~on~}\partial\cM_{\tiN}.
\]
The maximum principle for the Laplace-Beltrami operator~\cite{Aubin82}
implies that
$0<\inf_{\partial\cM}\hat{\phi}_{\tiD}\leqs\phi_{\tiD}(x)$, a.e. in
$\cM$. Fix Robin boundary data functionals
\[
\hat \phi_{\tiN}^{*} \in  W^{-\frac{1}{2},2}(\partial\cM_{\tiN},0),
\qquad
\hat\biw_{\tiIN}^{*} \in W^{-\frac{1}{p},p}(\partial\cM_{\tiIN},1);
\]
Given the extension $\phi_{\tiD}$ of the Dirichlet data
$\hat\phi_{\tiD}$ chosen above, fix any two functions $\phi_1$,
$\phi_2 \in L^{\infty}\cap W^{1,2}$, with the property that
$0<\phi_1\leqs\phi_2$ and satisfying $\phi_{\tiD}\in
[\phi_1,\phi_2]\cap W^{1,2}$.  Introduce the non-principal part
operators including the Robin boundary conditions,
\begin{align}
\label{WF-def-f}
&f: [\phi_1,\phi_2]\subset L^2 \times \biW^{1,p} 
\to W^{-1,2}_{\tiD},&
f(\phi,\biw)(\un\varphi) 
&:=  f_{\tiF}(\phi,\biw)(\un\varphi)
- \hat \phi_{\tiN}^{*}(\Tr_{\tiN}\un\varphi),\\
\label{WF-def-If}
&\bif : [\phi_1,\phi_2] \subset L^2 \to \biW^{-1,p}_{\tiID},&
\bif(\phi)(\un\bomega) 
&:= \bif_{\tiIF}(\phi)(\bomega)
- \hat \biw_{\tiIN}^{*}(\Tr_{\tiIN}\un\bomega),
\end{align}
where $f_{\tiF}$ and $\bif_{\tiIF}$ are given by
Eqs.~(\ref{WF-x0})-(\ref{WF-x1}). Introduce the affine spaces
$A^{1,2}$ and $\biA^{1,p}$, which include the Dirichlet boundary
conditions, as follows,
\begin{align}
\label{WF-def-A}
A^{1,2} & := \phi_{\tiD}  + W^{1,2}_{\tiD} 
:= \{ \phi \in W^{1,2} : \phi-\phi_{\tiD}\in W^{1,2}_{\tiD}\},\\ 
\label{WF-def-IA}
\biA^{1,p} &:= \biw_{\tiID} + \biW^{1,p}_{\tiID} := 
\{ \biw \in \biW^{1,p} : \biw -\biw_{\tiID} \in \biW^{1,p}_{\tiID}\}.
\end{align}
Then, find elements $\phi\in [\phi_1,\phi_2]\cap A^{1,2}$ and
$\biw\in\biA^{1,p}$ solutions of
\begin{align}
\label{WF-LYs}
a_{\tiL}(\phi,\un\varphi) + f(\phi,\biw)(\un\varphi) &= 0
\qquad\forall \, \un\varphi\in W^{1,2}_{\tiD},\\
\label{WF-LYm}
a_{\tiIL}(\biw,\un\bomega) + \bif(\phi)(\un\bomega) &= 0
\qquad\forall\, \un\bomega\in \biW^{1,p'}_{\tiID}.
\end{align}

It will be convenient later on to express
Eqs.~(\ref{WF-LYs})-(\ref{WF-LYm}) in terms of operators instead of
bilinear forms. Introduce the operators
\begin{gather}
\label{WF-def-AL}
A_{\tiL}:W^{1,2}\to W^{-1,2}_{\tiD},\qquad
A_{\tiL}\phi(\un\varphi) := a_{\tiL}(\phi,\un\varphi),\\
\label{WF-def-AIL}
A_{\tiIL}:\biW^{1,p}\to \biW^{-1,p}_{\tiID},\qquad
A_{\tiIL}\biw(\un\bomega) := a_{\tiIL}(\biw,\un\bomega).
\end{gather}
Also recall that given any $\phi\in [\phi_1,\phi_2]$ and
$\biw\in\biW^{1,p}$ then $f(\phi,\biw)\in W^{-1,2}_{\tiD}$ and
$\bif(\phi)\in\biW^{-1,p}_{\tiID}$. Hence, the
Eqs.~(\ref{WF-LYs})-(\ref{WF-LYm}) written in terms of operators is
the following: Find elements $\phi\in [\phi_1,\phi_2]\cap A^{1,2}$
and $\biw\in\biA^{1,p}$ solutions of
\begin{align}
\label{WF-LYs1}
A_{\tiL}\phi + f(\phi,\biw) &= 0,\\
\label{WF-LYm1}
A_{\tiIL}\biw + \bif(\phi) &= 0.
\end{align}

\begin{lemma}
\label{L:CtoW}
Every smooth solution $\phi$, $\tbw$ of the classical problem with
Eqs.~(\ref{CF-LYs})-(\ref{CF-LYm}) is also a solution of the weak problem
with Eqs.~(\ref{WF-LYs})-(\ref{WF-LYm}).
\end{lemma}

\Proof {\it (Lemma~\ref{L:CtoW}.)~}
Given any smooth fields $\phi$, $\biw$ solutions of
Eqs.~(\ref{CF-LYs})-(\ref{CF-LYm}), then the proof consists in
multiplying these equations by test functions $\un\varphi\in
W^{1,2}_{\tiD}$ and $\un\bomega\in\biW^{1,p'}_{\tiID}$, respectively,
and then integrating by parts. In the case of Eq.~(\ref{CF-LYs}) one
gets
\begin{equation}
\label{LYs1}
(-\Delta \phi,\un\varphi) +
\bigl(F(\phi,\biw),\un\varphi\bigr) =0.
\end{equation}
The first term on the left hand side can be rewritten as follows,
\begin{align*}
\bigl( -\Delta\phi, \un\varphi\bigr)
&= (\nabla\phi,\nabla \un\varphi)
-\bigl(\Tr_{\tiN}(\bin \cdot \nabla\phi),
\Tr_{\tiN}\un\varphi\bigr)_{\tiN}\\
&=(\nabla\phi,\nabla\un\varphi)
+ \bigl([K \,\Tr_{\tiN}\phi - \hat \phi_{\tiN}], 
\Tr_{\tiN} \un\varphi \bigr)_{\tiN}\\
&= a_{\tiL}(\phi,\un\varphi) 
-(\hat \phi_{\tiN}, \Tr_{\tiN} \un\varphi)_{\tiN},
\end{align*}
which holds for all $\un\varphi\in W^{1,2}_{\tiD}$ where in the second
line we introduce the Robin boundary condition on
$\partial\cM_{\tiN}$, and note the integral on $\partial\cM_{\tiD}$
vanishes because the test function $\un\varphi$ vanishes on this part
of the boundary, in the third line we introduce the definition of the
bilinear form $a_{\tiL}$. Now, replace this expression into
Eq.~(\ref{LYs1}) and one obtains Eq.~(\ref{WF-LYs}). Finally, since
$\phi$ is a solution of the classical problem, it can be written as
$\phi =\phi_{\tiD} +\varphi$ for some smooth extension $\phi_{\tiD}$
of the boundary data $\hat\phi_{\tiD}$, therefore $\phi\in A^{1,2}$.
(Again, $\phi_{\tiD}$ can be constructed e.g. by harmonic extension.)
In the case of Eq.~(\ref{CF-LYm}) one gets
\begin{equation}
\label{LYm1}
\bigl( -\nabla\cdot (\cL\biw),\un\bomega\bigr)
+ \bigl(\IF(\phi),\un\bomega \bigr) =0. 
\end{equation}
The first term on the left hand side can be rewritten as follows,
\begin{align*}
\bigl( -\nabla\cdot (\cL\biw),\un\bomega\bigr)
&= \bigl(\cL\biw,\nabla \un\bomega\bigr)
-\bigl(\Tr_{\tiIN}[\bin \cdot (\cL\biw)],
\Tr_{\tiIN}\un\bomega\bigr)_{\tiIN}\\
&=\bigl(\cL\biw,\cL\un\bomega\bigr)
+ \bigl([\IK \Tr_{\tiIN}\biw - \hat \biw_{\tiIN}], 
\Tr_{\tiIN} \un\bomega \bigr)_{\tiIN}\\
&= a_{\tiIL}(\biw,\un\bomega) 
- \bigl(\hat \biw_{\tiIN}, \Tr_{\tiIN}\un\bomega 
\bigr)_{\tiIN},
\end{align*}
which holds for all $\un\bomega\in \biW^{1,p'}_{\tiID}$, where the
first term in the second line comes from the symmetries of $\cL$, and
the second term in that line comes from the Robin boundary conditions;
the definition of $a_{\tiIL}$ is used to obtain the third line. Now,
replace this expression into Eq.~(\ref{LYm1}) and one obtains
Eq.~(\ref{WF-LYm}). Finally, since $\biw$ is a solution of classical
problem, it can be written as $\biw =\biw_{\tiID} +\bomega$ for some
smooth extension $\biw_{\tiID}$ of the boundary data
$\hat\biw_{\tiID}$, therefore $\biw\in\biA^{1,p}$.
\qed

Let us recall here that the space $W^{1,2}_{\tiD}$ is an ordered
Banach space with order cone $W^{1,2}_{\tiD +}$ defined as follows:
\[
W^{1,2}_{\tiD +} := \{ \phi\in W^{1,2}_{\tiD} :  \phi\geqs 0 
\mbox{ a.e. in } \cM\}.
\]
The order relation is then $\phi\geqs\un\phi$ iff $\phi-\un\phi\in
W^{1,2}_{\tiD +}$. In the Appendix we discuss the main
properties of ordered Banach spaces. In particular, we show that the
order structure implied by $W^{1,2}_{\tiD +}$ can be translated to the
dual space $W^{-1,2}_{\tiD}$ as follows,
\[
W^{-1,2}_{\tiD +} := \bigl\{ \phi^{*} \in W^{-1,2}_{\tiD} :
\phi^{*}(\un\phi)\geqs 0 \quad \forall \, \un\phi\in W^{1,2}_{\tiD +}
\, \bigr \}.
\]
Given two ordered Banach spaces $X$, $X_{+}$ and $Y$, $Y_{+}$ an
operator $A: D_A\subset X\to Y$ satisfies the {\bf maximum principle}
iff for every elements $u$, $v\in D_A$ such that $Au-Av\in Y_{+}$ it
holds that $u-v\in X_{+}$. In the particular case that the operator
$(A,D_A)$ is linear, then it satisfies the maximum principle iff for
every element $u\in X_{+}$ such that $Au\in Y_{+}$ it holds that $u\in
X_{+}$. If an operator $A$ satisfies the maximum principle and is
invertible, then the inverse is a monotone increasing operator, a
result shown in the Appendix. This last property is useful to solve
nonlinear equations of the form $Au=f(u)$, in the case that there
exist sub- and super-solutions to that equation (see below for the
definition). In this case there is a well-known existence proof
technique that works for many equations of this type, and has been one
of the main techniques used previously for the Hamiltonian
constraint~\cite{rBjI04,jI95,jIvM96,dM05}.  While we will exploit the
fact that the construction of sub- and super-solutions can be done in
a very weak setting, the use of the existence proof based directly on
barriers requires additional regularity beyond what is needed for the
barrier construction.  This additional regularity assumption can be
avoided by combining barriers with variational techniques, which we do
in \Sec\ref{S:HC-ExVM}.

The following properties are of interest to us below. Firstly, in the
Appendix we review results from the literature showing that the
operator $A_{\tiL}$ defined above satisfies a maximum
principle. Secondly, we can show that there exist sub- and
super-solutions to Eq.~(\ref{WF-LYs1}). Given any function $u\in
W^{1,2}$, introduce the notation
\[
u^{+}:= \mbox{ess~max}\{u,0\},\qquad
u^{-}:=-\mbox{ess~min}\{u,0\}.
\]
An element $\phi_{-}\in W^{1,2}$ is called a {\bf sub-solution} of
Eq.~(\ref{WF-LYs1}) iff the function $\phi_{-}$ satisfies the
inequalities
\begin{equation}
\label{w-WF1-sub-sol}
(\phi_{\tiD}-\phi_{-})^{-} \in W^{1,2}_{\tiD} \mbox{~~~and~~~}
-\bigl[ A_{\tiL}\phi_{-} + f(\phi_{-},\biw)\bigr] 
\in W^{-1,2}_{\tiD +}.
\end{equation}
An element $\phi_{+} \in W^{1,2}$ is called a {\bf super-solution} of
Eq.~(\ref{WF-LYs1}) iff the scalar function $\phi_{+}$ satisfies the
inequalities
\begin{equation}
\label{w-WF1-super-sol}
(\phi_{\tiD}-\phi_{+})^{+} \in W^{1,2}_{\tiD} \mbox{~~~and~~~}
\bigl[ A_{\tiL} \phi_{+} + f(\phi_{+},\biw) \bigr]
\in W^{-1,2}_{\tiD +}.
\end{equation}
The sub and super-solutions of Eq.~(\ref{WF-LYs1}) may depend on the
choice of the vector field $\biw$ that appears in the functional
$a_{w}^{*}$. A sub-solution $\phi_{-}$ of Eq.~(\ref{WF-LYs1}) is
called {\bf global} iff Eq.~(\ref{w-WF1-sub-sol}) holds for every
vector field $\biw\in\biW^{1,p}$ solution of the momentum constraint
Eq.~(\ref{WF-LYm1}) with any source function $\phi$ satisfying $(\phi-
\phi_{-})\in W^{1,2}_{\tiD+}$, and it is called {\bf local} iff it is
not global. Analogous definitions are introduced for super-solutions.
While it will be sufficient to derive only local sub- and
super-solutions to produce the existence and uniqueness results for
the Hamiltonian constraint using variational methods in
\Sec\ref{S:HC-ExVM} and using barrier methods in
\Sec\ref{S:WF-Existence-LB}, proving results for the coupled system
rests critically on deriving {\em global} sub- and super-solutions for
this coupled system; we come back to this in \Sec\ref{S:CS}.

\section{The momentum constraint}
\label{S:MC}

In this section we fix a particular scalar function $\phi\in
L^{\infty}$ and consider the momentum constraint
equation~(\ref{WF-LYm1}) for the vector valued function
$\biw\in\biW^{1,2}$. The result is a linear elliptic system of
equations for this variable $\biw$.  We first develop the weak
formulation of the momentum constraint more precisely in
\Sec\ref{S:SMC}.  In \Sec\ref{S:GKI} we establish generalized Korn
inequalities for the conformal Killing operator on compact manifolds
with boundary under several boundary condition scenarios; the results
do not appear to be in the literature.  We then briefly summarize here
the main ideas for solving the Dirichlet-Robin problems for the
momentum constraint equation, for an appropriately given $\phi$.  We
use two different methods, namely variational
methods~\cite{Jost-LiJost98,Struwe96,Zeidler-I}, and Riesz-Schauder
theory for compact operators~\cite{Wloka87}. Both methods yield
essentially the same results, since the momentum constraint equation
is linear in the variable $\biw$.

The variational approach is taken in \Sec\ref{S:MC-ExVM} when the
Dirichlet part of the boundary is non-empty, giving existence and
uniqueness of weak solutions to the momentum constraint in $W^{1,2}$.
The weak assumptions on the data do not allow for the use of standard
techniques to establish additional regularity.  While the variational
approach has no real advantage over Riesz-Schauder theory for the
momentum constraint, it will give us some insight in its use for the
Hamiltonian constraint, for which it will be critical.  In addition,
some of the supporting results are of interest in their own right, so
we include the analysis using variational methods here along with the
Riesz-Schauder arguments.  In \Sec\ref{S:MC-WP} we establish existence
and uniqueness of solutions to the Dirichlet-Robin problem for the
momentum constraint equation using Riesz-Schauder theory for compact
operators.  The literature on Riesz-Schauder theory for systems of
elliptic equations is not so clearly presented as it is for scalar
equations, so we summarize it here. The main ideas in this method
include establishing a G\aa rding inequality for a bilinear form
associated to the principal part of the equation in the appropriate
function spaces, and then transforming the problem into one involving
a Fredholm operator.
Finally, regularity of solutions to the momentum constraint is
discussed briefly in \Sec\ref{S:MC-RS}.

\subsection{Weak formulation}
\label{S:SMC}

Let $(\cM, h)$ be a 3-dimensional Riemannian manifold, where $\cM$ is
a smooth, compact manifold with Lipschitz boundary $\partial\cM$, and
$h\in C^2(\overline\cM ,2)$ is a positive definite metric. Introduce
the bilinear form
\begin{equation}
\label{MC-def-aIL}
a_{\tiIL} : \biW^{1,2} \times \biW^{1,2} \to \R,\qquad
a_{\tiIL}(\biw,\un\biw) := (\cL\biw,\cL\un\biw) 
+ \bigl(\IK  \Tr_{\tiIN}\biw, \Tr_{\tiIN} \un\biw \bigr)_{\tiIN},
\end{equation}
where the Robin tensor field $\IK\in
L^{\infty}(\partial\cM_{\tiIN},2)$ is symmetric and satisfies the
bound
\begin{equation}
\label{MC-cttk}
\hat \ttK\, \|\Tr_{\tiIN}\biw\|^2_{\tiIN} \leqs 
(\IK \Tr_{\tiIN}\biw, \Tr_{\tiIN}\biw)_{\tiIN},
\qquad \forall \biw\in \biW^{1,2},
\end{equation}
and where $\hat\ttK$ is a non-negative constant. Fix the functionals
$\bib_{\tau}^{*}$, $\bib_{j}^{*} \in \biW^{-1,2}_{\tiID}$.  Fix a
function $\phi\in L^{\infty}$ and introduce the linear functional
\begin{equation}
\bif_{\phi\tiF} \in \biW^{-1,2}_{\tiID},\qquad
\label{MC-x0}
\bif_{\phi\tiF} :=  (\bib_{\tau} \phi^6)^{*} + \bib_{j}^{*},
\end{equation}
We used the subscript $\phi$ in $\bif_{\phi\tiF}$ to emphasize that
$\phi$ is not a variable of the problem. The functional
$\bif_{\phi\tiF}$ is a generalization of the functional $\IF$ defined
in Eq.~(\ref{CF-def-IF}).

The {\bf weak Dirichlet-Robin boundary value formulation} for the
momentum constraint is the following: Fix Dirichlet and Robin
boundary data
\begin{equation}
\label{MC-BD}
\hat \biw_{\tiID} \in W^{\frac{1}{2},2}(\partial\cM_{\tiID},1),
\quad
\hat \biw_{\tiN}^{*} \in W^{-\frac{1}{2},2}(\partial\cM_{\tiIN},1),
\end{equation}
and introduce an extension $\biw_{\tiID}$ of the Dirichlet boundary
data as described in \Sec\ref{S:WF}; Introduce the non-principal part
operator including the Robin boundary conditions,
\begin{equation}
\bif_{\phi} \in \biW^{-1,2}_{\tiID},\qquad
\label{MC-def-bf}
\bif_{\phi}(\un\bomega) :=  \bif_{\phi\tiF}(\un\bomega)
- \hat \biw_{\tiIN}^{*}(\Tr_{\tiIN}\un\bomega),
\end{equation}
where $\bif_{\phi\tiF}$ is given by Eq.~(\ref{MC-x0}); Let
$\biA^{1,2}$ be the affine space given in Eq.~(\ref{WF-def-IA}) for the
case $p=2$; Then, find an element $\biw\in\biA^{1,2}$ solution of
\begin{equation}
\label{MC-LYm}
a_{\tiIL}(\biw,\un\bomega) + \bif_{\phi}(\un\bomega) = 0
\qquad \forall\,\un\bomega\in \biW^{1,2}_{\tiID}.
\end{equation}

It is convenient to express Eq.~(\ref{MC-LYm}) in terms of operators
instead of bilinear forms. Introduce the operator
\[
A_{\tiIL} : \biW^{1,2}\to \biW^{-1,2}_{\tiID},\qquad
A_{\tiIL}\biw(\un\bomega) := a_{\tiIL}(\biw,\un\bomega).
\]
Hence, Eq.~(\ref{MC-LYm}) written in terms of operators is the
following: find an element $\biw\in\biA^{1,2}$ solution of
\begin{equation}
\label{MC-LYm1}
A_{\tiIL}\biw + \bif_{\phi} = 0.
\end{equation}

\begin{lemma}
\label{L:M-CtoW}
Every smooth solution $\tbw$ of the classical Eq.~(\ref{CF-LYm}) for a
given smooth function $\phi$ is also a solution of the
Eqs.~(\ref{MC-LYm}).
\end{lemma}

\Proof {\it (Lemma~\ref{L:M-CtoW}.)~}
The proof is similar to the proof of Lemma~\ref{L:CtoW}, and we do not
reproduce it here. 
\qed

\subsection{Generalized Korn's inequalities}
\label{S:GKI}

The {\bf Korn inequalities} are a fundamental step in proving
existence of solutions to the linearized displacement-traction
equations in elasticity. The inequalities involve the Killing operator
$\ell :\biW^{1,2} \to L^2(\cM,2)$ with action $(\ell \biu)_{ab}
:=\nabla_a u_b + \nabla_b u_a$. There are two main inequalities,
called ``without'' or ``with boundary conditions'', which can be
described in terms of the bilinear form $a_{\ell}
:\biW^{1,2}\times\biW^{1,2}\to\R$ with action $a_{\ell}(\biu ,\biv):=
(\ell\biu ,\ell\biv)$. The former inequality says that the bilinear
form $a_{\ell}$ satisfies {\bf G\aa rding's inequality}, that is,
there exists $k_0>0$ such that
\[
k_0 \, \|\biu\|_{1,2}^2 \leqs \|\biu\|^2 + \|\ell\biu\|^2
\qquad\forall\, \biu \in \biW^{1,2}.
\]
The latter inequality says that the bilinear form $a_{\ell}$ is {\bf
coercive} in the space $\biW^{1,2}_{\tiID}$ in the case that
$\mbox{meas}(\partial\cM_{\tiID})\neq\emptyset$, that is, there exists
a constant $k_0>0$ such that
\[
k_0 \, \|\biu\|_{1,2}^2 \leqs \|\ell\biu\|^2
\qquad\forall\, \biu \in \biW^{1,2}_{\tiID}.
\]
These inequalities were first established in the case that the
manifold $\cM\subset\R^3$ and the metric $h_{ab}$ is the Euclidean
metric~\cite{aK1908,aK1909}, with new proofs given in~\cite{kF47}. A
review of elasticity theory is nicely presented in~\cite{Ciarlet97}
with Korn's inequalities discussed on Volume~II, pages~10-13. See
also~\cite{jN81}. Both types of Korn's inequalities for the Killing
operator have been generalized to Riemannian manifolds
in~\cite{wCjJ02}.

We just mention here that the G\aa rding type inequality on the
particular case of the spaces $W^{1,2}_0(\cM,n)$ can be proven for a
general class of bilinear forms called strongly
elliptic. See~\cite{Zeidler-IIA}, exercise 22.7b, page 396. A bilinear
form $a: W_0^{1,2}(\cM,n)\times W_0^{1,2}(\cM,n)\to\R$ with action
\begin{align*}
a(u,v)&= \int_{\cM} a_{ac_1\cdots c_nbd_1\cdots d_n}
\nabla^a u^{c_1\cdots c_n} \nabla^b v^{d_1\cdots d_n} \,dx\\
&\quad +\int_{\cM} b_{c_1\cdots c_nd_1\cdots d_n}
u^{c_1\cdots c_n} v^{d_1\cdots d_n} \,dx
\end{align*}
is {\bf strongly elliptic} iff there exists a positive constant
$\alpha_0$ such that
\[
a_{ac_1\cdots c_nbd_1\cdots d_n} \zeta^a\zeta^b 
u^{c_1\cdots c_n} u^{d_1\cdots d_n} \geqs\alpha_0\,
\zeta_a\zeta^a\, u_{c_1\cdots c_n} u^{c_1\cdots c_n}
\]
for all vectors $\zeta \in \R^3$ and all tensors $u_{c_1\cdots
c_n}\in\R^{3n}$. An example of a strongly elliptic form is the
bilinear form $a_{\ell}$.

The role played in elasticity theory by the Killing operator $\ell$ is
played in the momentum constraint Eq.~(\ref{WF-LYm}) by the conformal
Killing operator $\cL$, which is defined in~Eq.(\ref{CF-def-CK}).
Inequalities similar to those satisfied by the Killing operator can be
obtained for the conformal Killing operator, called here {\bf
generalized Korn's inequalities}.  First notice that the bilinear form
$a_{\cL} :\biW^{1,2}_{0}\times\biW^{1,2}_0\to\R$ given by
$a_{\cL}(\biu,\biv) = (\cL\biu,\cL\biv)$ is strongly elliptic, as the
following calculation shows:
\begin{gather*}
\bigl[\zeta^a u^c + \zeta^c u^a 
-\frac{2}{3} h^{ac} (\zeta_d u^d)\bigr]
\bigl[\zeta_a u_c + \zeta_c u_a 
-\frac{2}{3} h_{ac} (\zeta_e u^e)\bigr] 
\\
=2 (\zeta_a\zeta^a)(u_bu^b) 
+ \frac{2}{3}\, (\zeta_au^a)^2 
\geqs 2 (\zeta_a\zeta^a)(u_bu^b). 
\end{gather*}
Hence, a G\aa rding type inequality is satisfied by the bilinear form
$a_{\cL}$ on the Hilbert space $\biW^{1,2}_0$. However, this space is
too small in our case where we need the same inequality on the space
$\biW^{1,2}$. In addition, later we will need the coercivity type
inequality for the bilinear form $a_{\cL}$ on the space
$\biW^{1,2}_{\tiID}$.

We first review the generalized Korn inequality without boundary
conditions, which has been proven in~\cite{sD06} in the case where
$\cM\subset\R^n$, with $n\geqs 3$, and $h_{ab}$ is the Euclidean
metric. It is also shown in~\cite{sD06} that the inequality does not
hold for $n=2$ where the null space of the conformal Killing operator
is infinite dimensional. It is also mentioned in that article that the
same arguments given in~\cite{wCjJ02} imply that the generalized Korn
inequality without boundary conditions also holds on a Riemannian
manifold. We summarize these ideas in the following result.
\begin{lemma}
\label{L:GKI}{\bf (G\aa rding's inequality for $\cL$)}
Let $(\cM,h_{ab})$ be a 3-dimensional, compact, Riemannian manifold,
with Lipschitz boundary, and with a metric $h\in
C^2(\overline\cM,2)$. Then, there exists a positive constant $k_0$
such that the following inequality holds
\begin{equation}
\label{GKI}
k_0 \,\|\tbu\|^2_{1,2} \leqs \|\tbu\|^2 + \|\cL\tbu\|^2
\qquad \forall \tbu \in \tbW^{1,2}.
\end{equation}
\end{lemma}

\Proof {\it (Lemma~\ref{L:GKI}.)~}
See~\cite{sD06} for the proof.
\qed

Using Lemma~\ref{L:GKI} it is not
difficult to establish that the same type of inequality is satisfied
by the bilinear form $a_{\tiIL}$.
\begin{corollary}
\label{C:GI}{\bf(G\aa rding's inequality for $a_{\tiIL}$)}
Let $(\cM,h_{ab})$ be a 3-dimensional, compact, Riemannian manifold,
with Lipschitz boundary and with a metric $h\in
C^2(\overline\cM,2)$. Let $a_{\tiIL}$ be the bilinear form defined in
Eq.~(\ref{MC-def-aIL}) for any tensor $\IK\in
L^{\infty}(\partial\cM_{\tiIN},2)$. Then, there exists a positive
constant $k_1$ such that the following inequality holds
\begin{equation}
\label{GI}
k_1 \,\|\tbu\|^2_{1,2} \leqs \|\tbu\|^2 + a_{\tiIL}(\tbu,\tbu)
\qquad \forall \tbu \in \tbW^{1,2}.
\end{equation}
\end{corollary}

\Remark 
This result holds for both cases $\partial\cM_{\tiID}\neq\emptyset$
and $\partial\cM_{\tiID}=\emptyset$, and also notice that the Robin
tensor field $\IK$ is arbitrary; we do not require this tensor to be
positive definite.

\Proof {\it (Corollary~\ref{C:GI}.)~}
The definition of the bilinear form in Eq. (\ref{MC-def-aIL}) implies
\[
a_{\tiIL}(\biu,\biu) = \|\cL\biu\|^2
+ (\IK \Tr_{\tiIN} \biu,\Tr_{\tiIN}\biu)_{\tiIN}
\quad \forall \biu\in \biW^{1,2}.
\]
Recalling that $\IK \in L^{\infty}(\partial\cM_{\tiIN},2)$, then
the second term on the right hand side can be bounded as follows:
\begin{align*}
(\IK \Tr_{\tiIN} \biu,\Tr_{\tiIN}\biu)_{\tiIN}
&\leqs \|\IK\|_{\infty} \,\| \Tr_{\tiIN}\biu\|^2_{\tiIN}\\
&\leqs \|\IK\|_{\infty} \,\|\biu\| ~\|\nabla \biu\|\\
&\leqs  \frac{1}{2} \|\IK\|_{\infty}\, \Bigl(
\frac{1}{\epsilon^2} \|\biu\|^2 + \epsilon^2 \,\|\nabla\biu\|^2 \Bigr),
\end{align*}
for every non-zero number $\epsilon$. Let $\tilde k_1 :=
\|\IK\|_{\infty}/2$, and then compute
\begin{align*}
a_{\tiIL}(\biu,\biu) &\geqs \|\cL\biu\|^2
-\tilde k_1 \Bigl( \frac{1}{\epsilon^2} \,\|\biu\|^2 
+ \epsilon^2\|\nabla\biu\|^2 \Bigr)\\
&\geqs -\|\biu\|^2 + k_0 \|\biu\|^2_{1,2}
-\tilde k_1 \Bigl( \frac{1}{\epsilon^2}\, \|\biu\|^2 
+ \epsilon^2 \|\biu\|^2_{1,2} \Bigr)\\
&\geqs -\bigl(1+\frac{\tilde k_1}{\epsilon^2} \bigr) 
\|\biu\|^2 + (k_0 -\tilde k_1\epsilon^2) \|\biu\|^2_{1,2}.
\end{align*}
Choose the number $\epsilon$ such that $k_0-\tilde k_1\epsilon^2 =
k_0/2$, then
\[
a_{\tiIL}(\biu,\biu) \geqs
-\bigl(1+\frac{3\tilde k_1^2}{2k_0} \bigr) \|\biu\|^2
+ \frac{k_0}{2}\, \|\biu\|^2_{1,2},
\]
and so,
\begin{align*}
 \frac{k_0}{2}\, \|\biu\|^2_{1,2} &\leqs
\bigl(1+\frac{3\tilde k_1^2}{2k_0} \bigr) \|\biu\|^2
+ a_{\tiIL}(\biu,\biu)\\
&\leqs \bigl(1+\frac{3\tilde k_1^2}{2k_0} \bigr) \Bigl[
\|\biu\|^2 + a_{\tiIL}(\biu,\biu) \Bigr].
\end{align*}
Divide by $\bigl(1+\frac{3\tilde k_1^2}{2k_0} \bigr)$ and set
$k_1 =  \frac{k_0^2}{(2k_0+ 3\tilde k_1^2 )}$
and the result is the inequality (\ref{GI}).\qed

We have not found in the literature the generalized Korn inequality
with boundary conditions on only part of the manifold boundary,
neither in Euclidean space nor in an arbitrary Riemannian
manifold. This type of inequality is crucial in \Sec\ref{S:MC-ExVM},
so we proceed to establish this result. 
\begin{lemma}
\label{L:GKIbc}{\bf (Coercivity of $\cL$)}
Let $(\cM,h_{ab})$ be a 3-dimensional, compact, Riemannian manifold,
with Lipschitz boundary such that $\mbox{meas}(\partial\cM_{\tiID})
>0$, and the metric $h\in C^2(\overline\cM,2)$. Then, there exists a
positive constant $k_0$ such that the following inequality holds
\begin{equation}
\label{GKIbc}
k_0 \,\|\tbu\|^2_{1,2} \leqs \|\cL\tbu\|^2
\qquad \forall \tbu \in \tbW^{1,2}_{\tiID}.
\end{equation}
\end{lemma}

\Proof {\it (Lemma~\ref{L:GKIbc}.)~}
The proof has two main parts: The first one is to show that the null
space of the operator $\cL:\biW^{1,2}_{\tiID}\to L^2(\cM,2)$ is
trivial when $\mbox{meas}(\partial\cM_{\tiID}) > 0$; the second part
uses the G\aa rding type inequality satisfied by the operator $\cL$
and presented in Lemma~\ref{L:GKI} together with a well known argument
by contradiction to show Eq.~(\ref{GKIbc}).

The first part mentioned above also consists of two steps. We first
step is to show that any vector field belonging to the null space of
$\cL$, vectors called conformal Killing vectors, must satisfy a
particular set of ordinary differential equations (ODE). Indeed,
assume that $u^a$ is a conformal Killing vector, so $\cL\biu=0$, and
introduce the fields
\[
\alpha_{ab}:= \nabla_{[a}u_{b]},\qquad
\beta := \nabla_au^a,\qquad
\gamma_a := \nabla_a \beta,
\]
where we introduced the notation $\nabla_{[a}u_{b]} := (\nabla_au_b
-\nabla_bu_a)/2$, and similarly we will denote $\nabla_{(a}u_{b)} :=
(\nabla_au_b +\nabla_bu_a)/2$. A straightforward albeit long 
computation commuting derivatives shows that a conformal Killing
vector $u^a$ and its derivatives introduced above must satisfy the
following equations,
\begin{align}
\label{CKV1}
\nabla_a u_b &= \alpha_{ab} + \frac{1}{3} \,\beta h_{ab},\\
\label{CKV2}
\nabla_a \beta & = \gamma_a,\\
\label{CKV3}
\nabla_a \alpha_{bc} &= -R_{bca}{}^d u_d 
+ \frac{2}{3} \, \gamma_{[b}h_{c]a},\\
\label{CKV4}
\nabla_a \gamma_b &= -3 u^c\nabla_c L_{ab} - 2 \beta L_{ab}
- 6 R_{c(a}\alpha_{b)}{}^c,
\end{align}
where the tensor $R_{abc}{}^d$ is the Riemann tensor of the metric
connection $\nabla_a$, the tensor $R_{ab}=R_{acb}{}^c$ is the Ricci
tensor, and we have introduced the tensor $L_{ab} := R_{ab}-R
h_{ab}/4$. The first two equations above are the definitions of the
fields $\alpha_{ab}$, $\beta$ and $\gamma_a$. The other two equations
are obtained by commuting second and third derivatives of the
conformal Killing vector $u^a$. They are generalizations of the
well-known formulas for Killing vectors (where $\beta=0$,
$\gamma_a=0$) which can be found for example in~\cite{Wald84},
page~443. These formulas in the case of Lorentzian metrics have been
used in~\cite{rG69}. Contract Eqs.~(\ref{CKV1})-(\ref{CKV4}) on index
$a$ with any vector field $v^a$, and the result is a system of ODE for
the fields $u^a$, $\alpha_{ab}$, $\beta$ and $\gamma_a$. From this
system of ODE we conclude the following: If these four fields vanish
at a single point in $\cM$, then they vanish identically on $\cM$.

The second step is to show the following: If $u^a$ is a conformal
Killing vector that vanishes on a two-dimensional hypersurface
$\partial\cM_{\tiID}$ with $\mbox{meas} (\partial\cM_{\tiID}) >0$,
then the vector $u^a$, and the fields $\alpha_{ab}$, $\beta$ and
$\gamma_a$ vanish at any point on the hypersurface
$\partial\cM_{\tiID}$. This statement and the conclusion of the
paragraph above will imply that $u^a$ vanishes identically on the
manifold $\cM$. Denote by $n^a$ the unit vector field normal to the
tangent space at each point in the manifold $\partial\cM_{\tiID}$, and
introduce the first and second fundamental forms of the hypersurface
$\partial\cM_{\tiID}$ as follows,
\[
l_{ab} := h_{ab} - n_an_b,\qquad
\kappa_{ab} := -l_a{}^c\nabla_cn_b,
\]
where the tensor $\kappa_{ab}$ is symmetric. Denote by $D_a$ the
Levi-Civita connection associated with the two-metric $l_{ab}$ defined
on the 2-dimensional hypersurface $\partial\cM_{\tiID}$, so the
connection satisfies the property that $D_al_{bc}=0$. Extend the
vector field $n^a$ to a neighborhood of the hypersurface
$\partial\cM_{\tiID}$ in the manifold $\cM$ as the tangent vector
solution to the geodesic equation $n^a\nabla_an^b=0$ with initial data
$n^a$ on $\partial\cM_{\tiID}$. Hence, the resulting vector field
satisfies $n^a\nabla_an^b=0$ in a neighborhood of the hypersurface
$\partial\cM_{\tiID}$. Then, decompose the conformal Killing field
$u^a$ as follows $u^a = u_n n^a +\hat u^a$, with $n_a\hat u^a=0$.
Denote by $|_{\tiID}$ evaluation at the hypersurface
$\partial\cM_{\tiID}$, then the condition $u^a|_{\tiID}=0$ implies
\[
u_n|_{\tiID}=0,\qquad \hat u^a|_{\tiID}=0.
\]
The latter condition means that $(D_au_n)|_{\tiID}=0$ and $(D_a\hat
u^b)|_{\tiID}=0$, while the latter equation together with the equation
$n^an^b (\cL u)_{ab}=0$, which also holds on $\partial\cM_{\tiID}$,
imply that $(\nabla_nu_n)|_{\tiID}=0$, where we use the notation
$\nabla_n := n^a\nabla_a$. Therefore, from expression $\nabla_au^a=
\nabla_n u_n + D_a\hat u^a + \kappa_a{}^a u_n$ we then conclude that
\[
\beta|_{\tiID}=0.
\]
The equation $n^al_c{}^b (\cL u)_{ab}=0$ implies, after a short
calculation, that the equation $l_c{}^b\nabla_n\hat u_b +
\kappa_{cb}\hat u^b + D_cu_n =0$ holds in the manifold $\cM$, and so
it also holds on the hypersurface $\partial\cM_{\tiID}$.  This result
together with our previous results establish the condition
$(l_c{}^b\nabla_n\hat u_b)|_{\tiID}=0$. The decomposition
\[
\nabla_{[a}u_{b]} = D_{[a}\hat u_{b]} + n_{[b} \kappa_{a]c}\hat u^c 
+ n_{[b}D_{a]}u_n + n_{[a}l_{b]}{}^c \nabla_n \hat u_c
\]
and our previous results then imply that 
\[
\alpha_{ab}|_{\tiID} =0.
\]
We still have to show that the vector field $\gamma_a=\nabla_a\beta$
vanishes on the hypersurface $\partial\cM_{\tiID}$. Since $\beta$ is a
scalar field and vanishes on $\partial\cM_{\tiID}$, we conclude that
the field $(D_a\beta)|_{\tiID}=0$, which implies $(D_a\nabla_n
u_n)|_{\tiID}=0$. We now only need to compute the field
$(\nabla_n\beta)|_{\tiID}$. The identity $n^a
l_c{}^b\nabla_{[a}\nabla_{b]} u_n =0$ when evaluated on the
hypersurface $\partial\cM_{\tiID}$ together with our previous results
imply the equation $(D_c\nabla_nu_n)|_{\tiID} = (l_c{}^b
\nabla_nD_bu_n)|_{\tiID}$. But we just showed that the left hand side
vanishes, and so then does the right hand side $(l_c{}^b
\nabla_nD_bu_n)|_{\tiID}=0$. Finally, from the equation
$n^al_d{}^b\bigl(2\nabla_{[a}\nabla_{b]} \hat u_c -R_{abc}{}^e\hat
u_e\bigr)=0$ evaluated on the hypersurface $\partial\cM_{\tiID}$ we
conclude that
\[
(l_c{}^al_d{}^b \nabla_nD_a\hat u_b)|_{\tiID} - 
\bigl[D_c(l_d{}^b\nabla_n\hat u_b)\bigr]|_{\tiID}=0.
\]
The result $(l_c{}^b\nabla_n\hat u_b)|_{\tiID}=0$ implies that the
second term on the left hand side above vanishes, so we conclude that
$(l_c{}^a l_d{}^b\nabla_nD_a\hat u_b)|_{\tiID} =0$, and from this
equation one gets $\bigl[\nabla_n(D_a\hat u^a)\bigr]|_{\tiID}=0$.
Therefore, in order to show that the field $(\nabla_n\beta)|_{\tiID}$
vanishes we only have left to prove that the field
$(\nabla_n\nabla_nu_n)|_{\tiID}$ vanishes. That this is the case
follows from the equation $\nabla_n\bigl[n^an^b(\cL u)_{ab}\bigr]=0$,
which holds in the manifold $\cM$, and so on the hypersurface
$\partial\cM_{\tiID}$, and an explicit computation shows that
$(\nabla_n\nabla_nu_n)|_{\tiID}=0$. We then conclude that
\[
\gamma_a|_{\tiID}=0.
\]
Let us now recall that the ODE equations obtained by contracting
Eqs.~(\ref{CKV1})-(\ref{CKV4}) on index $a$ with any vector field
$v^a$ are homogeneous on the fields $u^a$, $\alpha_{ab}$, $\beta$ and
$\gamma_a$ with vanishing initial data on the hypersurface
$\partial\cM_{\tiID}$. The solution vanish identically in a
neighborhood of this hypersurface $\partial\cM_{\tiID}$ in the
manifold $\cM$. Repeating this procedure we conclude that the
conformal Killing vector field vanishes identically in $\cM$. This
result establishes that the null space of the operator $\cL$ is
trivial on the space $\biW^{1,2}_{\tiID}$.

We now consider the second part of the proof of Lemma~\ref{L:GKIbc}
using a well-known argument by contradiction. Assume that there exists
a sequence $\{\biu_n\}\subset\biW^{1,2}_{\tiID}$ such that
\[
\|\biu_n\|_{1,2}=1,\quad\mbox{and}\quad
\lim_{n\to\infty}\|\cL \biu\|  =0.
\]
The sequence $\{\biu_n\}$ is bounded in $\biW^{1,2}_{\tiID}$ which is
a reflexive Banach space, so there exists a subsequence, also denoted
as $\{\biu_n\}$, such that
\[
\biu_n \wto \biu_0 \quad\mbox{in}\quad \biW^{1,2}_{\tiID},
\quad\mbox{and}\quad
\biu_n \to \biu_0 \quad\mbox{in}\quad \biL^2,
\]
the latter statement following from the imbedding
$\biW^{1,2}_{\tiID}\subset\biL^2$ being compact.  So $\{\biu_n\}$ is a
Cauchy sequence in $\biL^2$, and by assumption the sequence
$\{\cL\biu_n\}\subset\biL^2$ is also a Cauchy sequence. The G\aa rding
inequality in Lemma~\ref{L:GKI} implies that
\[
k_0 \,\|\biu_n-\biu_m\|_{1,2}^2 \leqs 
\|\biu_n-\biu_m\|^2 + \|\cL\biu_n -\cL\biu_m\|^2 \to 0
\quad \mbox{as}\quad n,m\to \infty,
\]
and so the sequence $\{\biu_n\}$ is also a Cauchy sequence in
$\biW^{1,2}_{\tiID}$. We then conclude that 
\[
\biu_n\to\biu_0\quad\mbox{in}\quad\biW^{1,2}_{\tiID}\RI
\cL\biu_n \to 0=\cL\biu_0.
\]
But the null space of the operator $\cL$ is trivial on the space
$\biW^{1,2}_{\tiID}$, therefore we conclude that the element
$\biu_0=0$. However, this leads us to a contradiction from the
hypothesis that $\|\biu_n\|_{1,2}=1$ which implies that
$\|\biu_0\|_{1,2}=1$ so the element $\biu_0\neq 0$. Therefore, such
sequence $\{\biu_n\}$ does not exist, which then establishes the
Lemma.\qed

\begin{corollary}
\label{C:coercive}{\bf(Coercivity of $a_{\tiIL}$)}
Let $(\cM,h_{ab})$ be a 3-dimensional, compact, Riemannian manifold,
with Lipschitz boundary such that $\mbox{meas}(\partial\cM_{\tiID})
>0$, and the metric $h\in C^2(\overline\cM,2)$. Let $a_{\tiIL}$ be the
bilinear form defined in Eq.~(\ref{MC-def-aIL}), and assume that the
Robin tensor $\IK\in L^{\infty} (\partial\cM_{\tiIN},2)$ is positive
definite.  Then, there exists a positive constant $k_1$ such that the
following inequality holds
\begin{equation}
\label{coercive}
k_1 \,\|\tbu\|^2_{1,2} \leqs  a_{\tiIL}(\tbu,\tbu)
\qquad \forall \tbu \in \tbW^{1,2}_{\tiID}.
\end{equation}
\end{corollary}

\Proof {\it (Corollary~\ref{C:coercive}.)~}
Since the Robin tensor $\IK$ is positive definite, then the result
is straightforward from Eq.~(\ref{GKIbc}), due to the following
inequalities,
\begin{align*}
k_0 \,\|\biu\|^2_{1,2} &\leqs \|\cL\biu\|^2\\
&\leqs  \|\cL\biu\|^2
+ \hat \ttK\, \|\Tr_{\tiIN}\biu\|^2_{\tiIN}\\
&\leqs  \|\cL\biu\|^2 
+ (\IK \Tr_{\tiIN}\biu, \Tr_{\tiIN}\biu)_{\tiIN}\\
&\leqs a_{\tiIL}(\biu,\biu),
\qquad \forall \biu \in \biW^{1,2}_{\tiID}.
\end{align*}
This inequality establishes the Corollary.\qed

\Remark
It can be shown that the result in the Corollary~\ref{C:coercive}
remains valid if the Robin tensor field $\IK$ is slightly negative
definite.

\subsection{Results using variational methods}
\label{S:MC-ExVM}

The momentum constraint Eq.~(\ref{MC-LYm}) can be written as the Euler
condition for stationarity of a real-valued functional on a Banach space.
Direct methods in the calculus of variations can be used to find the 
points that minimize this functional in the Banach space in the case 
that the hypersurface $\partial\cM_{\tiID}\neq\emptyset$. The main concepts
needed from the calculus of variations are summarized in the Appendix,
where we also explain the part of the notation used in this Section.
Since the momentum constraint is linear, we will achieve similar (in
fact, slightly more general) results in \Sec\ref{S:MC-WP} using
Riesz-Schauder Theory.  However, the presentation here is a guide for
our variational treatment of the Hamiltonian constraint equation in
\S\ref{S:HC-ExVM}, and the results we assemble in this section are of
interest in their own right.

Let $a_{\tiIL} :\biW^{1,2}_{\tiID}\times\biW^{1,2}_{\tiID}\to\R$ be a
bilinear form with action defined in Eq.~(\ref{MC-def-aIL}), and fix
the functionals $\bib_{\tau}^{*}$,
$\bib_{j}^{*}\in\biW^{-1,2}_{\tiID}$. Let $\biw_{\tiID}\in\biW^{1,2}$
be the extension of the Dirichlet data $\hat\biw_{\tiID}$, and
$\hat\biw_{\tiIN}^{*}$ be the Robin data functional, both data defined
in Eq.~(\ref{MC-BD}).  Introduce the functional
\begin{equation}
\label{VM-def-ttJ}
J_{\tiIL}:\biW^{1,2}_{\tiID}\to\R,
\qquad
J_{\tiIL}(\bomega):= \frac{1}{2}\, a_{\tiIL}(\bomega,\bomega) 
+ G_{\phi}(\bomega),
\end{equation}
where the functional $G_{\phi}$ is given by
\[
G_{\phi}(\bomega) := g_{\phi}(\biw_{\tiID}+\bomega) 
- \hat\biw_{\tiIN}^{*}(\Tr_{\tiIN}\bomega)
+ a_{\tiIL}(\biw_{\tiID},\bomega),
\]
with the functional $g_{\phi}(\biw)$ having the form
\begin{equation}
\label{VM-def-ttg}
g_{\phi}(\biw) := (\bib_{\tau}\phi^6)^{*}(\biw)
+ \bib_{j}^{*}(\biw),
\end{equation}
and we will use the notation $\biw=\biw_{\tiID}+\bomega$.

\begin{theorem}{\bf (Existence of a minimizer)}
\label{T:VM-MC-EM}
Let $J_{\tiIL} :\tbW^{1,2}_{\tiID}\to\R$ be the functional 
defined in Eq.~(\ref{VM-def-ttJ}). Assume that the hypersurface
$\partial\cM_{\tiID}\neq\emptyset$, and fix an extension of the
Dirichlet boundary data $\tbw_{\tiID}\in\tbW^{1,2}$ and the Robin
boundary data $\hat\tbw_{\tiIN}^{*}\in W^{-\frac{1}{2},2}
(\partial\cM_{\tiIN},1)$. Fix the functionals $\tbb_{\tau}^{*}$,
$\tbb_{j}^{*}\in\tbW^{-1,2}_{\tiID}$, and the tensor $\IK\in
L^{\infty}(\partial\cM_{\tiIN},2)$ satisfying the inequality in
Eq.~(\ref{MC-cttk}) with $\hat\ttK\geqs 0$.  Then, there exists a
unique element $\bomega\in\tbW^{1,2}_{\tiID}$ minimizer of the
functional $J_{\tiIL}$ on $\tbW^{1,2}_{\tiID}$, that is,
\[
J_{\tiIL}(\bomega) = \inf_{\un\bomega\in\tbW^{1,2}_{\tiID}}
J_{\tiIL}(\un\bomega).
\]
\end{theorem}

\Proof {\it (Theorem~\ref{T:VM-MC-EM}.)~}
We start showing that the functional $J_{\tiIL}$ is coercive, and the
first step is the following inequality
\[
J_{\tiIL}(\bomega) \geqs \frac{1}{2}\, a_{\tiIL}(\bomega,\bomega)
-|G_{\phi}(\bomega)|.
\]
The linear terms in $G_{\phi}$ can be bounded as follows: recall the
notation $\biw = \biw_{\tiID}+\bomega$, then
\begin{align*}
|g_{\phi}(\biw)| &\leqs \bigl|(\bib_{\tau}\phi^6)^{*}(\biw)\bigr| 
+ \bigl| \bib_{j}^{*}(\biw)|\\
&\leqs \Bigl[ \|\phi\|_{\infty}^{6} \,\|\bib_{\tau}^{*}\|_{-1,2}
\, + \|\bib_{j}^{*}\|_{-1,2} \Bigr] \, \|\biw\|_{1,2},
\end{align*}
introducing the constant $c_{g} :=
\big[\|\phi\|_{\infty}^{6}\,\|\bib_{\tau}^{*}\|_{-1,2}\,
+\|\bib_{j}^{*}\|_{-1,2}\bigr]/2$, we then obtain
\begin{align}
\label{VM-MC-ineq1}
|g_{\phi}(\biw)| &\leqs 2 c_{g} \,\|\biw\|_{1,2}
\leqs \frac{1}{\epsilon}\,  c_{g}^2 
+ \epsilon\,\|\biw\|_{1,2}^2,
\leqs \frac{1}{\epsilon}\,  c_{g}^2 
+ 2\epsilon \,\|\biw_{\tiID}\|_{1,2}^2 
+ 2\epsilon\,\|\bomega\|_{1,2}^2,
\end{align}
where $\epsilon$ is any positive constant. The second term in
the functional $G_{\phi}$ can be bounded as follows
\begin{align}
\nonumber
\bigl|- \hat\biw_{\tiIN}^{*}(\Tr_{\tiIN}\bomega)\bigr| &\leqs
\|\hat\biw_{\tiIN}^{*}\|_{-\frac{1}{2},2,\tiIN}\, 
\|\Tr_{\tiIN}\bomega\|_{\frac{1}{2},2,\tiIN}\\
\nonumber
&\leqs c_0 \,\|\hat\biw_{\tiIN}^{*}\|_{-\frac{1}{2},2,\tiIN}\, 
\|\bomega\|_{1,2}\\
\label{VM-MC-ineq2}
&\leqs \frac{c_0^2}{2\epsilon} \,
\|\hat\biw_{\tiIN}^{*}\|_{-\frac{1}{2},2,\tiIN}^2 
+ \frac{\epsilon}{2}\, \|\bomega\|_{1,2}^2,
\end{align}
where $c_0$ is a positive constant. The third term in $G_{\phi}$ can
be bounded as follows
\begin{align}
\nonumber
| a_{\tiIL}(\biw_{\tiID},\bomega)| &\leqs c_{\tiIL} \,
\|\biw_{\tiID}\|_{1,2}\,\|\bomega\|_{1,2}\\
\label{VM-MC-ineq3}
&\leqs \frac{c_{\tiIL}^2}{2\epsilon} \,\|\biw_{\tiID}\|_{1,2}^2 
+ \frac{\epsilon}{2}\, \|\bomega\|_{1,2}^2.
\end{align}
By adding the inequalities in
Eqs.~(\ref{VM-MC-ineq1})-(\ref{VM-MC-ineq3}) we obtain the bound on
$G_{\phi}$,
\begin{gather}
\label{VM-MC-ineq-G}
|G_{\phi}(\bomega)| \leqs c_{\tiG} 
+ 3\epsilon \,\|\bomega\|_{1,2}^2,\\
\nonumber
c_{\tiG}:= \frac{1}{\epsilon}\,  \Bigl[ c_{g}^2
+ 2\epsilon^2 \,\|\biw_{\tiID}\|_{1,2}^2
+ \frac{c_0^2}{2}\,\|\hat\biw_{\tiIN}^{*}\|_{-\frac{1}{2},2,\tiIN}^2 
+ \frac{ c_{\tiIL}^2}{2}\,\|\biw_{\tiID}\|_{1,2}^2 \Bigr].
\end{gather}
We now consider the bilinear form $a_{\tiIL}$. First, the tensor $\IK$
satisfies the inequality in Eq.~(\ref{MC-cttk}) with $\hat\ttK\geqs 0$,
so we have
\[
a_{\tiIL}(\bomega,\bomega) \geqs (\cL\bomega,\cL\bomega).
\]
The assumption that the hypersurface
$\partial\cM_{\tiID}\neq\emptyset$ and the generalized Korn inequality
in Lemma~\ref{L:GKIbc} imply that there exists a positive constant
$k_0$ such that $(\cL\bomega,\cL\bomega)\geqs
k_0\,\|\bomega\|_{1,2}^2$, which together with the inequality above
imply
\begin{equation}
\label{VM-MC-ineq-aIL}
a_{\tiIL}(\bomega,\bomega) \geqs k_0\,\|\bomega\|_{1,2}^2.
\end{equation}
Therefore, from the
inequalities~(\ref{VM-MC-ineq-G})-(\ref{VM-MC-ineq-aIL}) we obtain
\[
J_{\tiIL}(\bomega) \geqs (k_0-3\epsilon) \|\bomega\|_{1,2}^2 - c_{\tiG}.
\]
Choosing $\epsilon$ small enough we have established that $J_{\tiIL}$
is coercive in $\biW^{1,2}_{\tiID}$.

We now show that the functional $J_{\tiIL}$ is \lscw. Let
$\{\bomega_n\}\subset\biW^{1,2}_{\tiID}$ be a sequence such that
$\bomega_n\wto\bomega_0$ in $\biW^{1,2}_{\tiID}$, which then implies
that $\bomega_n\to\bomega_0$ in $\biL^2$.  We again start with the
functional $G_{\phi}$, which is linear on its variable
$\bomega\in\biW^{1,2}_{\tiID}$, therefore it is continuous under weak
convergence (by definition of weak convergence). So it is also \lscw,
and the following equation holds
\[
G_{\phi}(\bomega_0) = \liminf_{n\to\infty} \,G_{\phi}(\bomega_n).
\]
We only have to show that the functional $\bomega\mapsto
a_{\tiIL}(\bomega,\bomega)$ given in Eq.~(\ref{MC-def-aIL}) is
also~\lscw.  The first term in the bilinear form $a_{\tiIL}$ defines a
norm in $\biW^{1,2}_{\tiID}$, since the generalized Korn inequality
given in Lemma~\ref{L:GKIbc} and the fact that the conformal Killing
operator is bounded in $\biW^{1,2}_{\tiID}$ imply that there
exist positive constants $k_0$, $K_0$ such that
\[
k_0 \,\|\bomega\|_{1,2}^2 \leqs \|\cL\bomega\|^2 
\leqs K_0\,\|\bomega\|_{1,2}^2,\qquad
\forall \, \bomega\in \biW^{1,2}_{\tiID}.
\]
This last inequality means that the map $\bomega\mapsto\|\cL\bomega\|$
defines a norm in $\biW^{1,2}_{\tiID}$, and so it is \lscw, a result
proven in the Appendix. Therefore, the following inequality holds,
\[
\|\cL\bomega_0\|^2 \leqs \liminf_{n\to\infty}\, \|\cL\bomega_n\|^2.
\]
The second term in the definition of the bilinear form $a_{\tiIL}$
contains the two-index tensor $\IK$, which is positive definite and
symmetric, therefore the function
\[
\bomega \mapsto
(\IK\Tr_{\tiIN}\bomega,\Tr_{\tiIN}\bomega)_{\tiIN}
\]
is a continuous and convex functional, and so it is \lscw, a result
also proven in the Appendix. We then conclude that the functional
$J_{\tiIL}$ is \lscw. Therefore, Theorem~\ref{T:VM-M} in the Appendix
in the case $U=X$ shows that there exists a minimizer for $J_{\tiIL}$
in $\biW^{1,2}_{\tiID}$.

The uniqueness of the minimizer is a consequence of the strict
convexity of the functional $J_{\tiIL}$, which is a general result
that, once again, is established in the Appendix. We have to show that
for all non-zero $\hat\bomega$, $\un\bomega\in\biW^{1,2}_{\tiID}$ and
all $t\in (0,1)$ holds
\[
J_{\tiIL}\bigl(t\hat\bomega+(1-t)\un\bomega\bigr) 
< tJ_{\tiIL}(\hat\bomega) + (1-t) J_{\tiIL}(\un\bomega),
\]
or equivalently, as it is explained in the Appendix, we only have to
show that for all non-zero $\tilde\bomega$,
$\un\bomega\in\biW^{1,2}_{\tiID}$ holds
\[
DJ_{\tiIL}(\un\bomega)(\tilde\bomega)< 
J_{\tiIL}(\tilde\bomega+\un\bomega) - J_{\tiIL}(\un\bomega).
\]
A straightforward calculation shows that for all $\tilde\bomega$,
$\un\bomega\in\biW^{1,2}_{\tiID}$ holds
\begin{align*}
J_{\tiIL}(\tilde\bomega+\un\bomega) - J_{\tiIL}(\un\bomega)
&= \frac{1}{2}\; a_{\tiIL}(\tilde\bomega,\tilde\bomega)
+a_{\tiIL}(\un\bomega,\tilde\bomega)+G_{\phi}(\tilde\bomega)\\
&= DJ_{\tiIL}(\un\bomega)(\tilde\bomega)
+\frac{1}{2}\,a_{\tiIL}(\tilde\bomega,\tilde\bomega)\\
&\geqs DJ_{\tiIL}(\un\bomega)(\tilde\bomega)
+\frac{1}{2}\,(\cL\tilde\bomega,\cL\tilde\bomega)\\
&> DJ_{\tiIL}(\un\bomega)(\tilde\bomega),
\end{align*}
where the symmetry of the Robin two-tensor field $\IK$ is used to
establish the first line, and the last line is obtained from the
generalized Korn's inequality Eq.~(\ref{GKIbc}). Therefore, the
functional $J_{\tiIL}$ is strictly convex, hence,
Theorem~\ref{T:VM-mcf} in the Appendix implies that the minimizer
$\bomega$ is unique. This establishes the Theorem.\qed

The next result shows that the minimum $\bomega$ of the functional
$J_{\tiIL}$ on the space $\biW^{1,2}_{\tiID}$ found in
Theorem~\ref{T:VM-MC-EM} is a solution of the Euler equation
$DJ_{\tiIL}(\bomega)=0$.
\begin{theorem}{\bf (Momentum constraint)}
\label{T:VM-MC-ES}
Assume the hypotheses in Theorem~\ref{T:VM-MC-EM}. Then, the
functional $J_{\tiIL}$ is G\^ateaux differentiable on
$\tbW^{1,2}_{\tiID}$ and the minimizer $\bomega\in \tbW^{1,2}_{\tiID}$
is solution of the Euler equation
\[
DJ_{\tiIL}(\bomega)(\un\bomega)=0,\qquad
\forall\, \un\bomega\in\tbW^{1,2}_{\tiID},
\]
where the equation above is the momentum constraint
Eq.~(\ref{MC-LYm}).
\end{theorem}

\Proof {\it (Theorem~\ref{T:VM-MC-ES}.)~}
It is straightforward to verify that the functional $J_{\tiIL}$ is
G\^ateaux differentiable, and its derivative at an arbitrary element
$\hat\bomega\in\biW^{1,2}_{\tiID}$ is given by
\[
DJ_{\tiIL}(\hat\bomega)(\un\bomega)= 
a_{\tiIL}(\hat\bomega,\un\bomega) + G_{\phi}(\un\bomega)
= a_{\tiIL}(\hat\biw,\un\bomega) + \bif_{\phi}(\un\bomega),
\]
with $\hat\biw:=\biw_{\tiID}+\hat\bomega$. Therefore, the G\^ateaux
derivative $DJ_{\tiIL}$ is the left hand side in Eq.~(\ref{MC-LYm}).
Let $\bomega\in\biW^{1,2}_{\tiID}$ be the minimizer of the functional
$J_{\tiIL}$ on the space $\biW^{1,2}_{\tiID}$. Then the following
inequality holds,
\begin{equation}
\label{VM-MC-ES-ineq1}
DJ_{\tiIL}(\bomega)(\un\bomega)\geqs 0,\qquad
\forall \, \un\bomega\in \biW^{1,2}_{\tiID}.
\end{equation}
For the proof, write down the G\^ateaux derivative of the functional
$J_{\tiIL}$ at the minimizer $\bomega$,
\[
DJ_{\tiIL}(\bomega)(\un\bomega) = \lim_{t\to 0^{+}} \bigl[
J_{\tiIL}(\bomega+t\un\bomega)-J_{\tiIL}(\bomega) \bigr].
\]
The element $\bomega$ is a minimizer of $J_{\tiIL}$, so
$J_{\tiIL}(\bomega+t\un\bomega)\geqs J_{\tiIL}(\bomega)$, which
establishes Eq.~(\ref{VM-MC-ES-ineq1}). This
Eq.~(\ref{VM-MC-ES-ineq1}) holds for $-\un\bomega$, so we conclude
that $DJ_{\tiIL}(\bomega)=0$. This establishes the Theorem.\qed

\subsection{Results using Riesz-Schauder theory}
\label{S:MC-WP}

We present here a proof of existence and uniqueness of solutions of
the weak Dirichlet-Robin boundary value problem for the momentum
constraint Eq.~(\ref{MC-LYm1}). The proof is based on the
Riesz-Schauder theory for compact operators, see~\cite{Wloka87}. The
proof is more general than the one given in \Sec\ref{S:MC-ExVM}
because it includes the case where $\mbox{meas}(\partial\cM_{\tiID})
=0$, that is, the pure Robin case.
Riesz-Schauder theory was used for the momentum constraint 
in \cite{mH01a} to develop an approximation theory 
and corresponding error estimates for numerical approximations.

\begin{theorem}{\bf (Momentum constraint)}
\label{T:w-MC}
Consider the weak formulation for the momentum constraint
Eq.~(\ref{MC-LYm1}). Assume that the Robin tensor field $\IK$
satisfies Eq.~(\ref{MC-cttk}) with positive constant $\hat\ttK$. Then,
there exists a unique solution $\tbw \in \tbA^{1,2}$ to the momentum
constraint Eq.~(\ref{MC-LYm1}), and there exist positive constants
$c_1$ and $c_2$ such that the following estimate holds,
\begin{equation}
\label{w-MC-est}
\|\tbw\|_{1,2} \leqs \|\phi\|_{\infty}^6  \,
\|\tbb_{\tau}^{*}\|_{-1,2}
+ \|\tbb_{j}^{*}\|_{-1,2}
+c_1 \, \|\hat\tbw_{\tiIN}^{*}\|_{-\frac{1}{2},2,\tiIN}
+c_2 \, \|\tbw_{\tiID}\|_{1,2}.
\end{equation}
\end{theorem}

\Proof {\it (Theorem~\ref{T:w-MC}.)~}
First translate the problem from the affine space $\biA^{1,2}$ into a
problem on the vector space $\biW^{1,2}_{\tiID}$ with the change of
variable $\biw = \biw_{\tiID} + \bomega$. Then, Eq.~(\ref{MC-LYm1})
has the form: Find $\bomega\in\biW^{1,2}_{\tiID}$ solution of
\begin{equation}
\label{MC-eq1}
A_{\tiIL}\bomega +G_{\phi} =0
\end{equation}
where $G_{\phi}(\un\bomega) :=\bif_{\phi}(\un\bomega) +
a_{\tiIL}(\biw_{\tiID},\un\bomega)$, and we now consider the operator
$A_{\tiIL} :\biW^{1,2}_{\tiID}\to\biW^{-1,2}_{\tiID}$. To find a
solution of Eq.~(\ref{MC-eq1}) is equivalent to show that this
operator $A_{\tiIL}$ is invertible. Lemma~\ref{L:GKI} says that the
conformal Killing operator $\cL$ satisfies G\aa rding's inequality
Eq.~(\ref{GKI}). This implies that the bilinear form $a_{\tiIL}$ also
satisfies a G\aa rding inequality, which was proven in
Corollary~\ref{C:GI}. Then, Theorem~\ref{T:GI-FT} in the Appendix
implies that the operator $A_{\tiIL}$ is Fredholm with index
zero. That means $\dim N_{A_{\tiIL}} = \codim R_{A_{\tiIL}}$, which
can be described saying that the operator $A_{\tiIL}$ is bijective iff
it is injective. This property is described in the PDE literature as
``uniqueness implies existence''. So, in order to show that
$A_{\tiIL}$ is invertible we only have to show that its null space is
trivial. Consider an element $\biu\in\biW^{1,2}_{\tiID}$ such that
$A_{\tiIL}\biu=0$. In particular $A_{\tiIL}\biu(\biu)=0$, which is
equivalent to
\[
0 = \|\cL\biu\|^2 + (\IK \Tr_{\tiIN}\biu,\Tr_{\tiIN}\biu)_{\tiIN}
\geqs \|\cL\biu\|^2 + \hat \ttK \,\|\Tr_{\tiIN}\biu\|^2_{\tiIN}.
\]
Both terms must vanish, since the tensor $\IK$ is strictly positive
definite and so the constant $\hat \ttK$ is positive. From the first
term one obtains that $\biu$ is a conformal Killing vector, and from
the second term together with $\biu\in\biW^{1,2}_{\tiID}$ one obtains
that $\Tr\biu=0$ on the whole boundary $\partial\cM$ and so,
$\biu\in\biW^{1,2}_{0}$. Therefore, $\biu =0$ in the manifold $\cM$
since the bilinear form $a_{\tiIL}$ is strongly elliptic. So, the
null space of the operator $A_{\tiIL}$ is trivial, and then
$A_{\tiIL}$ is invertible. Finally, it is not difficult to check that
the estimate given in Eq.~(\ref{GI-FT-est}) of the Appendix applied to
$\bomega =\biw -\biw_{\tiID}$ implies the estimate on $\biw$ given in
Eq.~(\ref{w-MC-est}). This establishes the Theorem.\qed

\Remark
In the case that the hypersurface $\partial\cM_{\tiID}\neq\emptyset$
the Robin tensor $\IK$ need not to be strictly positive. There exists
a unique solution to the momentum constraint in the case that the
constant $\hat\ttK$ in Eq.~(\ref{MC-cttk}) is slightly negative, that
is, $\hat\ttK > -\hat\ttK_0$ for small enough $\hat\ttK_0>0$. The
proof uses the coercivity of the conformal Killing operator $\cL$, the
inequality~(\ref{GKIbc}) in Lemma~\ref{L:GKIbc}, instead of the G\aa
rding inequality. It can be shown that the assumption that $\hat \ttK$
is negative but not too negative implies that the bilinear form
$a_{\tiIL}$ itself is strictly positive on
$\biW^{1,2}_{\tiID}$. Recalling that the linear form $\hat\bif_{\phi}
:\biW^{1,2}_{\tiID}\to\R$ is bounded, then the Riesz representation
Theorem says that there exists a unique $\biu\in\biW^{1,2}_{\tiID}$
solution to the weak problem with Eq.~(\ref{MC-LYm}). In terms of the
operator $A_{\tiIL}$, this statement means that $A_{\tiIL}
:\biW^{1,2}_{\tiID}\to\biW^{-1,2}_{\tiID}$ is invertible.

\subsection{Regularity of solutions}
\label{S:MC-RS}

In this Section we state without proof regularity results, which can
be obtained from the literature, and are applied to the weak solutions
of the momentum constraint.

\begin{theorem}{\bf (Regularity $\tbW^{1,p}$)}
\label{T:MC-E-Reg1}
Assume the hypotheses in Theorem~\ref{T:w-MC}, and in addition assume
that the boundary set $\partial\cM$ is $C^{1,1}$. Assume that the
source functional $\tbf_{\phi\tiF}$ and the boundary data satisfy the
following conditions,
\[
\tbb_{\tau}^{*},\tbb_{j}^{*}\in \tbW^{-1,p},\quad
\tbw_{\tiID}\in \tbW^{1,p},\quad
\hat\tbw_{\tiIN}^{*}\in W^{-\frac{1}{p},p}(\partial\cM_{\tiIN},1),\quad
\quad p\geqs 2,
\]
then, the solution $\tbw$ to the momentum constraint
Eq.~(\ref{MC-LYm1}) satisfies that $\tbw\in\tbW^{1,p}$ and there exist
positive constants $c_1$ and $c_2$ such that the following estimate
holds,
\begin{equation}
\label{w-MC-est2}
\|\tbw\|_{1,p} \leqs \|\phi\|_{\infty}^6  \,
\|\tbb_{\tau}^{*}\|_{-1,p}
+ \|\tbb_{j}^{*}\|_{-1,p}
+c_1 \, \|\hat\tbw_{\tiIN}^{*}\|_{-\frac{1}{p},p,\tiIN}
+c_2 \, \|\tbw_{\tiID}\|_{1,p}.
\end{equation}
\end{theorem}

\Proof {\it (Theorem~\ref{T:MC-E-Reg1}.)~}
We only describe a sketch of the proof. See for
example~\cite{kG89}. See also~\cite{Campanato80} for interior
estimates only, Theorems in \Sec7 and \Sec8. These results can be
extended up to the boundary for smooth enough boundaries.\qed

We also present here a result from~\cite{Ciarlet97}, stating higher
regularity of the weak solution of the momentum constraint
Eq.~(\ref{MC-LYm1}) in the case that the data and the source function
also possess additional regularity.

\begin{theorem}{\bf (Regularity $\tbW^{2,p}$)}
\label{T:MC-E-Reg2}
Assume the hypotheses in Theorem~\ref{T:w-MC}, and in addition assume
that the boundary set $\partial\cM$ is $C^{2}$. Assume that the source
functionals have the form $\tbb_{\tau}^{*}(\un\bomega) =
(\tbb_{\tau},\un\bomega)$, and $\tbb_{j}^{*}(\un\bomega) =
(\tbb_{j},\un\bomega)$, while the boundary data have the form
$\hat\tbw_{\tiIN}^{*}(\Tr_{\tiIN}\un\bomega) =
(\hat\tbw_{\tiIN},\Tr_{\tiIN}\un\bomega)_{\tiIN}$, for all
$\un\bomega\in\tbW^{1,2}_{\tiID}$. If the following conditions hold
\[
\tbw_{\tiID}\in \tbW^{2,p},\quad
\hat\tbw_{\tiIN}\in W^{\frac{1}{p'},p}(\partial\cM_{\tiIN},1),\quad
\tbb_{\tau}, \tbb_{j}\in \tbL^p,
\quad p\geqs \frac{6}{5},
\]
then, the solution $\tbw$ to the momentum constraint
Eq.~(\ref{MC-LYm1}) satisfies that $\tbw\in\tbW^{2,p}$ and there exist
positive constants $c_1$ and $c_2$ such that the following estimate
holds,
\begin{equation}
\label{w-MC-est3}
\|\tbw\|_{2,p} \leqs \|\phi\|_{\infty}^6  \,
\|\tbb_{\tau}\|_{p}
+ \|\tbb_{j}\|_{p}
+c_1 \, \|\hat\tbw_{\tiIN}\|_{\frac{1}{p'},p,\tiIN}
+c_2 \, \|\tbw_{\tiID}\|_{2,p}.
\end{equation}
\end{theorem}

\Proof {\it (Theorem~\ref{T:MC-E-Reg2}.)~}
We only describe a sketch of the proof, which
follows~\cite{Ciarlet97}, Vol. II, page 296. It is based on the fact
that the momentum constraint bilinear form $a_{\tiIL}$ is strongly
elliptic and satisfies the supplementary and complementing conditions
given in~\cite{aAaDlN64}.\qed

\section{The Hamiltonian constraint}
\label{S:HC}

In this section we fix a particular functional $a_{w}^{*}$ in an
appropriate space and we then look for weak solutions only of the
Hamiltonian constraint Eq.~(\ref{WF-LYs1}). We first develop the weak
formulation more precisely in \Sec\ref{S:WF-HC}, and as in
\Sec\ref{S:WF} we assume the weakest regularity of the equation
coefficients such that the equation itself is well-defined.  As was
the case for the momentum constraint, we will be able to use
variational methods to obtain existence (and when possible,
uniqueness) results for the Hamiltonian constraint in this weakest
setting.  First, we establish some preliminary results on generalized local 
and global barriers (constant sub- and super-solutions) for weak solutions
in \Sec\ref{S:WF-LB}.  The term local means that the barrier does not
depend on the coefficient $a_{w}^{*}$, while global means the barrier
does depend on this coefficient. We summarize the generalized local and 
global barriers in \Sec\ref{S:HC-SB}. 
In \Sec\ref{S:HC-apriori}, we establish some related
{\em a~priori} $L^{\infty}$-bounds on any $W^{1,2}$-solution to the
Hamiltonian constraint.
In \Sec\ref{S:HC-ExVM}, we then use the
barriers from \Sec\ref{S:WF-LB}, together with a variational argument,
to establish existence, and when possible uniqueness, of solutions to
the Hamiltonian constraint in the weakest possible setting of
$L^{\infty}\cap W^{1,2}$.  Due to the lack of
G\^ateaux-differentiability of the nonlinearity in $W^{1,2}$, the
connection between the energy used for the variational argument and
the Hamiltonian constraint as its Euler condition is non-trivial, and
is established through several Lemmas.  In \Sec\ref{S:WF-Existence-LB}
we give a second (non-variational) argument for existence, using a
barriers approach as in most of the earlier work~\cite{jI95,jIvM96},
which requires additional regularity on the equation coefficients.
Regularity of solutions is discussed briefly in \Sec\ref{S:HC-Reg}.

The results obtained using variational methods in \Sec\ref{S:HC-ExVM}
can be viewed as lowering the regularity of the recent result of
Maxwell on ``rough'' CMC solutions in $W^{k,2}$ for $k>3/2$ down to
$L^{\infty} \cap W^{1,2}$.  We note that the barrier-based
existence results for the Hamiltonian constraint equation in
\Sec\ref{S:WF-Existence-LB}, and the compactness argument in
\Sec\ref{S:CS-Ex} giving existence for the coupled non-CMC system,
require higher regularity on the equation coefficients. However, we
still end up with some non-CMC results for the coupled system in
weaker settings and in more general physical situations than have been
previously obtained.  These additional assumptions are clearly stated
in those sections.

\subsection{Weak formulation}
\label{S:WF-HC}

Let $(\cM, h)$ be a 3-dimensional Riemannian manifold, where $\cM$ is
a smooth, compact manifold with Lipschitz boundary $\partial\cM$, and
$h\in C^2(\overline\cM ,2)$ is a positive definite metric. Introduce
the bilinear form
\begin{equation}
\label{HC-def-aL}
a_{\tiL} : W^{1,2} \times W^{1,2} \to \R,\qquad
a_{\tiL}(\phi,\un\phi) 
:= (\nabla\phi,\nabla\un\phi) 
+ (K\, \Tr_{\tiN} \phi,\Tr_{\tiN}\un\phi)_{\tiN},
\end{equation}
where the Robin function $K\in L^{\infty} (\partial\cM_{\tiN},0)$
satisfies the bound
\begin{equation}
\label{HC-cttk}
\hat \ttk\, \|\Tr_{\tiN}\phi\|^2_{\tiN} \leqs 
(K \Tr_{\tiN}\phi, \Tr_{\tiN}\phi)_{\tiN},
\qquad \forall \phi\in W^{1,2},
\end{equation}
with $\hat\ttk$ being a non-negative constant. Fix the functionals
\begin{equation}
\label{HC-coeff}
a_{\tau}^{*} \in W^{-1,2}_{\tiD},\quad
a_{\rho}^{*} \in W^{-1,2}_{\tiD+},\quad
a_{w}^{*} \in W^{-1,2}_{\tiD}.
\end{equation}
The assumption on the background metric implies that the function
$a_{\tiR}$ is continuous on the manifold $\overline\cM$, so the
functional $a_{\tiR}^{*}\in W^{-1,2}_{\tiD}$ given by
\begin{equation}
\label{HC-Coeff-aR}
a_{\tiR}^{*}(\un\varphi):= (a_{\tiR},\un\varphi),\qquad
\forall\, \un\varphi\in W^{1,2}_{\tiD}
\end{equation}
is well-defined. Given any two functions $\phi_1$, $\phi_2
\in L^{\infty}$ with $0<\phi_1\leqs\phi_2$, define the interval
\[
[\phi_1,\phi_2] 
:= \{ \phi \in L^{\infty} : \phi_1 \leqs \phi \leqs \phi_2\},
\]
which is a closed, bounded set in $L^{\infty}$, and also in
$L^2$. Introduce the nonlinear operator
\begin{gather}
\nonumber
f_{w\tiF}: [\phi_1,\phi_2]\subset L^2 \to W^{-1,2}_{\tiD},\\
\label{HC-x0}
f_{w\tiF}(\phi) :=  
  (a_{\tau} \phi^5)^{*} + (a_{\tiR} \phi)^{*}
- (a_{\rho} \phi^{-3})^{*} - (a_{w} \phi^{-7})^{*}.
\end{gather}
We used the subscript $\biw$ in $f_{w\tiF}$ to emphasize that $\biw$
is not a variable for the analysis of the Hamiltonian constraint in
isolation from the momentum constraint. The functional $f_{w\tiF}$ is
the generalization of the functional $F$ defined in
Eq.~(\ref{CF-def-F}). We remark that the operator defined in
Eq.~(\ref{HC-x0}) is continuous but {\em not G\^ateaux
differentiable}. It has G\^ateaux derivatives only along directions in
$L^{\infty}$, not in the whole space $L^2$. This property of the
functional $f_{w\tiF}$ will introduce some technical complexity in the
use of variational methods for the Hamiltonian constraint (see
\Sec\ref{S:HC-ExVM}).

The {\bf weak Dirichlet-Robin boundary value formulation} for the
Hamiltonian constraint is the following: Fix Dirichlet and Robin
boundary data
\begin{equation}
\label{HC-BD}
0 < \mbox{ess}\inf_{\cM_{\tiD}}\hat\phi_{\tiD} \leqs 
\hat\phi_{\tiD} \in L^{\infty}(\partial\cM_{\tiD},0) \cap 
W^{\frac{1}{2},2}(\partial\cM_{\tiD},0),
\quad
\hat \phi_{\tiN}^{*} \in W^{-\frac{1}{2},2}(\partial\cM_{\tiN},0);
\end{equation}
Introduce an extension $\phi_{\tiD}$ of the Dirichlet boundary data as
explained in \Sec\ref{S:WF}, in particular, given $\hat\phi_{\tiD}>0$
on $\partial \cM_{\tiD}$, we can use the Laplace-Beltrami operator to
harmonically extend $\hat\phi_{\tiD}$ to $\phi_{\tiD}$ such that
$\phi_{\tiD}>0$ a.e. in $\cM$.  Given such extension function
$\phi_{\tiD}$, fix any two functions $\phi_{1}$, $\phi_{2} \in
L^{\infty}\cap W^{1,2}$, with the property that
$0<\phi_{1}\leqs\phi_{2}$ and such that $\phi_{\tiD}\in
[\phi_{1},\phi_{2}]\cap W^{1,2}$; Introduce the non-principal part
operator including the Robin boundary conditions,
\begin{equation}
\label{HC-def-f}
f_{w}: [\phi_1,\phi_2]\subset L^2 \to W^{-1,2}_{\tiD},\qquad
f_{w}(\phi)(\un\varphi) :=  f_{w\tiF}(\phi)(\un\varphi)
- \hat \phi_{\tiN}^{*}(\Tr_{\tiN}\un\varphi),
\end{equation}
where the functional $f_{w\tiF}$ is given by Eq.~(\ref{HC-x0}); Let
$A^{1,2}$ be the affine space defined in Eq.~(\ref{WF-def-A}), which
includes the Dirichlet boundary condition; Then, find an element
$\phi\in [\phi_{1},\phi_{2}]\cap A^{1,2}$ solution of the equation
\begin{equation}
\label{HC-LYs}
a_{\tiL}(\phi,\un\varphi) + f_{w}(\phi)(\un\varphi) = 0
\qquad \forall\, \un\varphi\in W^{1,2}_{\tiD}.
\end{equation}

As was the case earlier for analysis of the momentum constraint,
it is convenient to express Eq.~(\ref{HC-LYs}) in terms of operators
instead of bilinear forms. Introduce the operator
\[
A_{\tiL} : W^{1,2}\to W^{-1,2}_{\tiD},\qquad
A_{\tiL}\phi(\un\varphi) := a_{\tiL}(\phi,\un\varphi).
\]
Also recall that, if given any $\phi\in [\phi_{1},\phi_{2}]$, then
$f_{w}(\phi)\in W^{-1,2}_{\tiD}$. Hence, Eq.~(\ref{HC-LYs}) written in
terms of operators is the following: find an element $\phi\in
[\phi_{1},\phi_{2}]\cap A^{1,2}$ solution of
\begin{equation}
\label{HC-LYs1}
A_{\tiL}\phi + f_{w}(\phi) = 0.
\end{equation}

\begin{lemma}
\label{L:pw-HC-CtoW}
Given a smooth vector field $\tbw$, every smooth function $\phi$
solution of the classical Dirichlet-Robin boundary value formulation
for the Hamiltonian constraint Eq.~(\ref{CF-LYs}) is also a solution
of the weak formulation with Eq.~(\ref{HC-LYs}) corresponding to the
equation coefficients and Robin data function given by the following
expressions, which hold for all $\un\varphi\in W^{1,2}_{\tiD}$,
\begin{align*}
a_{\tau}^{*}(\un\varphi)&:=(a_{\tau},\un\varphi),&
a_{\rho}^{*}(\un\varphi)&:=(a_{\rho},\un\varphi),\\
a_{w}^{*}(\un\varphi)&:=(a_{w},\un\varphi),&
\hat\phi_{\tiN}^{*}(\Tr_{\tiN}\un\varphi) 
&:= (\hat\phi_{\tiN},\Tr_{\tiN}\un\varphi)_{\tiN}.
\end{align*}
\end{lemma}

\Proof {\it (Lemma~\ref{L:pw-HC-CtoW}.)~}
The proof is similar to the proof of Lemma~\ref{L:CtoW} and it is not
reproduced here. 
\qed

Given any function $u\in W^{1,2}$, recall the notation
\[
u^{+}:= \mbox{ess~max}\{u,0\},\qquad
u^{-}:=-\mbox{ess~min}\{u,0\}.
\]
An element $\phi_{-}\in W^{1,2}$ is called a {\bf sub-solution} of
Eq.~(\ref{HC-LYs1}) iff the function $\phi_{-}$ satisfies the
inequalities
\begin{equation}
\label{w-HC1-sub-sol}
(\phi_{\tiD}-\phi_{-})^{-} \in W^{1,2}_{\tiD} \mbox{~~~and~~~}
-\bigl[ A_{\tiL}\phi_{-} + f_{w}(\phi_{-})\bigr] 
\in W^{-1,2}_{\tiD +}.
\end{equation}
An element $\phi_{+} \in W^{1,2}$ is called a {\bf super-solution} of
Eq.~(\ref{HC-LYs1}) iff the scalar function $\phi_{+}$ satisfies the
inequalities
\begin{equation}
\label{w-HC1-super-sol}
(\phi_{\tiD}-\phi_{+})^{+} \in W^{1,2}_{\tiD} \mbox{~~~and~~~}
\bigl[ A_{\tiL} \phi_{+} + f_{\biw}(\phi_{+}) \bigr]
\in W^{-1,2}_{\tiD +}.
\end{equation}
The sub and super-solutions of Eq.~(\ref{HC-LYs1}) may depend on the
choice of $a_{w}^{*}$. A sub-solution is called {\bf global} iff
Eq.~(\ref{w-HC1-sub-sol}) holds for every functional $a_{w}^{*}\in
W^{-1,2}_{\tiD}$, and it is called {\bf local} iff it is not
global. 

\subsection{Global and local barriers}
\label{S:WF-LB}

In this section we show that there exist sub- and super-solutions to
the Hamiltonian constraint equation~(\ref{HC-LYs1}) for different
assumptions on the equation coefficients. The results in
Lemmas~\ref{L:HC-LSp}-\ref{L:HC-GSb-s+} are generalizations to the
weak problem of the barriers found in~\cite{jIvM96} in the case of
closed manifolds, scalar curvature $R=-1$, and equation coefficients
with higher regularity. The main idea of this generalization is to
look at candidates for sub- and super-solutions only among the constant
functions, and not among all functions in $[\phi_1,\phi_2]\subset
L^{\infty}$, where $0<\phi_1\leqs\phi_2$. This type of approach is
reasonable, since in the smooth coefficient case there exist sub- and
super-solutions which are indeed constants.

Assume that the background metric $h$ belongs to $C^2 (\overline\cM
,2)$, then the Ricci scalar of curvature $R$ is a continuous function
on the manifold $\overline\cM$. Introduce the constants
\begin{align}
\label{WF-LB-tR-up}
a_{\tau}^{\tiwedge} &:= \sup_{0\neq \un\varphi\in W^{1,2}_{\tiD +}}
\frac{a_{\tau}^{*}(\un\varphi)}{~\|\un\varphi\|_{1,2}},&
a_{\tiR}^{\tiwedge} &:=\sup_{0\neq \un\varphi\in W^{1,2}_{\tiD +}}
\frac{(|a_{\tiR}|,\un\varphi)}{~\|\un\varphi\|_{1,2}},\\
\label{WF-LB-rw-up}
a_{\rho}^{\tiwedge} &:= \sup_{0\neq \un\varphi\in W^{1,2}_{\tiD +}}
\frac{a_{\rho}^{*}(\un\varphi)}{~\|\un\varphi\|_{1,2}},&
a_{w}^{\tiwedge} &:= \sup_{0\neq \un\varphi\in W^{1,2}_{\tiD +}}
\frac{a_{w}^{*}(\un\varphi)}{~\|\un\varphi\|_{1,2}}.
\end{align}
In order that the Lemmas below also hold for the particular case when
the Ricci scalar $R$ vanishes identically, we introduce the constant
$\bar{a}_{\tiR}^{\tiwedge} := 1+ a_{\tiR}^{\tiwedge}$. Given a
two-index tensor $\sigma\in L^{p}(\cM,2)$ and a vector field
$\biw\in\biW^{1,p}$, with $p=12/5$, introduce the functionals
$a_{\sigma}^{*}$ and $a_{\cL w}^{*}$ given by
$a_{\sigma}^{*}(\un\varphi) = (\sigma^2,\un\varphi)/8$ and $a_{\cL
w}^{*}(\un\varphi) =\bigl((\cL\biw)^2,\un\varphi\bigr)/8$, where
$\un\phi\in W^{1,2}_{\tiD}$. Now introduce the further constants
\begin{align*}
a_{\sigma}^{\tiwedge} &:= \sup_{0\neq \un\varphi\in W^{1,2}_{\tiD +}}
\frac{a_{\sigma}^{*}(\un\varphi)}{~\|\un\varphi\|_{1,2}},&
K^{\tiwedge} &:= \sup_{0\neq \un\varphi\in W^{1,2}_{\tiD +}}
\frac{(K,\Tr_{\tiN}\un\varphi)}{~\|\un\varphi\|_{1,2}},\\
\hat\phi_{\tiN}^{\tiwedge} &:= 
\sup_{0\neq \un\varphi\in W^{1,2}_{\tiD +}}
\frac{\hat\phi_{\tiN}^{*}(\Tr_{\tiN}\un\varphi)}{~\|\un\varphi\|_{1,2}},&
\phi_{\tiD}^{\tiwedge} &:= \sup_{\cM} \phi_{\tiD},
\end{align*}
where we recall that the function $\phi_{\tiD}$ is the harmonic
extension of the Dirichlet boundary data $\hat\phi_{\tiD}$ discussed
in \Sec\ref{S:WF}. In an analogous way, switching $\sup$ to $\inf$,
introduce the quantities $a_{\tau}^{\tivee}$, $a_{\tiR}^{\tivee}$,
$a_{\rho}^{\tivee}$, $a_{w}^{\tivee}$, $a_{\sigma}^{\tivee}$,
$K^{\tivee}$, $\hat\phi_{\tiN}^{\tivee}$ and $\phi_{\tiD}^{\tivee}$.
\begin{lemma}{\bf(Local super-solution $R$ bounded)}
\label{L:HC-LSp}
Consider the weak formulation for the Hamiltonian constraint given in
\Sec\ref{S:WF-HC}. Assume that the constants $a_{\tau}^{\tivee}$,
$K^{\tivee}$ are positive, and denote by $\phi_{w+}$ the constant
\begin{equation}
\label{HC-LSp}
\phi_{w+} := \max\, \Bigl\{ 
1, \; \Bigl[\frac{\bar{a}_{\tiR}^{\tiwedge} + a_{\rho}^{\tiwedge} 
+ a_{w}^{\tiwedge}}{a_{\tau}^{\tivee}}\Bigr]^{1/4},
\; \frac{\hat\phi_{\tiN}^{\tiwedge}}{K^{\tivee}},
\; \phi_{\tiD}^{\tiwedge} \Bigr\}.
\end{equation}
Then, $\phi_{w+}$ is a local super-solution of Eq.~(\ref{HC-LYs1}).
\end{lemma}

\Proof {\it (Lemma~\ref{L:HC-LSp}.)~}
We look for a super-solution among the constant functions.  Therefore,
let $\phi_0$ be any constant in $[\phi_1,\phi_2]$, with $\phi_1>0$,
then the following inequalities hold for every element $\un\varphi\in
W^{1,2}_{\tiD+}$,
\begin{align*}
f_{w\tiF}(\phi_0)(\un\varphi) 
&= ( a_{\tau} \phi_0^{5})^{*}(\un\varphi) 
+ ( a_{\tiR} \phi_0)^{*}(\un\varphi)
- (a_{\rho}\phi_0^{-3})^{*}(\un\varphi)
- ( a_{w} \phi_0^{-7})^{*}(\un\varphi)\\
&= a_{\tau}^{*}(\un\varphi) \, \phi_0^{5} 
+ a_{\tiR}^{*}(\un\varphi) \,\phi_0
- a_{\rho}^{*}(\un\varphi) \,\phi_0^{-3}
- a_{w}^{*}(\un\varphi) \,\phi_0^{-7}\\
&\geqs \bigl( a_{\tau}^{\tivee} \, \phi_0^5 
- \bar{a}_{\tiR}^{\tiwedge}\, \phi_0 
- a_{\rho}^{\tiwedge}\,\phi_0^{-3}
- a_{w}^{\tiwedge}\, \phi_0^{-7} \bigr) 
\,\|\un\varphi\|_{1,2}.
\end{align*}
Introduce the polynomial on $\phi_0$ given by
\begin{equation}
\label{HC-def-q}
q(\phi_0) := a_{\tau}^{\tivee} \, \phi_0^5 
- \bar{a}_{\tiR}^{\tiwedge}\, \phi_0 
- a_{\rho}^{\tiwedge}\,\phi_0^{-3}
- a_{w}^{\tiwedge}\, \phi_0^{-7}.
\end{equation}
The assumptions that the constants $a_{\tau}^{\tivee}$ and
$\bar{a}_{\tiR}^{\tiwedge}$ are strictly positive, while the constants
$a_{\rho}^{\tiwedge}$ and $a_{w}^{\tiwedge}$ are non-negative imply
that there exists a unique positive root of this polynomial. The proof
consists of three steps. First, there exists at least one positive
root of the polynomial $q$, because for $\phi_0$ large enough
$q(\phi_0)$ is positive, and for $\phi_0$ close to zero from positive
values $q(\phi_0)$ is negative, as it can be seen from the following
expression,
\[
q(\phi_0) = \phi_0^{-7} \bigl[
 a_{\tau}^{\tivee} \, \phi_0^{12} 
- \bar{a}_{\tiR}^{\tiwedge}\, \phi_0^8 
- a_{\rho}^{\tiwedge}\,\phi_0^{4}
- a_{w}^{\tiwedge} \bigr],
\]
where the term between brackets becomes negative for small enough
$\phi_0$. Second, this positive root is unique, since the function $q$
is increasing for all $\phi_0 > \alpha_0 := [\bar{a}_{\tiR}^{\tiwedge}
/ (5a_{\tau}^{\tivee})]^{1/4}$ (the proof is to verify that $q'>0$ for
$\phi_0 >\alpha_0$); and the function $q$ satisfies the inequality
$q(\phi_0)\leqs r(\phi_0) := a_{\tau}^{\tivee}\,\phi_0^5
-\bar{a}_{\tiR}^{\tiwedge}\,\phi_0$ for all positive numbers $\phi_0$.
Since $r(\alpha_1)=0$ for $\alpha_1 := [\bar{a}_{\tiR}^{\tiwedge} /
a_{\tau}^{\tivee} ]^{1/4}$, and $\alpha_1>\alpha_0$ (so the root of
the polynomial $q$ must belong to the interval where $q$ is
increasing), we then conclude that the root of the polynomial $q$ is
unique. Denote by $\bar\phi_0$ the unique positive root of the
polynomial $q$. Since $\bar\phi_0 >\alpha_0$, then $q(\phi_0) >
q(\bar\phi_0) =0$ for any $\phi_0 >\bar\phi_0$. The idea now is to
find an upper bound for the root $\bar\phi_0$. The result is going to
be the first two expressions on the right hand side in
Eq.~(\ref{HC-LSp}); the remaining two terms on the right hand side of
Eq.~(\ref{HC-LSp}) will account for the boundary contributions.

In the case that $\bar\phi_0\leqs 1$ (which could be verified, for
example, by explicit evaluation), then choose a candidate for
super-solution to be $\tilde\phi_{w+} =1$. In the case that
$\bar\phi_0 > 1$, then there exists an upper bound for this root, as
can be seen from the following argument. Given any $\phi_0
\geqs 1$, then the following inequalities hold,
\[
(\phi_0)^{n+1} \geqs 1 \RI 
-(\phi_0)^{-(n+1)} \geqs -1 \RI 
-(\phi_0)^{-n} \geqs -\phi_0, 
\]
therefore, this inequality for $n=3$ and $n=7$ implies that for all
$\phi_0\geqs 1$ holds
\[
q(\phi_0) \geqs s(\phi_0) := a_{\tau}^{\tivee}\,(\phi_0)^5 
- \bigl(\bar{a}_{\tiR}^{\tiwedge} + a_{\rho}^{\tiwedge} 
+ a_{w}^{\tiwedge} \bigr) \phi_0.
\]
This new polynomial $s$ vanishes at 
\[
\bar\phi_1 := \Bigl[\frac{ \bar{a}_{\tiR}^{\tiwedge} 
+ a_{\rho}^{\tiwedge} + a_{w}^{\tiwedge} }{a_{\tau}^{\tivee}}
\Bigr]^{1/4},
\]
and the inequality $\bar\phi_1>\alpha_0$ says that $\bar\phi_1$
belongs to the interval where the polynomial $q$ is
increasing. Therefore, we have that
\[
q(\bar\phi_1) \geqs s(\bar\phi_1) =0 
\mbox{~~and~~} q(\bar\phi_0)=0 \RI
\bar\phi_1 \geqs \bar\phi_0.
\]
So, in this case $\bar\phi_0>1$, choose the candidate for
super-solution to be $\tilde\phi_{w+}=\bar\phi_1$. Then, introducing
the constant
\[
\tilde\phi_{w+} := \max \Bigl\{ 
1, \Bigl[\frac{\bar{a}_{\tiR}^{\tiwedge} + a_{\rho}^{\tiwedge} 
+ a_{w}^{\tiwedge}}{a_{\tau}^{\tivee}} \Bigr]^{1/4}
\Bigr\},
\]
we have established that the following inequality holds
\begin{equation}
\label{HC-LSp-1}
f_{w\tiF}(\phi_0)(\un\varphi) \geqs 0 
\qquad \forall \, \un\varphi \in W^{1,2}_{\tiD +},
\qquad \forall \, \phi_0\geqs\tilde\phi_{w+}.
\end{equation}

We now account for the boundary contributions.
The definitions of $\hat\phi_{\tiN}^{\tiwedge}$ and $K^{\tivee}$
imply that for any constant $\phi_0$ holds
\[
\bigl( [K\phi_0 -\hat\phi_{\tiN}], \Tr_{\tiN}\un\varphi \bigr)_{\tiN}
\geqs (K^{\tivee} \phi_0 -\hat\phi_{\tiN}^{\tiwedge})
\, \|\un\varphi\|_{1,2}
\qquad  \forall \, \un\varphi \in W^{1,2}_{\tiD +}.
\]
In particular, defining the constant
\[
\overline \phi_{w+} := \max \, \Bigl\{ 
1, \; \Bigl[\frac{\bar{a}_{\tiR}^{\tiwedge} + a_{\rho}^{\tiwedge} 
+ a_{w}^{\tiwedge}}{a_{\tau}^{\tivee}}\Bigr]^{1/4},
\; \frac{\hat\phi_{\tiN}^{\tiwedge}}{K^{\tivee}}
 \Bigr\},
\]
follows that $\overline\phi_{w+}\geqs\tilde\phi_{w+}$ and the
following inequality holds
\begin{equation}
\label{HC-LSp-2}
\bigl( [K \phi_{0} -\hat\phi_{\tiN}], 
\Tr_{\tiN}\un\varphi \bigr)_{\tiN} \geqs 0
\qquad \forall\, \phi_0\geqs \overline\phi_{w+}.
\end{equation}
Adding Eqs.~(\ref{HC-LSp-1}) and~(\ref{HC-LSp-2}) we conclude that
\[
(K \phi_0,\Tr_{\tiN}\un\varphi)_{\tiN} +
f_{w}(\phi_0)(\un\varphi) \geqs 0
\qquad \forall \,\un\varphi\in W^{1,2}_{\tiD +}, 
\qquad \forall \,\phi_0\geqs \overline\phi_{w+}.
\]
Recalling now that any constant $\phi_0$ satisfies $(\nabla\phi_0
,\nabla\un\varphi)=0$, we conclude that
\[
A_{\tiL}\phi_0(\un\varphi) + f_w(\phi_0)(\un\varphi) \geqs 0
\qquad  \forall \un\varphi\in W^{1,2}_{\tiD +},
\qquad \forall \,\phi_0\geqs \overline\phi_{w+}.
\]
Finally, introduce the constant $\phi_{w+}$ as given by
Eq.~(\ref{HC-LSp}). In particular, this constant satisfies
$(\phi_{\tiD} -\phi_{w+})^{+} =0$ and
$\phi_{w+}\geqs\overline\phi_{w+}$, so the following inequality holds,
\[
(\phi_{\tiD}-\phi_{w+})^{+}\in W^{1,2}_{\tiD +} \mbox{~~~and~~~} 
\big[A_{\tiL}\phi_{w+} + f_{w}(\phi_{w+}) \bigr]
\in W^{-1,2}_{\tiD +},
\]
which establishes that $\phi_{\biw+}$ is a super-solution of
Eq.~(\ref{HC-LYs1}).\qed

We now find a global super-solution for the Hamiltonian and momentum
constraint Eq.~(\ref{WF-LYs1})-(\ref{WF-LYm1}), where global means
that the super-solution is independent of the vector field $\biw$
solution of the momentum constraint Eq.~(\ref{WF-LYm1}). This global
super-solution is a generalization suitable to our weak setting of the
super-solution derived in~\cite{jIvM96}. We use the same idea as in
\Sec\ref{S:WF-LB}, that is, we look for super-solutions only among
the constant functions.  The ``near-CMC'' assumption on the trace of
the extrinsic curvature made in \cite{jIvM96} to construct the
super-solution is still present here, although in weaker norms.

\begin{lemma}{\bf(Global super-solution $R$ bounded)}
\label{L:CS-GSp}
Consider the weak formulation for the Hamiltonian and momentum
constraints given in \Sec\ref{S:WF}. Assume that the numbers
$(a_{\tau}^{\tivee}-\ttK_1)$ and $K^{\tivee}$ are positive, where the
constant $\ttK_1$ is defined in Eq.~(\ref{CS-def-K1}). Denote by
$\phi_{+}$ the constant
\begin{equation}
\label{CS-GSp}
\phi_{+} := \max\, \Bigl\{ 
1, \; \Bigl[\frac{\bar{a}_{\tiR}^{\tiwedge} + a_{\rho}^{\tiwedge} 
+ \ttK_2}{a_{\tau}^{\tivee}-\ttK_1}\Bigr]^{1/4},
\; \frac{\hat\phi_{\tiN}^{\tiwedge}}{K^{\tivee}},
\; \hat\phi_{\tiD}^{\tiwedge} \Bigr\},
\end{equation}
with the constant $\ttK_2$ defined in Eq.~(\ref{CS-def-K2}). Then, the
constant $\phi_{+}$ is a global super-solution of
Eqs.~(\ref{WF-LYs1})-(\ref{WF-LYm1}).
\end{lemma}

\Proof {\it (Lemma~\ref{L:CS-GSp}.)~}
Consider the weak formulation in \Sec\ref{S:WF}. Let $\phi_0$ be any
constant in $[\phi_1,\phi_2]$ with $\phi_1>0$, then the following
inequalities hold,
\begin{align}
\nonumber
f_{\tiF}(\phi_0,\biw)(\un\varphi) 
&= ( a_{\tau}^{*} \phi_0^{5})(\un\varphi) 
+ ( a_{\tiR}^{*} \phi_0)(\un\varphi)
- (a_{\rho}^{*}\phi_0^{-3})(\un\varphi)
- ( a_{w}^{*} \phi_0^{-7})(\un\varphi)\\
\nonumber
&= a_{\tau}^{*}(\un\varphi) \, \phi_0^{5} 
+ a_{\tiR}^{*}(\un\varphi) \,\phi_0
- a_{\rho}^{*}(\un\varphi) \,\phi_0^{-3}
- a_{w}^{*}(\un\varphi) \,\phi_0^{-7}\\
\label{CS-ineq1}
&\geqs \bigl( a_{\tau}^{\tivee} \, \phi_0^5 
- \bar{a}_{\tiR}^{\tiwedge}\, \phi_0 
- a_{\rho}^{\tiwedge}\,\phi_0^{-3}
- a_{w}^{\tiwedge}\, \phi_0^{-7} \bigr) 
\,\|\un\varphi\|_{1,2}.
\end{align}
The number $a_{w}^{\tiwedge}$ in the last term is bounded when $\biw$
is solution of the momentum constraint Eq.~(\ref{WF-LYm}) with any
source $\phi\in [\phi_1,\phi_2]$. For the proof, start with the
definition of $a_{w}^{\tiwedge}$ in \Sec\ref{S:WF-LB}, where
$a_{w}^{*}(\un\varphi) = (a_{w},\un\varphi)$ with $a_{w}\in
L^{6/5}$ and $\un\varphi\in W^{1,2}_{\tiD}$. Then, the following
inequalities hold
\[
(a_{w},\un\varphi) \leqs \|a_{w}\|_{\frac{6}{5}} \, \|\un\varphi\|_{6}
\leqs c_s \,\|a_{w}\|_{\frac{6}{5}} \, \|\un\varphi\|_{1,2},
\]
where we used the imbedding $W^{1,2}\subset L^6$, and where $c_s$ is the
positive imbedding constant that relates the norm of these spaces. The
inequality above holds for all $\un\varphi\in W^{1,2}_{\tiD+}$, and in
particular holds for the supremum in that space, hence
\[
a_{w}^{\tiwedge} \leqs c_s\, \|a_{w}\|_{\frac{6}{5}}.
\]
Now, the definition of the function $a_{w}$, standard inequalities
and the notation $p=12/6$ show that,
\[
\|a_{w}\|_{\frac{6}{5}} = \frac{1}{8} \, \|\sigma + \cL\biw\|_{p}^2
\leqs \frac{1}{4} \,\Bigl( \|\sigma\|_{p}^2 
+ \|\cL\biw\|_{p}^2 \Bigr)
\leqs \frac{1}{4} \,\Bigl( \|\sigma\|_{p}^2 
+ c_{\cL}\, \|\biw\|_{1,p}^2 \Bigr).
\]
In \Sec\ref{S:MC-WP} and \Sec\ref{S:MC-RS} it is shown that there
exist positive constants $c_1$ and $c_2$ such that
\[
\|\biw\|_{1,p} \leqs \|\phi\|_{\infty}^6 
\,\|\bib_{\tau}^{*}\|_{-1,p}
+ \|\bib_{j}^{*}\|_{-1,p} 
+c_1 \, \|\hat\biw_{\tiIN}^{*}\|_{-\frac{1}{p},p,\tiIN}
+c_2 \, \|\biw_{\tiID}\|_{1,p}.
\]
Then, the bound for the number $a_{w}^{\tiwedge}$ can be written as
\begin{equation}
\label{CS-aw-bound}
a_{w}^{\tiwedge} \leqs 
\ttK_1 \, \|\phi\|_{\infty}^{12} + \ttK_2,
\end{equation}
where the constants $\ttK_1$ and $\ttK_2$ are given by
\begin{align}
\label{CS-def-K1}
\ttK_1 &:= 4c_sc_{\cL} \|\bib_{\tau}^{*}\|_{-1,p}^2,\\
\label{CS-def-K2}
\ttK_2 &:= \frac{c_s}{4}\; \|\sigma\|_{p}^2  
+ c_s c_{\cL} \,\Bigl( \|\bib_{j}^{*}\|_{-1,p}^2
+  c_1^2 \,\|\hat\biw_{\tiIN}^{*}\|_{-\frac{1}{p},p,\tiIN}^2 
+ c_2^2 \, \|\biw_{\tiID}\|_{1,p}^2 \Bigr).
\end{align}
Introducing this expression in Eq.~(\ref{CS-ineq1}), one finds
that for all $\phi_0$ constant and all $\phi$, both in
$[\phi_1,\phi_2]$, it holds that
\[
f_{\tiF}(\phi_0,\biw) \geqs 
\Bigl[ a_{\tau}^{\tivee} \, \phi_0^5 
- \bar{a}_{\tiR}^{\tiwedge}\, \phi_0 
- a_{\rho}^{\tiwedge}\,\phi_0^{-3}
- \bigl( \ttK_1 \, \|\phi\|_{\infty}^{12} + \ttK_2 \bigr)
 \phi_0^{-7} \Bigr]\,\|\un\varphi\|_{1,2}.
\]
Now, evaluate this expression at $\phi_0=\phi=\phi_2$. The result is
\begin{equation}
\label{CS-ineq2}
f_{\tiF}(\phi_2,\biw) \geqs  \Bigl[
\bigl(a_{\tau}^{\tivee}-\ttK_1\bigr) \, \phi_{2}^{5} 
- \bar{a}_{\tiR}^{\tiwedge} \, \phi_{2}
- a_{\rho}^{\tiwedge} \, \phi_{2}^{-3}  
- \ttK_2 \,\phi_{2}^{-7} \Bigr]\,\|\un\varphi\|_{1,2}.
\end{equation}
The assumption in Lemma~\ref{L:CS-GSp} implies that
$a_{\tau}^{\tivee}-\ttK_1 >0$, so the polynomial 
\[
\tilde q(\phi_2) :=
\bigl(a_{\tau}^{\tivee}-\ttK_1\bigr) \, \phi_{2}^{5} 
- \bar{a}_{\tiR}^{\tiwedge} \, \phi_{2}
- a_{\rho}^{\tiwedge} \, \phi_{2}^{-3}  
- \ttK_2 \,\phi_{2}^{-7}
\]
has the same form as the polynomial $q$ introduced in
Eq.~(\ref{HC-def-q}). Therefore, the analysis done on the polynomial
$q$ in the proof of Lemma~\ref{L:HC-LSp} holds for the polynomial
$\tilde q$, in particular, $\tilde q$ has a unique positive root
$\bar\phi_0$.  The remainder of the proof involves finding an upper
bound for $\bar\phi_0$, and the argument is almost identical to the
one given in the proof of Lemma~\ref{L:HC-LSp}.

In the case that $\bar\phi_0\leqs 1$ (which could be verified, for
example, by explicit evaluation), then choose a candidate for
super-solution to be $\tilde\phi_{+} =1$. In the case that
$\bar\phi_0 > 1$, then there exists an upper bound for this root, as
can be seen from the following argument. Given any $\phi_0
\geqs 1$, then the following inequalities hold,
\[
(\phi_0)^{n+1} \geqs 1 \RI 
-(\phi_0)^{-(n+1)} \geqs -1 \RI 
-(\phi_0)^{-n} \geqs -\phi_0,
\]
therefore, this inequality for $n=3$ and $n=7$ implies that for all
$\phi_0\geqs 1$ holds
\[
\tilde q(\phi_0) \geqs \tilde s(\phi_0) := 
\bigl( a_{\tau}^{\tivee}- \ttK_1\bigr) \,(\phi_0)^5 
- \bigl(\bar{a}_{\tiR}^{\tiwedge} + a_{\rho}^{\tiwedge} 
+ \ttK_2 \bigr) \phi_0.
\]
This new polynomial $\tilde s$ vanishes at the point
\[
\bar\phi_1 := \Bigl[\frac{ \bar{a}_{\tiR}^{\tiwedge} 
+ a_{\rho}^{\tiwedge} + \ttK_2 }{a_{\tau}^{\tivee}- \ttK_1} 
\Bigr]^{1/4},
\]
which belongs to the interval where the polynomial $\tilde q$ is
increasing. Therefore, we have that
\[
\tilde q(\bar\phi_1) \geqs \tilde s(\bar\phi_1) =0 \RI
\bar\phi_1 \geqs \bar\phi_0.
\]
So, in this case $\bar\phi_0>1$, choose the candidate for
super-solution to be $\tilde\phi_{+}=\bar\phi_1$. Then, introducing
the constant
\[
\tilde\phi_{+} := \max \Bigl\{ 
1, \Bigl[\frac{\bar{a}_{\tiR}^{\tiwedge} + a_{\rho}^{\tiwedge} 
+ \ttK_2}{a_{\tau}^{\tivee} - \ttK_1} \Bigr]^{1/4}
\Bigr\},
\]
we have established that the following inequality holds
\begin{equation}
\label{CS-GSp-1}
f_{\tiF}(\phi_2,\biw)(\un\varphi) \geqs 0 
\qquad \forall \, \un\varphi \in W^{1,2}_{\tiD +},
\qquad \forall \, \phi_2\geqs\tilde\phi_{+},
\end{equation}
and for all vector field $\biw\in\biW^{1,p}$ solution of the momentum
constraint Eq.~(\ref{WF-LYm1}) with source function $\phi\in
[0,\phi_2]$.

As in the proof of Lemma~\ref{L:HC-LSp}, what remains is to account for
the boundary contributions.
The definitions of $\hat\phi_{\tiN}^{\tiwedge}$ and $K^{\tivee}$
imply that for any constant $\phi_2$ holds
\[
\bigl( [K\phi_2 -\hat\phi_{\tiN}], \Tr_{\tiN}\un\varphi \bigr)_{\tiN}
\geqs (K^{\tivee} \phi_2 -\hat\phi_{\tiN}^{\tiwedge})
\, \|\un\varphi\|_{1,2}
\qquad  \forall \, \un\varphi \in W^{1,2}_{\tiD +}.
\]
In particular, defining the constant
\[
\overline \phi_{+} := \max \, \Bigl\{ 
1, \; \Bigl[\frac{\bar{a}_{\tiR}^{\tiwedge} + a_{\rho}^{\tiwedge} 
+ \ttK_2 }{a_{\tau}^{\tivee} - \ttK_1}\Bigr]^{1/4},
\; \frac{\hat\phi_{\tiN}^{\tiwedge}}{K^{\tivee}}
 \Bigr\},
\]
follows that $\overline\phi_{+}\geqs\tilde\phi_{+}$ and the following
inequality holds
\begin{equation}
\label{CS-GSp-2}
\bigl( [K \phi_2 -\hat\phi_{\tiN}], 
\Tr_{\tiN}\un\varphi \bigr)_{\tiN} \geqs 0
\qquad \forall\, \phi_2\geqs \overline\phi_{+}.
\end{equation}
Adding Eqs.~(\ref{CS-GSp-1}) and~(\ref{CS-GSp-2}) we conclude that
\[
(K \phi_2,\Tr_{\tiN}\un\varphi)_{\tiN} +
f(\phi_2,\biw)(\un\varphi) \geqs 0
\qquad \forall \,\un\varphi\in W^{1,2}_{\tiD +}, 
\qquad \forall \,\phi_2\geqs \overline\phi_{+},
\]
and for all $\biw\in \biW^{1,p}$ solution of the momentum constraint
Eq.~(\ref{WF-LYm1}) with source function $\phi\in [0,\phi_2]$.
Recalling now that any constant $\phi_2$ satisfies $(\nabla\phi_2
,\nabla\un\varphi)=0$, we conclude that
\[
A_{\tiL}\phi_2(\un\varphi) + f(\phi_2,\biw)(\un\varphi) \geqs 0
\qquad  \forall \un\varphi\in W^{1,2}_{\tiD +},
\qquad \forall \,\phi_2\geqs \overline\phi_{+}.
\]
Finally, employ now the constant $\phi_{+}$ as given by
Eq.~(\ref{CS-GSp}). This constant satisfies the conditions
$(\hat\phi_{\tiD} -\phi_{+})^{+} =0$ and
$\phi_{+}\geqs\overline\phi_{+}$, so the following inequality holds,
\[
(\phi_{\tiD}-\phi_{+})^{+}\in W^{1,2}_{\tiD+} \mbox{~~~and~~~} 
\big[A_{\tiL}\phi_{+} + f(\phi_{+},\biw) \bigr]
\in W^{-1,2}_{\tiD+},
\]
for all $\biw\in \biW^{1,p}$ solution of the momentum constraint
Eq.~(\ref{WF-LYm1}) with source $\phi\in [0,\phi_{+}]$. This
establishes that $\phi_{+}$ is a global super-solution of
Eqs.~(\ref{WF-LYs1})-(\ref{WF-LYm1}).\qed

Consider now the particular case of a background metric $h\in
C^2(\overline\cM,2)$ having a strictly negative Ricci scalar of
curvature, that is, 
\begin{equation}
\label{HC-GSb-R-}
-a_{\tiR}^{\tiwedge} 
\leqs  \frac{(a_{\tiR},\un\varphi)}{~\|\un\varphi\|_{1,2}} \leqs 
- a_{\tiR}^{\tivee} < 0
\qquad \forall \un\varphi\in W^{1,2}_{\tiD +}.
\end{equation}
In this case it is possible to obtain a global sub-solution of
Eq.~(\ref{HC-LYs1}).
\begin{lemma}{\bf(Global sub-solution for $R<0$)}
\label{L:HC-GSb-R-}
Consider the weak formulation for the Hamiltonian constraint given in
\Sec\ref{S:WF-HC}. Assume that the constants $a_{\tau}^{\tiwedge}$,
$K^{\tiwedge}$, $\hat\phi_{\tiN}^{\tivee}$, and $\phi_{\tiD}^{\tivee}$
are positive, and the Ricci scalar $R$ satisfies
Eq.~(\ref{HC-GSb-R-}). Denote by $\phi_{-}$ the constant
\begin{equation}
\label{HC-GSb-R-1}
\phi_{-} := \min\, \Bigl\{ 
\Bigl( \frac{a_{\tiR}^{\tivee}}{a_{\tau}^{\tiwedge}} \Bigr)^{1/4},
\; \frac{\hat\phi_{\tiN}^{\tivee}}{K^{\tiwedge}},
\; \phi_{\tiD}^{\tivee} \Bigr\}.
\end{equation}
Then, $\phi_{-}$ is a global sub-solution of Eq.~(\ref{HC-LYs1}).
\end{lemma}

\Proof {\it (Lemma~\ref{L:HC-GSb-R-}.)~}
We look for the sub-solution among the constant functions. Therefore,
let $\phi_0$ be any constant in $[\phi_1,\phi_2]$, with $\phi_1>0$,
then the following inequalities hold for every element $\un\varphi\in
W^{1,2}_{\tiD+}$,
\begin{align*}
f_{w\tiF}(\phi_0)(\un\varphi) &= (a_{\tau}\phi_0^5)^{*}(\un\varphi)
+(a_{\tiR}\phi_0)^{*}(\un\varphi) 
- (a_{\rho}\phi_0^{-3})^{*}(\un\varphi) 
- (a_{w} \phi_0^{-7})^{*}(\un\varphi)\\
&\leqs (a_{\tau}\phi_0^5)^{*}(\un\varphi) 
+ (a_{\tiR}\phi_0)^{*}( \un\varphi)\\
&\leqs (a_{\tau},\un\varphi) \, \phi_0^5
+ (a_{\tiR}, \un\varphi)\, \phi_0\\
&\leqs \Bigl[ a_{\tau}^{\tiwedge} \,\phi_0^5
- a_{\tiR}^{\tivee} \,\phi_0 \Bigr] \,\|\un\varphi\|_{1,2},
\end{align*}
where we used that both functionals $a_{\rho}^{*}$ and $a_w^{*}$
belong to the space $W^{-1,2}_{\tiD+}$, and the number
$\phi_0>0$. Introduce the polynomial
\[
q(\phi_0) := a_{\tau}^{\tiwedge} \,\phi_0^5 
- a_{\tiR}^{\tivee} \,\phi_0.
\]
There exists a unique positive root for $q$ given by the number
$\tilde\phi_{-} :=\bigl( a_{\tiR}^{\tivee} /
a_{\tau}^{\tiwedge}\bigr)^{1/4}$, and for all $0<\phi_0<
\tilde\phi_{-}$ the corresponding values $q(\phi_0)$ are
negative. Therefore, the following inequality holds
\begin{equation}
\label{HC-GSb-1}
f_{w\tiF}(\phi_0)(\un\varphi) \leqs 0 
\qquad \forall \, \un\varphi\in W^{1,2}_{\tiD +}, 
\qquad \forall \, \phi_0 \in(0, \tilde\phi_{-}],
\qquad \forall\,a_{w}^{*}\in\biW^{-1,2}_{\tiD+}.
\end{equation}

We now account for the boundary contributions. The definitions of the
numbers $\hat\phi_{\tiN}^{\tivee}$ and $K^{\tiwedge}$ imply that for
any constant $\phi_0$, it holds that
\[
\bigl( [K\phi_0 -\hat\phi_{\tiN}], \Tr_{\tiN}\un\varphi \bigr)_{\tiN}
\leqs (K^{\tiwedge} \phi_0 -\hat\phi_{\tiN}^{\tivee})
\, \|\un\varphi\|_{1,2}
\qquad  \forall \, \un\varphi \in W^{1,2}_{\tiD +}.
\]
In particular, defining the constant
\[
\overline \phi_{-} := \min \,\Bigl\{ 
\Bigl( \frac{a_{\tiR}^{\tivee}}{a_{\tau}^{\tiwedge}} \Bigr)^{1/4},
\; \frac{\hat\phi_{\tiN}^{\tivee}}{K^{\tiwedge}}
 \Bigr\},
\]
it follows that $0 <\overline\phi_{-}\leqs\tilde\phi_{-}$ and the
following inequality holds
\begin{equation}
\label{HC-GSb-2}
\bigl( [K \phi_{0} -\hat\phi_{\tiN}], 
\Tr_{\tiN}\un\varphi \bigr)_{\tiN} \leqs 0,
\qquad \forall\, \phi_0\in(0, \overline\phi_{-}].
\end{equation}
Adding Eqs.~(\ref{HC-GSb-1}) and~(\ref{HC-GSb-2}) we conclude that
\[
(K \phi_0,\Tr_{\tiN}\un\varphi)_{\tiN} +
f_{w}(\phi_0)(\un\varphi) \leqs 0
\qquad \forall \,\un\varphi\in W^{1,2}_{\tiD +} 
\qquad \forall \,\phi_0\in (0,\overline\phi_{-}].
\]
Recalling now that any constant $\phi_0$ satisfies $(\nabla\phi_0
,\nabla\un\varphi)=0$, we conclude that
\[
A_{\tiL}\phi_0(\un\varphi) + f_w(\phi_0)(\un\varphi) \leqs 0
\qquad  \forall \un\varphi\in W^{1,2}_{\tiD +}
\qquad \forall \, \phi_0\in (0,\overline\phi_{-}].
\]
Finally, introduce the constant $\phi_{-}$ as given by
Eq.~(\ref{HC-GSb-R-1}). In particular, this constant satisfies
$(\phi_{\tiD} - \phi_{-})^{-} =0$ and $0
<\phi_{-}\leqs\overline\phi_{-}$, so the following inequality holds,
\[
(\phi_{\tiD}-\phi_{-})^{-}\in W^{1,2}_{\tiD} \mbox{~~~and~~~} 
-\big[A_{\tiL}\phi_{-} + f_{w}(\phi_{-}) \bigr]
\in W^{-1,2}_{\tiD +},
\qquad \forall\, a_{w}^{*} \in \biW^{-1,2}_{\tiD+},
\]
which establishes that $\phi_{-}$ is a global sub-solution of
Eq.~(\ref{HC-LYs1}).\qed

In the case that the Ricci scalar of curvature is non-negative, then
it is not clear whether a constant and positive sub-solution to
Eq.~(\ref{HC-LYs1}) exists. The latter exists when the conformally
rescaled matter energy density $\rho$ satisfies the condition
$a_{\rho}^{\tivee}>0$. This result is summarized in the following two
Lemmas below.

\begin{lemma}{\bf (Global sub-solution for $R\geqs 0$ and 
$a_{\rho}^{\tivee} >0$)}
\label{L:HC-GSb-r+}
Consider the weak formulation for the Hamiltonian constraint given in
\Sec\ref{S:WF-HC}. Assume that the constants $a_{\tau}^{\tiwedge}$,
$a_{\rho}^{\tivee}$, $\hat\phi_{\tiN}^{\tivee}$ $K^{\tiwedge}$, and
$\phi_{\tiD}^{\tivee}$ are positive. Let $\phi_{-}$ be the constant
\begin{equation}
\label{HC-GSb-r+}
\phi_{-} := 
\min\, \Bigl\{ 
\left[\frac{1}{2 a_{\tau}^{\tiwedge}}\,
\Bigl(-a_{\tiR}^{\tiwedge} +\sqrt{
(a_{\tiR}^{\tiwedge})^2 + 4a_{\tau}^{\tiwedge}
a_{\rho}^{\tivee}}\,\Bigr)\right]^{1/4}, 
\; \frac{\hat\phi_{\tiN}^{\tivee}}{K^{\tiwedge}},
\; \phi_{\tiD}^{\tivee} \Bigr\}.
\end{equation}
Then, $\phi_{-}$ is a global sub-solution of Eq.~(\ref{HC-LYs1}).
\end{lemma}

\Proof {\it (Lemma~\ref{L:HC-GSb-r+}.)~}
We look for the sub-solution among the constant functions. Therefore,
let $\phi_0$ be any constant in $[\phi_1,\phi_2]$ with $\phi_1>0$,
then the following inequalities hold for every element $\un\varphi\in
W^{1,2}_{\tiD+}$,
\begin{align*}
f_{w\tiF}(\phi_0)(\un\varphi) &= 
  (a_{\tau}\phi_0^5)^{*}(\un\varphi)
+ (a_{\tiR}\phi_0)^{*}(\un\varphi) 
- (a_{\rho}\phi_0^{-3})^{*}(\un\varphi) 
- (a_{w}\phi_0^{-7})^{*}(\un\varphi)\\
&\leqs (a_{\tau}\phi_0^5)^{*}(\un\varphi)  
+ (a_{\tiR}\phi_0)^{*}(\un\varphi)
- (a_{\rho}\phi_0^{-3})^{*}(\un\varphi)\\
&\leqs a_{\tau}^{*}(\un\varphi) \, \phi_0^5
+ a_{\tiR}^{*}(\un\varphi)\, \phi_0
- a_{\rho}^{*}(\un\varphi)\, \phi_0^{-3} \\
&\leqs \Bigl[ a_{\tau}^{\tiwedge} \,\phi_0^5
+ a_{\tiR}^{\tiwedge} \,\phi_0 
-a_{\rho}^{\tivee}\, \phi_0^{-3} \Bigr] \,\|\un\varphi\|_{1,2},
\end{align*}
where the used the assumption that the functional $a_w^{*}$ is
non-negative. Introduce the polynomial
\[
q(\phi_0) := a_{\tau}^{\tiwedge} \,\phi_0^5
+ a_{\tiR}^{\tiwedge} \,\phi_0  -a_{\rho}^{\tivee}\, \phi_0^{-3},
\]
which is a non-decreasing function, because its derivative
\[
q'(\phi_0) = 5 \,a_{\tau}^{\tiwedge} \,\phi_0^4
+ a_{\tiR}^{\tiwedge}  + 3\, a_{\rho}^{\tivee}\, \phi_0^{-4},
\]
is strictly positive for non-zero $\phi_0$. Rewrite the polynomial $q$
as follows,
\begin{equation}
\label{HC-GSb-r+ineq1}
q(\phi_0) \leqs \phi_0^{-3}\, q_{\rho}(\phi_0) \quad
\mbox{~~with~~}\quad 
q_{\rho}(\phi_0) := a_{\tau}^{\tiwedge} \,\phi_0^{8}
+ a_{\tiR}^{\tiwedge} \,\phi_0^4 -a_{\rho}^{\tivee}.
\end{equation}
There exists a unique positive root $\phi_{\rho}$ of the
polynomial $q_{\rho}$ given by
\[
\phi_{\rho} = \left[ \frac{1}{2 a_{\tau}^{\tiwedge}}\,
\Bigl(-a_{\tiR}^{\tiwedge} + \sqrt{ (a_{\tiR}^{\tiwedge})^2 +  
4a_{\tau}^{\tiwedge}a_{\rho}^{\tivee}}\,\Bigr) \right]^{1/4}.
\]
Then, the inequality in Eq.~(\ref{HC-GSb-r+ineq1}) and the
non-decreasing property of $q$ imply that the polynomial $q$ satisfies
\[
q(\phi_0) \leqs 0 \qquad \forall \, 0< \phi_0 \leqs \phi_{\rho}.
\]
We then summarize the discussion above saying that the following
inequality holds
\begin{equation}
\label{HC-GSb-r+1}
f_{w\tiF}(\phi_0)(\un\varphi) \leqs 0 
\qquad \forall \, \un\varphi\in W^{1,2}_{\tiD +},
\qquad \forall \,\phi_0\in(0, \phi_{\rho}],
\end{equation}
and for all $a_{w}^{*}\in\biW^{-1,2}_{\tiD+}$.  From this point
forward, the proof is identical to the proof of
Lemma~\ref{L:HC-GSb-R-}. What remains is to account for the boundary
contributions.  The definitions of $\hat\phi_{\tiN}^{\tivee}$ and
$K^{\tiwedge}$ imply that for any constant $\phi_0$, it holds that
\[
\bigl( [K\phi_0 -\hat\phi_{\tiN}], \Tr_{\tiN}\un\varphi \bigr)_{\tiN}
\leqs (K^{\tiwedge} \phi_0 -\hat\phi_{\tiN}^{\tivee})
\, \|\un\varphi\|_{1,2}
\qquad  \forall \, \un\varphi \in W^{1,2}_{\tiD +}.
\]
In particular, defining the constant $\di \overline \phi_{-}
:=\min\,\bigl\{\phi_{\rho}
,\;\frac{\hat\phi_{\tiN}^{\tivee}}{K^{\tiwedge}}\bigr\}$, it follows
that $0 <\overline\phi_{-}\leqs\phi_{\rho}$, and the following
inequality holds
\begin{equation}
\label{HC-GSb-r+2}
\bigl( [K \phi_{0} -\hat\phi_{\tiN}], 
\Tr_{\tiN}\un\varphi \bigr)_{\tiN} \leqs 0,
\qquad \forall\, \phi_0\in (0,\overline\phi_{-}].
\end{equation}
Adding Eqs.~(\ref{HC-GSb-r+1}) and~(\ref{HC-GSb-r+2}) we conclude that
\[
(K \phi_0,\Tr_{\tiN}\un\varphi)_{\tiN} +
f_{w}(\phi_0)(\un\varphi) \leqs 0
\qquad \forall \,\un\varphi\in W^{1,2}_{\tiD +} 
\qquad \forall \, \phi_0\in(0, \overline\phi_{-}].
\]
Recalling now that any constant $\phi_0$ satisfies $(\nabla\phi_0
,\nabla\un\varphi)=0$, we conclude that
\[
A_{\tiL}\phi_0(\un\varphi) + f_w(\phi_0)(\un\varphi) \leqs 0
\qquad  \forall \un\varphi\in W^{1,2}_{\tiD +}
\qquad \forall \, \phi_0\in(0, \overline\phi_{-}].
\]
Finally, introduce the constant $\phi_{-}$ as given by
Eq.~(\ref{HC-GSb-r+}). In particular, this constant satisfies
$(\phi_{\tiD} - \phi_{-})^{-} =0$ and $0
<\phi_{-}\leqs\overline\phi_{-}$, so the following inequality holds,
\[
(\phi_{\tiD}-\phi_{-})^{-}\in W^{1,2}_{\tiD} \mbox{~~~and~~~} 
-\big[A_{\tiL}\phi_{-} + f_{w}(\phi_{-}) \bigr]
\in W^{-1,2}_{\tiD +},
\qquad \forall\, a_{w}^{*}\in W^{-1,2}_{\tiD+},
\]
which establishes that $\phi_{-}$ is a global sub-solution of
Eq.~(\ref{HC-LYs1}).\qed

Again in the case that the Ricci scalar of curvature $R$ is
non-negative there exists a sub-solution to the Hamiltonian and
momentum constraint Eqs.~(\ref{WF-LYs1})-(\ref{WF-LYm1}), when the
trace-free divergence-free two-index tensor $\sigma$ is big enough. By
big we mean that $a_{\sigma}^{\tivee} > \sigma_0$, with the positive
constant $\sigma_0$ given in Eq.~(\ref{WF-LB-def-s0}). This result
requires the near-CMC hypotheses present in Lemma~\ref{L:CS-GSp}, and
is summarized below.

\begin{lemma}{\bf (Global sub-solution for $R\geqs 0$ and 
$a_{\sigma}^{\tivee} > \sigma_0$)}
\label{L:HC-GSb-s+}
Consider the weak formulation for the Hamiltonian and momentum
constraints given in \Sec\ref{S:WF} and assume that the hypotheses in
Lemma~\ref{L:CS-GSp} hold. Assume that the constants
$a_{\tau}^{\tiwedge}$, $\hat\phi_{\tiN}^{\tivee}$ $K^{\tiwedge}$, and
$\phi_{\tiD}^{\tivee}$ are positive, while the constant
$a_{\sigma}^{\tivee} >\sigma_0$, with the positive constant $\sigma_0$
given in Eq.~(\ref{WF-LB-def-s0}). Denote by $\phi_{\sigma}$ the only
positive root of the polynomial $q_{\sigma}(x) := a_{\tau}^{\tiwedge}
x^3 + a_{\tiR}^{\tiwedge}x^2 - a_{\sigma}^{\tivee}/4$, where $x\in\R$.
Let $\phi_{-}$ be the constant
\begin{equation}
\label{HC-GSb-s+}
\phi_{-} := \min\, \Bigl\{ \phi_{\sigma}, 
\; \frac{\hat\phi_{\tiN}^{\tivee}}{K^{\tiwedge}},
\; \phi_{\tiD}^{\tivee} \Bigr\}. 
\end{equation}
Then, $\phi_{-}$ is a global sub-solution of Eq.~(\ref{WF-LYs1}).
\end{lemma}

\Proof {\it (Lemma~\ref{L:HC-GSb-s+}.)~}
Consider the weak formulation for the Hamiltonian and momentum
constraints given in \Sec\ref{S:WF}. The $a_{w}^{*}$ defined in
Eq.~(\ref{WF-dual-coeff}) belongs to the space $W^{-1,2}_{\tiD+}$,
since the function $a_{w}=(\sigma+\cL\biw)^2/8$ belongs to the space
$L^{p/2}$, with $p=12/5$. Given any positive number $\epsilon$, the
inequality $2|\sigma_{ab} (\cL w)^{ab}|\leqs\epsilon \sigma^2 +
(\cL\biw)^2/\epsilon$ implies that function $a_{w}$ satisfies the
following inequality,
\[
8a_{w} = \sigma^2+(\cL \biw)^2 + 2 \sigma_{ab}(\cL w)^{ab}\\
\geqs (1-\epsilon)\,\sigma^2 
- \Bigl( \frac{1}{\epsilon} -1\Bigr) \,(\cL \biw)^2,
\]
hence, for any number $\epsilon\in (0,1)$ the functional $a_{w}^{*}$
must fulfill the inequality
\[
a_{w}^{*}(\un\varphi)= (a_{w},\un\varphi) 
\geqs (1-\epsilon) \,a_{\sigma}^{\tivee} \,\|\un\varphi\|_{1,2}
- \frac{1}{8}\,\Bigl( \frac{1}{\epsilon} -1\Bigr) \, 
\|\cL \biw\|_{p}^2\,\|\un\varphi\|_{6}
\qquad\forall\ \un\varphi\in W^{1,2}_{\tiD+}.
\]
For every vector field $\biw\in\biW^{1,p}$ solution of the momentum
constraint Eq.~(\ref{WF-LYm1}) with source function $\phi$ holds the
inequality in Eq.~(\ref{w-MC-est2}), therefore there exist positive
constants $c_{\cL}$, $c_1$ and $c_2$ such that
\[
\|\cL \biw\|_{p}^2 \leqs 4c_{\cL}^2 \Bigl[
\|\phi\|_{\infty}^{12} \,\|\bib_{\tau}^{*}\|_{-1,p}^2 
+\|\bib_j^{*}\|_{-1,p}^2 
+ c_1^2\,\|\hat\biw_{\tiIN}^{*}\|_{-\frac{1}{p},p,\tiIN}^2
+ c_2^2\,\|\biw_{\tiID}\|_{1,p}^2 \Bigr].
\]
Hence, for all source functions $\phi\in [0,\phi_{+}]$, where the
constant $\phi_{+}$ is the positive super-solution found in
Lemma~\ref{L:CS-GSp}, holds the inequality
\[
\|\cL\biw\|_{p}^2 \leqs 4 c_{\cL}^2 \Bigl[ 
\phi_{+}^{12} \,\|\bib_{\tau}^{*}\|_{-1,p}^2 
+\|\bib_j^{*}\|_{-1,p}^2 
+ c_1^2\,\|\hat\biw_{\tiIN}^{*}\|_{-\frac{1}{p},p,\tiIN}^2
+ c_2^2\,\|\biw_{\tiID}\|_{1,p}^2 \Bigr].
\]
Introducing the positive constant $c_t$ defined by the inequality
$\|\un\varphi\|_6\leqs c_t\,\|\un\varphi\|_{1,2}$, and the constant
$\sigma_0$ given by
\begin{equation}
\label{WF-LB-def-s0}
\sigma_0 := 2c_t c_{\cL}^2 \Bigl[ 
\phi_{+}^{12} \,\|\bib_{\tau}^{*}\|_{-1,p}^2 
+\|\bib_j^{*}\|_{-1,p}^2 
+ c_1^2\,\|\hat\biw_{\tiIN}^{*}\|_{-\frac{1}{p},p,\tiIN}^2
+ c_2^2\,\|\biw_{\tiID}\|_{1,p}^2 \Bigr],
\end{equation}
we obtain that
\[
a_{w}^{*}(\un\varphi) \geqs 
\Bigl[ (1-\epsilon) \,a_{\sigma}^{\tivee}
- \frac{\sigma_0}{4}\,\Bigl( \frac{1}{\epsilon} -1\Bigr) \Bigr]\, 
\|\un\varphi\|_{1,2}.
\]
Choose the number $\epsilon =1/2$, then we get
$a_{w}^{*}(\un\varphi)\geqs (1/2)\bigl( a_{\sigma}^{\tivee}
-\sigma_0/2\bigr)\,\|\un\varphi\|_{1,2}$. By assumption, we know that
$a_{\sigma}^{\tivee} >\sigma_0$, therefore, we conclude that
\begin{equation}
\label{WF-LB-ineq-01}
a_{w}^{*}(\un\varphi) \geqs 
\frac{1}{4} \, a_{\sigma}^{\tivee} \, \|\un\varphi\|_{1,2}
\qquad \forall \,\un\varphi \in W^{1,2}_{\tiD+}.
\end{equation}
Having established the inequality above we now start looking for the
sub-solution among the constant functions. Therefore, let $\phi_0$ be
any constant in $[\phi_1,\phi_{+}]$ with $0<\phi_1\leqs\phi_{+}$, then
the following inequalities hold for every element $\un\varphi\in
W^{1,2}_{\tiD+}$,
\begin{align*}
f_{w\tiF}(\phi_0)(\un\varphi) &= 
  (a_{\tau}\phi_0^5)^{*}(\un\varphi)
+ (a_{\tiR}\phi_0)^{*}(\un\varphi) 
- (a_{\rho}\phi_0^{-3})^{*}(\un\varphi) 
- (a_{w}\phi_0^{-7})^{*}(\un\varphi)\\
&\leqs a_{\tau}^{*}(\un\varphi) \, \phi_0^5
+ a_{\tiR}^{*}(\un\varphi)\, \phi_0
- a_{\rho}^{*}(\un\varphi)\, \phi_0^{-3} 
- a_{w}^{*}(\un\varphi)\, \phi_0^{-7}\\
&\leqs \Bigl[ a_{\tau}^{\tiwedge} \,\phi_0^5
+ a_{\tiR}^{\tiwedge} \,\phi_0 
-a_{\rho}^{\tivee}\, \phi_0^{-3} 
-\frac{1}{4}\,a_{\sigma}^{\tivee}\,\phi_{0}^{-7} \Bigr] 
\,\|\un\varphi\|_{1,2},
\end{align*}
where the used Eq.~(\ref{WF-LB-ineq-01}) to obtain the last line
above. Introduce the polynomial
\[
q(\phi_0) := a_{\tau}^{\tiwedge} \,\phi_0^5
+ a_{\tiR}^{\tiwedge} \,\phi_0  -a_{\rho}^{\tivee}\, \phi_0^{-3}
-\frac{1}{4}\,a_{\sigma}^{\tivee}\,\phi_0^{-7},
\]
which is a non-decreasing function, because its derivative
\[
q'(\phi_0) = 5 \,a_{\tau}^{\tiwedge} \,\phi_0^4
+ a_{\tiR}^{\tiwedge}  + 3\, a_{\rho}^{\tivee}\, \phi_0^{-4}
+ \frac{7}{4}\, a_{\sigma}^{\tivee}\,\phi_0^{-8},
\]
is strictly positive for non-zero $\phi_0$. Rewrite the polynomial $q$
as follows
\begin{equation}
\label{HC-GSb-s+ineq2}
q(\phi_0) = \phi_0^{-7}\, q_{\sigma}(\phi_0) \quad
\mbox{~~with~~}\quad 
q_{\sigma}(\phi_0) := a_{\tau}^{\tiwedge} \,\phi_0^{12}
+ a_{\tiR}^{\tiwedge} \,\phi_0^8 - \frac{1}{4}\,a_{\sigma}^{\tivee}.
\end{equation}
This polynomial $q_{\sigma}$ has a unique positive root
$\phi_{\sigma}$, which exists because $q_{\sigma}(0) <0$ and
$\lim_{\phi_0\to \infty}q_{\sigma}(\phi_0) = \infty$, while the root
is unique because the polynomial $q_{\sigma}$ is an increasing
function for positive $\phi_0$. So, Eq.~(\ref{HC-GSb-s+ineq2}) implies
that $q(\phi_{\sigma})=0$, and together with the property of the
polynomial $q$ being non-decreasing, we conclude that
\[
q(\phi_0) \leqs 0 \qquad \forall \, 0< \phi_0 \leqs\phi_{\sigma}.
\]
We then summarize the discussion above saying that the following
inequality holds
\begin{equation}
\label{HC-GSb-s+1}
f_{w\tiF}(\phi_0)(\un\varphi) \leqs 0 
\qquad \forall \, \un\varphi\in W^{1,2}_{\tiD +},
\qquad \forall \,\phi_0\in(0, \phi_{\sigma}],
\end{equation}
and for all vector field $\biw\in\biW^{1,p}$ solution of the momentum
constraint Eq.~(\ref{WF-LYm1}) with source function $\phi\in
[0,\phi_{+}]$, where the constant $\phi_{+}$ is the super-solution
found in Lemma~\ref{L:CS-GSp}. From this point forward, the proof is
identical to the proof of Lemma~\ref{L:HC-GSb-R-}. What remains is to
account for the boundary contributions.  The definitions of
$\hat\phi_{\tiN}^{\tivee}$ and $K^{\tiwedge}$ imply that for any
constant $\phi_0$, it holds that
\[
\bigl( [K\phi_0 -\hat\phi_{\tiN}], \Tr_{\tiN}\un\varphi \bigr)_{\tiN}
\leqs (K^{\tiwedge} \phi_0 -\hat\phi_{\tiN}^{\tivee})
\, \|\un\varphi\|_{1,2}
\qquad  \forall \, \un\varphi \in W^{1,2}_{\tiD +}.
\]
In particular, defining the constant $\di \overline \phi_{-}
:=\min\,\bigl\{\phi_{\sigma}
,\;\frac{\hat\phi_{\tiN}^{\tivee}}{K^{\tiwedge}}\bigr\}$, it then follows
that $0 <\overline\phi_{-}\leqs\phi_{\sigma}$, and the following
inequality holds
\begin{equation}
\label{HC-GSb-s+2}
\bigl( [K \phi_{0} -\hat\phi_{\tiN}], 
\Tr_{\tiN}\un\varphi \bigr)_{\tiN} \leqs 0,
\qquad \forall\, \phi_0\in (0,\overline\phi_{-}].
\end{equation}
Adding Eqs.~(\ref{HC-GSb-s+1}) and~(\ref{HC-GSb-s+2}) we conclude that
\[
(K \phi_0,\Tr_{\tiN}\un\varphi)_{\tiN} +
f_{w}(\phi_0)(\un\varphi) \leqs 0
\qquad \forall \,\un\varphi\in W^{1,2}_{\tiD +} 
\qquad \forall \, \phi_0\in(0, \overline\phi_{-}].
\]
Recalling now that any constant $\phi_0$ satisfies $(\nabla\phi_0
,\nabla\un\varphi)=0$, we conclude that
\[
A_{\tiL}\phi_0(\un\varphi) + f_w(\phi_0)(\un\varphi) \leqs 0
\qquad  \forall \un\varphi\in W^{1,2}_{\tiD +}
\qquad \forall \, \phi_0\in(0, \overline\phi_{-}].
\]
Finally, introduce the constant $\phi_{-}$ as given by
Eq.~(\ref{HC-GSb-s+}). In particular, this constant satisfies
$(\phi_{\tiD} - \phi_{-})^{-} =0$ and $0
<\phi_{-}\leqs\overline\phi_{-}$, so the following inequality holds,
\[
(\phi_{\tiD}-\phi_{-})^{-}\in W^{1,2}_{\tiD} \mbox{~~~and~~~} 
-\big[A_{\tiL}\phi_{-} + f_{w}(\phi_{-}) \bigr]
\in W^{-1,2}_{\tiD +},
\]
and for all vector field $\biw\in\biW^{1,p}$ solution of the momentum
constraint Eq.~(\ref{WF-LYm1}) with source function $\phi\in
[0,\phi_{+}]$, where the constant $\phi_{+}$ is the super-solution
found in Lemma~\ref{L:CS-GSp}. This establishes that $\phi_{-}$ is a
global sub-solution of Eq.~(\ref{WF-LYs1}) in the interval
$[0,\phi_{+}]$.\qed

\subsection{Summary on barriers}
\label{S:HC-SB}

We present in this short Section a summary of the various results we
have obtained in \Sec\ref{S:WF-LB} for weak sub- and
super-solutions. We state the constant sub- and super-solutions for
each value of the Ricci scalar $R$. We do not state again the
definition of the various constants that define the sub- and
super-solutions, which can be found at the beginning of
\Sec\ref{S:WF-LB}. Similarly, we do not state again the assumptions on
the coefficients and data in the Hamiltonian and momentum constraint
equations needed to construct the barriers. These assumptions can be
found in \Sec\ref{S:WF-LB}, or they can be read out directly from the
barriers expressions, because they are the sufficient conditions that
guarantee that the barriers are positive and finite numbers.
\begin{align*}
&R\mbox{~~bounded}&&\left\{
\begin{aligned} 
\phi_{w+} &:= \max\, \Bigl\{ 
1, \; \Bigl[\frac{\bar{a}_{\tiR}^{\tiwedge} + a_{\rho}^{\tiwedge} 
+ a_{w}^{\tiwedge}}{a_{\tau}^{\tivee}}\Bigr]^{1/4},
\; \frac{\hat\phi_{\tiN}^{\tiwedge}}{K^{\tivee}},
\; \phi_{\tiD}^{\tiwedge} \Bigr\},\\
\phi_{+} &:= \max\, \Bigl\{ 
1, \; \Bigl[\frac{\bar{a}_{\tiR}^{\tiwedge} + a_{\rho}^{\tiwedge} 
+ \ttK_2}{a_{\tau}^{\tivee}-\ttK_1}\Bigr]^{1/4},
\; \frac{\hat\phi_{\tiN}^{\tiwedge}}{K^{\tivee}},
\; \hat\phi_{\tiD}^{\tiwedge} \Bigr\},
\end{aligned}\right. \\
&R<0&\phi_{-} &:= \min\, \Bigl\{ 
\Bigl( \frac{a_{\tiR}^{\tivee}}{a_{\tau}^{\tiwedge}} \Bigr)^{1/4},
\; \frac{\hat\phi_{\tiN}^{\tivee}}{K^{\tiwedge}},
\; \phi_{\tiD}^{\tivee} \Bigr\},\\
&R\geqs 0&\phi_{-} &:= \left\{
\begin{aligned}
&\min\, \Bigl\{ \phi_{\rho}, 
\; \frac{\hat\phi_{\tiN}^{\tivee}}{K^{\tiwedge}},
\; \phi_{\tiD}^{\tivee} \Bigr\} 
\mbox{~~if~~} a_{\rho}^{\tivee} >0,
\quad a_{\sigma}^{\tivee}\geqs 0,\\
&\min\, \Bigl\{ \phi_{\sigma},
\; \frac{\hat\phi_{\tiN}^{\tivee}}{K^{\tiwedge}},
\; \phi_{\tiD}^{\tivee} \Bigr\} 
\mbox{~~if~~} a_{\rho}^{\tivee} \geqs 0,
\quad a_{\sigma}^{\tivee}>\sigma_0.\\
\end{aligned} \right. 
\end{align*}
Regarding the sub-solution for Ricci scalar $R\geqs 0$, only the
constant $\phi_{\rho}$ has been given explicitly by the expression
$\phi_{\rho}:=\left[\Bigl(-a_{\tiR}^{\tiwedge} +\sqrt{
(a_{\tiR}^{\tiwedge})^2 + 4a_{\tau}^{\tiwedge}
a_{\rho}^{\tivee}}\,\Bigr)/(2 a_{\tau}^{\tiwedge})\right]^{1/4}$,
while the constant $\phi_{\sigma}$ has not been given explicitly, but
it has been proven that $\phi_{\sigma}$ is a finite, positive number.

\subsection{{\em A~priori} $L^{\infty}$-bounds on $W^{1,2}$-solutions}
\label{S:HC-apriori}

We now establish some related {\em a~priori} $L^{\infty}$-bounds on
any $W^{1,2}$-solution to the Hamiltonian constraint equation.
Although such results are standard for semi-linear scalar problems
with monotone nonlinearities (for example, see~\cite{jJ85}), the
nonlinearity appearing in the Hamiltonian constraint becomes
non-monotone when $R$ becomes negative.
Nonetheless, we are able to obtain {\em a~priori} $L^{\infty}$-bounds
on solutions to the Hamiltonian constraint in all cases including
the non-monotone case.  The results are based on a new
abstract result (Lemma~\ref{L:ape} in the Appendix) which holds for
general semi-linear problems in ordered Banach spaces, under very weak
assumptions on the nonlinearity (see the second assumption in
parts~\ref{ape-i} and~\ref{ape-ii} in Lemma~\ref{L:ape}).  Monotone
nonlinearities have the required property, but the property is much 
weaker than monotonicity and is satisfied for more general nonlinearities 
such as the one appearing in the Hamiltonian constraint.
The results here generalize the {\em a~priori} $L^{\infty}$-bounds on 
weak solutions appearing previously in an earlier set of unpublished 
notes\footnote{
M.~Holst,
{\em Weak solutions to the {Einstein} constraint equations on manifolds 
with boundary}.
Notes from the 2002--2003 Caltech Visitors Program in the Numerical
Simulation of Gravitational Wave Sources.
}
and in a thesis\footnote{
J.~Kommemi,
{\em Variational methods for weak solutions to the {Einstein}
{Hamiltonian} constraint on finite domains with boundary}.
Honors Thesis, Department of Mathematics, UC San Diego, 2007.
}
to the weakest possible assumptions on the
coefficients appearing in the Hamiltonian nonlinearity.
\begin{theorem}{\bf({\em A~priori} bounds on $W^{1,2}$-solutions)}
\label{T:ape}
Consider the weak formulation for the Hamiltonian constraint given in
\Sec\ref{S:WF-HC}, and assume the hypotheses in Lemma~\ref{L:HC-LSp},
in Lemma~\ref{L:HC-GSb-R-} and either in Lemma~\ref{L:HC-GSb-r+} or in
Lemma~~\ref{L:HC-GSb-s+}.  Assume that given a functional
$a_w^* \in W_{\tiD+}^{-1,2}$, there exists a solution $\phi\in
W^{1,2}$ of Hamiltonian constraint Eq.~(\ref{HC-LYs1}).  Then, there
exists positive numbers $\phi_{\tivee}$, $\phi_{\tiwedge}$ with
$\phi_{\tivee} \leqs
\phi_{\tiwedge}$, such that
\begin{equation}
\label{apriori-bounds}
0 < \phi_{\tivee} \leqs \phi \leqs \phi_{\tiwedge},
\quad \quad a.e.~in~ \cM.
\end{equation}
\end{theorem}

\Proof {\it (Theorem~\ref{T:ape}.)~}
Let $\phi_{\tivee}$ and $\phi_{\tiwedge}$ be the (constant) sub-
and super-solutions, respectively, given in
Lemmas~\ref{L:HC-LSp}-\ref{L:HC-GSb-s+}. In the proofs of these Lemmas
it is established that the function $f_{w}$ given in
Eq.~(\ref{HC-def-f}) is monotone increasing for its argument
$\tilde\phi$ satisfying $\tilde\phi\geqs \phi_{\tiwedge}$ and for
$0<\tilde\phi\leqs\phi_{\tivee}$. Therefore, Lemma~\ref{L:ape} implies
that Eq.~(\ref{apriori-bounds}) holds. This establishes the
Theorem.\qed

\subsection{Results using variational methods}
\label{S:HC-ExVM}

The Hamiltonian constraint Eq.~(\ref{HC-LYs1}) can be written as the
Euler condition for stationarity of a real-valued functional in a 
Banach space. Direct
methods in the calculus of variations can be used to find the points that 
minimize this functional in particular types of closed sets in Banach
spaces: closed sets under weak convergence, which we denote here as
\closedw. When the \closedw set is chosen appropriately, and the
barriers found in \Sec\ref{S:WF-LB} are incorporated into the
argument, we can show that the minimum is actually a solution of the
Euler condition for the functional, that is, of the Hamiltonian
constraint equation. The solution found with this approach requires
fewer regularity assumptions on the data, and the resulting solution
is weaker (has less regularity) than the solution found using barrier
methods.
The variational structure was exploited in \cite{mH01a} to develop an
approximation theory and corresponding error estimates for 
numerical approximations to the Hamiltonian constraint.

Let $a_{\tiL}: W^{1,2}_{\tiD}\times W^{1,2}_{\tiD}\to\R$ be a bilinear
form with action defined in Eq.~(\ref{HC-def-aL}). Let $a_{\tau}^{*}$,
$a_{\tiR}^{*}$, $a_{\rho}^{*}$, and $a_{w}^{*}$ the functionals
defined in Eq.~(\ref{HC-coeff})-(\ref{HC-Coeff-aR}). Let
$\phi_{\tiD}\in L^{\infty}\cap W^{1,2}$ be the positive, harmonic
extension of the Dirichlet data function $\hat\phi_{\tiD}$ discussed
in \Sec\ref{S:WF}, and $\hat\phi_{\tiN}^{*}$ be the Robin data
functional, both data defined in Eq.~(\ref{HC-BD}).  Finally, let
$\phi_1$, $\phi_2\in L^{\infty}$ be functions satisfying
$0<\phi_1\leqs\phi_2$, but otherwise arbitrary, and denote by $U
:=\bigr( [\phi_1-\phi_{\tiD},\phi_2-\phi_{\tiD}]\cap
W^{1,2}_{\tiD}\bigr)\subset W^{1,2}_{\tiD}$. Then, introduce the
functional
\begin{equation}
\label{VM-def-J}
J_{\tiL}: U\subset W^{1,2}_{\tiD} \to\overline\R,\qquad
J_{\tiL}(\varphi) := \frac{1}{2}\, a_{\tiL}(\varphi,\varphi) 
+ G_w(\varphi),
\end{equation}
where the functional $G_{w}$ is given by
\[
G_{w}(\varphi) := g_{w}(\phi_{\tiD}+\varphi) 
- \hat\phi_{\tiN}^{*}(\Tr_{\tiN}\varphi)
+ a_{\tiL}(\phi_{\tiD},\varphi),
\]
with the functional $g_{w}(\phi)$ having the form
\begin{equation}
\label{VM-def-g}
g_{w}(\phi) := \frac{1}{6}\, (a_{\tau}\phi^5)^{*}(\phi)
+ \frac{1}{2}\, (a_{\tiR}\phi)^{*}(\phi)
+ \frac{1}{2}\, (a_{\rho}\phi^{-3})^{*}(\phi)
+ \frac{1}{6}\, (a_{w}\phi^{-7})^{*}(\phi);
\end{equation}
while the definition $\overline\R:= \R\cup \{+\infty\}\cup\{-\infty\}$
is explained in the Appendix.

\begin{theorem}{\bf (Existence of a minimizer)}
\label{T:VM-EM}
Consider the functional $J_{\tiL}: U\subset
W^{1,2}_{\tiD}\to\overline\R$ defined in Eq.~(\ref{VM-def-J}). Fix a
positive extension $\phi_{\tiD}\in L^{\infty}\cap W^{1,2}$ of the
Dirichlet boundary data, and fix the Robin boundary data
$\hat\phi_{\tiN}^{*}\in W^{-\frac{1}{2},2}
(\partial\cM_{\tiN},0)$. Fix the functionals $a_{\tau}^{*}$,
$a_{\rho}^{*}$, $a_{w}^{*}$, and the function $K$ satisfying the
inequalities $a_{\tau}^{\tivee} > 0$, $a_{\rho}^{\tivee}\geqs 0$,
$a_{w}^{\tivee}\geqs 0$, $K^{\tivee} > 0$. Then, there exists an
element $\varphi\in U$ minimizer of the functional $J_{\tiL}$ in $U$,
that is,
\[
J_{\tiL}(\varphi) = \inf_{\un\varphi\in U}J_{\tiL}(\un\varphi).
\]
Furthermore, if the Ricci scalar $R$ is non-negative in $\overline
\cM$, then the minimizer $\varphi$ is unique.
\end{theorem}

\Proof {\it (Theorem~\ref{T:VM-EM}.)~}
The set $U$ is closed in $W^{1,2}$, as the following argument shows:
By contradiction, assume that there exists a sequence $\{\phi_n\}\in
U$ such that $\phi_n\to\phi_0$ in $W^{1,2}$ but $\phi_0\notin U$. In
particular $\phi_n\to\phi_0$ in $L^2$ and the inequality
$\bigl|\|\phi_n\|-\|\phi_0\|\bigr|\leqs \|\phi_n-\phi_0\|$ implies
that $\|\phi_n\|\to\|\phi_0\|$. The assumption that the limiting
element $\phi_0\notin U$ implies that there exists a set $\cM_0\subset
\cM$ with $\mbox{meas}(\cM_0)\neq 0$ such that $\phi_0>\phi_2$ or
$\phi_0<\phi_1$. We consider here only the first case, the second one
is proven in an analogous way. Then, we have the following
inequalities,
\begin{align*}
\|\phi_n\|^2_{\cM} &= 
\|\phi_n\|^2_{\cM_0} 
+ \|\phi_n\|^2_{(\cM\backslash\cM_0)}\\
&\leqs \|\phi_2\|^2_{\cM_0} 
+ \|\phi_n\|^2_{(\cM\backslash\cM_0)},
\end{align*}
where we used the notation $\|\phi_n\|_{\cM_0}
:=\|\phi_n\|_{L^2(\cM_0)}$. Then taking the limit on both sides
\begin{align*}
\lim_{n\to\infty}\|\phi_n\|^2_{\cM} 
&\leqs \|\phi_2\|^2_{\cM_0} 
+ \|\phi_0\|^2_{(\cM\backslash\cM_0)}\\
&< \|\phi_0\|^2_{\cM_0} 
+ \|\phi_0\|^2_{(\cM\backslash\cM_0)} 
= \|\phi_0\|^2_{\cM},
\end{align*}
from which we conclude that $\lim_{n\to\infty}\|\phi_n\| <
\|\phi_0\|$, contradicting the assumption above that
$\lim_{n\to\infty}\|\phi_n\| =\|\phi_0\|$. Therefore, the set $U$ is
closed. In addition, the set $U$ is convex, therefore it is shown in
the Appendix that $U$ is \closedw.

We now show that the functional $J_{\tiL}$ is coercive. First notice
that the condition $\phi_1>0$ implies that the functional with values
$g_{w}(\phi)$ is continuous and bounded in $[\phi_1,\phi_2]\subset
L_{+}^{\infty}$, therefore there exists a positive constant $c_g$ such
that
\[
|g_{w}(\phi)|\leqs c_g,\qquad \forall \phi \in [\phi_1,\phi_2].
\]
Second, the Robin term is bounded as the following calculation shows
\begin{align*}
|\hat\phi_{\tiN}^{*}(\Tr_{\tiN}\varphi)| &\leqs
\|\hat\phi_{\tiN}^{*}\|_{-\frac{1}{2},2,\tiN}\,
\|\Tr_{\tiN}\varphi\|_{\frac{1}{2},2,\tiN}\\
&\leqs c_t\, \|\hat\phi_{\tiN}^{*}\|_{-\frac{1}{2},2,\tiN}\,
\|\varphi\|_{1,2}\\
&\leqs 
\frac{c_t^2}{2\epsilon}\,
\|\hat\phi_{\tiN}^{*}\|_{-\frac{1}{2},2,\tiN}^2
+ \frac{\epsilon}{2}\, \|\varphi\|_{1,2}^2
\qquad\forall \,\varphi\in W^{1,2},
\end{align*}
where $\epsilon$ is an arbitrary positive constant, and $c_t$ is a
positive constant that bounds the trace operator. Third, the Dirichlet
term is also bounded because the bilinear form $a_{\tiL}$ is bounded,
then the following inequalities hold,
\begin{align*}
a_{\tiL}(\phi_{\tiD},\varphi) 
&= (\nabla\phi_{\tiD},\nabla\varphi) 
+(K\Tr_{\tiN} \phi_{\tiD},\Tr_{\tiN}\varphi)_{\tiN}\\
&\leqs (1+c_t\,\|K\|_{\infty})\,
\|\phi_{\tiD}\|_{1,2}\, \|\varphi\|_{1,2}\\
&\leqs \frac{1}{2\epsilon} (1+c_t\,\|K\|_{\infty})^2\,
\|\phi_{\tiD}\|_{1,2}^2\, 
+\frac{\epsilon}{2} \,\|\varphi\|_{1,2}^2,
\qquad \forall \,\varphi\in W^{1,2}.
\end{align*}
These calculations show that 
\begin{gather*}
|G_{w}(\varphi)|\leqs c_{\tiG}(\epsilon) 
+ \epsilon\,\|\varphi\|_{1,2}^2,
\qquad \forall \,\varphi\in U\subset W^{1,2}_{\tiD},\\
c_{\tiG}(\epsilon) =c_g
+\frac{1}{2\epsilon}\,\Bigl[c_t^2\,
\|\hat\phi_{\tiN}^{*}\|_{-\frac{1}{2},2,\tiN}^2 
+(1+c_t\,\|K\|_{\infty})^2\|\phi_{\tiD}\|_{1,2}^2\Bigr].
\end{gather*}
Fourth, the following argument, similar to the one used in the proof
of Corollary~\ref{C:GI}, shows that $a_{\tiL}$ satisfies a G\aa rding
inequality. Start with a bound on the Robin term in $a_{\tiL}$,
\begin{align*}
(K\Tr_{\tiN}\varphi,\Tr_{\tiN}\varphi)_{\tiN} 
&\leqs \|K\|_{\infty}\, \|\Tr\varphi\|_{\tiN}^2\\
&\leqs c_t\,\|K\|_{\infty} \,\|\varphi\|\,\|\nabla\varphi\|\\
&\leqs \frac{c_t^2}{2\epsilon}\,\|K\|_{\infty}^2\,\|\varphi\|^2
+ \frac{\epsilon}{2}\,\|\nabla\varphi\|^2\\
&\leqs \frac{c_t^2}{2\epsilon}\,\|K\|_{\infty}^2\,\|\varphi\|^2
+ \frac{\epsilon}{2}\,\|\nabla\varphi\|^2_{1,2};
\end{align*}
then it is not difficult to derive the following inequality on
$a_{\tiL}$,
\begin{align*}
a_{\tiL}(\varphi,\varphi) &=
(\nabla \varphi,\nabla \varphi)
+ (K\Tr_{\tiN}\varphi,\Tr_{\tiN}\varphi)_{\tiN}\\
&\geqs \|\varphi\|_{1,2}^2 -\|\varphi\|^2 
- \frac{c_t^2}{2\epsilon}\,\|K\|_{\infty}^2\,\|\varphi\|^2
- \frac{\epsilon}{2}\,\|\nabla\varphi\|^2_{1,2}\\
&\geqs \bigl(1-\frac{\epsilon}{2}\bigr)\, \|\varphi\|_{1,2}^2 
- \bigl( 1 + \frac{c_t^2}{2\epsilon}\,\|K\|_{\infty}^2\bigr)
\,\|\varphi\|^2\\
&\geqs \bigl(1-\frac{\epsilon}{2}\bigr)\, \|\varphi\|_{1,2}^2 
- c_a(\epsilon),
\end{align*}
where $c_a(\epsilon) :=\bigl[ 1 +c_t^2\,\|K\|_{\infty}^2 /
(2\epsilon)\bigr] c_s\,\|\phi_2\|^2_{\infty}$ and $c_s$ is the
positive constant in the imbedding $L^{\infty}\subset L^2$. Finally,
the fifth step is to put all these inequalities together,
\begin{align*}
J_{\tiL}(\varphi) &\geqs 
\frac{1}{2}\bigl( 1-\frac{\epsilon}{2}\bigr) \;\|\varphi\|_{1,2}^2
-c_a(\epsilon) -c_{\tiG}(\epsilon) - \epsilon\,\|\varphi\|_{1,2}^2\\
&\geqs \bigl(\frac{1}{2}-\frac{5}{4}\,\epsilon\bigr)\, 
\|\varphi\|_{1,2}^2 -c_{\tiJ}(\epsilon),\qquad
\forall \varphi\in U\subset W^{1,2},
\end{align*}
where $c_{\tiJ}(\epsilon)= c_a(\epsilon)+ c_{\tiG}(\epsilon)$. By
choosing $\epsilon$ positive and small enough the inequality above
establishes that $J_{\tiL}$ is coercive. Therefore, $J_{\tiL}$ is
proper, and is also trivially bounded below by $-c_{\tiJ}(\epsilon)$.

We now show that the functional $J_{\tiL}$ is \lscw, and we do it term
by term on $J_{\tiL}$. We start with the term proportional to
$a_{\tiL}(\varphi,\varphi)$, which is \lscw by the following three
facts: First, the norm in a Banach space is a \lscw functional
(statement proved in the Appendix); Second, the compactness of the
imbedding $W^{1,2}\subset L^2$. These two facts together imply the
following: Given a sequence $\{\varphi_n\}\subset W^{1,2}$ such that
$\varphi_n\wto\varphi_0$ in $W^{1,2}_{\tiD}$, and so
$\varphi_n\to\varphi_0$ in $L^2$, we have that
\begin{align*}
(\nabla\varphi,\nabla\varphi) 
&= \|\varphi\|_{1,2}^2 -\|\varphi\|^2
\leqs \liminf_{n\to\infty}\bigl(
\|\varphi_n\|_{1,2}^2 - \|\varphi_n\|^2\bigr)
=\liminf_{n\to\infty}\|\nabla\varphi_n\|^2,
\end{align*}
which establishes that this term is \lscw. The third fact is that the
remaining term in the bilinear form $a_{\tiL}$ has the form
\[
(K\Tr_{\tiN}\varphi,\Tr_{\tiN}\varphi)_{\tiN} = 
(K,[\Tr_{\tiN}\varphi]^2)_{\tiN},
\]
which is continuous under strong convergence and convex (the latter
because the function $K>0$), therefore Lemma~\ref{L:VM-cf} in the
Appendix implies that this term is \lscw. These three facts then
establish that the functional $\varphi\mapsto
a_{\tiL}(\varphi,\varphi)/2$ is \lscw.  We now consider the remaining
terms in the functional $J_{\tiL}$ which are present in the functional
$G_{w}$. The terms in $G_{w}$ which are linear in the function
$\varphi$ are \lscw because they are continuous under weak
convergence, by definition of weak convergence. The nonlinear terms
are gathered together in the functional $g_{w}$, and all of them
except the term $(a_{\tiR}\phi)^{*}(\phi)$ are continuous and strictly
convex. Therefore, Lemma~\ref{L:VM-cf} implies they are \lscw. The
only remaining term, $(a_{\tiR}\phi)^{*}(\phi)$, where $a_{\tiR}^{*}$ is
not positive definite, is also \lscw, since the imbedding
$W^{1,2}\subset L^2$ is compact. The proof of this statement is the
following calculation: Let $\{\phi_n\}\subset W^{1,2}$ such that
$\phi_n\wto\phi_0$ in $W^{1,2}$ which then implies that
$\phi_n\to\phi_0$ in $L^2$.  Then we have
\begin{align*}
\bigl|(a_{\tiR}\phi_0)^{*}(\phi_0)-(a_{\tiR}\phi_n)^{*}(\phi_n)\bigr|
&= \bigl| \bigl(a_{\tiR}, (\phi_0^2-\phi_n^2)\bigr)\bigr|\\
&=\bigl| \bigl(a_{\tiR}, (\phi_0+\phi_n)(\phi_0-\phi_n)\bigr)\bigr|\\
&=\Bigl(\|a_{\tiR}\|_{\infty} \,\|\phi_0+\phi_n\|\Bigr)\, 
\|\phi_0-\phi_n\| \to 0  \mbox{~as~}n\to \infty,
\end{align*}
which then establishes that the functional $\phi\mapsto
(a_{\tiR}\phi)^{*}(\phi)$ is continuous under weak convergence, and so
\lscw. We conclude that the functional $J_{\tiL}$ is \lscw. Hence, all
hypotheses in Theorem~\ref{T:VM-M} in the Appendix are satisfied by
the functional $J_{\tiL}$, therefore there exists a function
$\varphi\in U$ minimizer of $J_{\tiL}$. Furthermore, the minimizer
$\varphi$ is unique in the case that the Ricci scalar $R$ is
non-negative in $\overline\cM$.  The reason is that in this case the
functional $G_{w}$ is strictly convex, since all terms in $G_{w}$ are
convex and at least one of them is strictly convex. Therefore,
Theorem~\ref{T:VM-mcf} implies that the minimizer $\varphi$ is unique.
This establishes the Theorem.\qed

We now show that the functional $J_{\tiL}$ defined in
Eq.~(\ref{VM-def-J}) is G\^ateaux differentiable on $U$ along
directions in $U$. We also show that this derivative can be extended
along all directions on $W^{1,2}_{\tiD}$, and that it coincides with
the Hamiltonian constraint operator defined by the left hand side of
Eq.~(\ref{HC-LYs1}).  This technical Lemma is critical to connecting
the minimizer of the functional $J_{\tiL}$ found in
Theorem~\ref{T:VM-EM} to solutions of the Hamiltonian constraint,
which we do below in Theorem~\ref{T:VM-ES}.

\begin{lemma}{\bf ($J_{\tiL}$ G\^ateaux differentiable)}
\label{L:VM-JGD}
The functional $J_{\tiL}: U\subset W^{1,2}_{\tiD}\to\overline\R$
defined in Eq.~(\ref{VM-def-J}) has G\^ateaux derivative
$DJ_{\tiL}(\tilde\varphi)(\un\varphi)$ for all $\tilde\varphi\in U$
along any direction $\un\varphi\in U$. Furthermore, the map
$DJ_{\tiL}(\tilde\varphi): U\to\R$ can be continuously extended for
every $\tilde\varphi\in U$ into a map $DJ_{\tiL}(\tilde\varphi)
:W^{1,2}_{\tiD}\to\R$, and this operator is precisely the left hand
side in the Hamiltonian constraint Eq.~(\ref{HC-LYs1}).
\end{lemma}

\Proof {\it (Lemma~\ref{L:VM-JGD}.)~}
The G\^ateaux derivative of the functional $J_{\tiL}$ defined in
Eq.~(\ref{VM-def-J}) can be computed term by term. By definition of
the G\^ateaux derivative it is clear that
\begin{align*}
DJ_{\tiL}(\tilde\varphi)(\un\varphi) &=
a_{\tiL}(\tilde\varphi,\un\varphi) 
+Dg_{w}(\phi_{\tiD}+ \tilde\varphi)(\un\varphi)
- \hat\phi_{\tiN}^{*}(\un\varphi) 
+ a_{\tiL}(\phi_{\tiD},\un\varphi)\\
&= a_{\tiL}(\tilde\phi,\un\varphi) 
+Dg_{w}(\tilde\phi)(\un\varphi)
- \hat\phi_{\tiN}^{*}(\un\varphi),
\end{align*}
where we introduced the notation $\tilde\phi :=\phi_{\tiD}
+\tilde\varphi$. By the definition of the functional $g_{w}$ given in
Eq.~(\ref{VM-def-g}), and by the definition of the Gelfand triple
structure described in the Appendix, it is possible to compute
$Dg_{w}$ term by term. For example, this calculation on the first term
is the following: Denote $g_{\tau}(\phi):= (1/6) (a_{\tau}\phi^5)^{*}
(\phi)$, then
\begin{align*}
Dg_{\tau}(\tilde\phi)(\un\varphi) &= \frac{1}{6}\,
\lim_{t\to 0^{+}} \frac{1}{t} \, \bigl[
(a_{\tau}(\tilde\phi+t\un\varphi)^5)^{*}(\tilde\phi+t\un\varphi)
- (a_{\tau}\tilde\phi^5)^{*}(\tilde\phi)\bigr]\\
&= \frac{1}{6}\,
\lim_{t\to 0^{+}} \lim_{n\to \infty} \frac{1}{t} \, \bigl[
\bigl(a_{\tau n},(\tilde\phi+t\un\varphi)^6 \bigr)
- \bigl(a_{\tau n},\tilde\phi^6\bigr) \bigr]\\
&= \frac{1}{6}\,
 \lim_{n\to \infty} \lim_{t\to 0^{+}} \frac{1}{t} \, \bigl[
\bigl(a_{\tau n},(\tilde\phi+t\un\varphi)^6\bigr)
- \bigl(a_{\tau n},\tilde\phi^6\bigr) \bigr]\\
&=  \lim_{n\to \infty} 
\bigl(a_{\tau n}\,\tilde\phi^5,\un\varphi\bigr)\\
&= (a_{\tau}\tilde\phi^5)^{*}(\un\varphi),
\end{align*}
where $\tilde\phi=\phi_{\tiD}+\tilde\varphi$, and this calculation
holds for all $\tilde\varphi$, $\un\varphi\in U\subset
W^{1,2}_{\tiD}$. The limits can be interchanged to obtain the third
line in the equations above because the sequence $(a_{\tau n},\phi^6)$
is uniformly bounded in the index $n$ for every element $\phi\in
[\phi_1,\phi_2]$, that is,
\[
|(a_{\tau n},\tilde{\phi}^6)| \leqs 
\|\phi_2\|_{\infty}^5 \,|(a_{\tau n},\tilde\phi)|
\leqs \|\phi_2\|_{\infty}^5 \,\|a_{\tau}^{*}\|_{-1,2}
\,\|\tilde \phi\|_{1,2}.
\]
We need the assumption $0<\phi_1\leqs\phi_2$ in order we can do the
same calculation above for the terms in the functional $g_{w}$ that
contain negative powers of the function $\tilde\phi$. Then, the
principle of uniform boundness can be extended from sequences of
linear functionals to the sequence that approximates the nonlinear
functional $g_{\tau}$ above. Therefore, the limits in the second line
of the expression for $Dg_{\tau}$ can be interchanged to obtain the
third line in that expression. See~\cite{DunfordSchwartz-I},
pages~52-53, and also see~\cite{ReedSimon80}, pages~80-81 for a proof
of the principle of uniform boundness. References about
generalizations of this principle can be found
in~\cite{DunfordSchwartz-I}, page~82. Let us return to the proof of
Lemma~\ref{L:VM-JGD}. The expression on the last line in the
inequalities above can be continuously extended for all $\un\varphi\in
W^{1,2}_{\tiD}$, as the following calculation shows
\[
(a_{\tau}\tilde\phi^5)^{*}(\un\varphi) 
\leqs \|\tilde\phi\|_{\infty}^5 \,|a_{\tau}^{*}(\un\varphi)|
\leqs \|\tilde\phi\|_{\infty}^5 \,\|a_{\tau}^{*}\|_{-1,2}
\,\|\un\varphi\|_{1,2}.
\]
Analogous calculations on the remaining terms in the functional
$g_{w}$ then show that
\[
Dg_{w}(\tilde\phi)(\un\varphi) = f_{w\tiF}(\tilde\phi)(\un\varphi),
\qquad \forall\, \un\varphi\in W^{1,2}_{\tiD}.
\]
Therefore, we have established that
\[
DJ_{\tiL}(\tilde\phi-\phi_{\tiD})(\un\varphi) =
A_{\tiL}\tilde\phi(\un\varphi) + f_{w}(\tilde\phi)(\un\varphi),
\]
for all $\tilde\phi\in U$ and all $\un\varphi\in W^{1,2}_{\tiD}$.
This last equation establishes the Lemma.\qed

So far the functions $\phi_1$ and $\phi_2$ that define the subset $U$
in Theorem~\ref{T:VM-EM} and Lemma~\ref{L:VM-JGD} can be any elements
in $L^{\infty}$, with only the condition that $0
<\phi_1\leqs\phi_2$. Theorem~\ref{T:VM-EM} above says that there
always exists a minimizer $\varphi$ of the functional $J_{\tiL}$ in
the set $U$. The following result says that if the functions $\phi_1$
and $\phi_2$ are sub- and super-solutions of the G\^ateaux derivative
$DJ_{\tiL}:U\subset W^{1,2}_{\tiD}\to W^{-1,2}_{\tiD}$, then the
minimizer is actually the solution of the Euler equation
$DJ_{\tiL}(\varphi)=0$, and is thus a weak solution to the Hamiltonian
constraint Eq.~(\ref{HC-LYs1}).

\begin{theorem}{\bf(Hamiltonian constraint)}
\label{T:VM-ES}
Assume the hypotheses given in Theorem~\ref{T:VM-EM}, and also assume
that either the constant $a_{\rho}^{\tivee} > 0$ or the constant
$a_{\sigma}^{\tivee} >\sigma_0$, where the positive constant
$\sigma_0$ is defined in Eq.~(\ref{WF-LB-def-s0}). Furthermore, assume
that the subset $U\subset W^{1,2}_{\tiD}$ is defined by
$\phi_1=\phi_{-}$ and $\phi_2=\phi_{+}$, where $\phi_{-}$ and
$\phi_{+}$ are any of the sub- and super-solutions of the Hamiltonian
constraint Eq.~(\ref{HC-LYs1}) found in \Sec\ref{S:WF-LB}. Then, the
minimizer $\varphi\in U$ found in Theorem~\ref{T:VM-EM} is a solution
of the Euler equation
\[
DJ_{\tiL}(\varphi)(\un\varphi)=0,\qquad
\forall\, \un\varphi\in W^{1,2}_{\tiD},
\]
where $DJ_{\tiL}$ is the G\^ateaux derivative of $J_{\tiL}$, and the
equation above is the Hamiltonian constraint Eq.~(\ref{HC-LYs1}).
\end{theorem}

\Proof {\it (Theorem~\ref{T:VM-ES}.)~}
Let $\phi_{-}$, $\phi_{+}$ be sub- and super-solutions of the
Hamiltonian constraint Eq.~(\ref{HC-LYs1}), respectively, and define
$U := [\phi_{-}-\phi_{\tiD}, \phi_{+}-\phi_{\tiD}]\cap
W^{1,2}_{\tiD}$. Let $\varphi\in U$ be a minimizer of $J_{\tiL}$ on
$U$, whose existence was established in Theorem~\ref{T:VM-EM}, and
denote $\phi:=\phi_{\tiD}+\varphi$. We first establish the following
result involving a minimizer $\varphi$ of the functional $J_{\tiL}$:
Given any $\psi\in W^{1,2}_{\tiD}$ such that $\varphi +t \psi\in U$
for small enough, positive number $t$, the following inequality holds,
\begin{equation}
\label{VM-ineq1}
DJ_{\tiL}(\varphi)(\psi) \geqs 0.
\end{equation}
For the proof, compute the G\^ateaux derivative of $J_{\tiL}$ at
$\varphi$ along $\psi$,
\begin{equation}
\label{VM-ineq1a}
DJ_{\tiL}(\varphi)(\psi) = \lim_{t\to 0^{+}} \frac{1}{t}\,
\bigl[ J_{\tiL}(\varphi +t\psi) - J_{\tiL}(\varphi) \bigr],
\end{equation}
which is well-defined because we assume that for $0<t$ small enough
the element $\varphi +t\psi\in U$. The function $\varphi$ is the
minimum of the functional $J_{\tiL}$ in the set $U$, so
$J_{\tiL}(\varphi+t\psi)\geqs J_{\tiL}(\varphi)$, which establishes
Eq.~(\ref{VM-ineq1}) when the limit $t\to 0^{+}$ is computed in
Eq.~(\ref{VM-ineq1a}).

We now use the inequality~(\ref{VM-ineq1}) to show that the function
$\varphi$ is solution of the Hamiltonian constraint
Eq.~(\ref{HC-LYs1}). Let $\zeta$ be any scalar function in the space
$C^{\infty}_{\tiD}(\overline\cM,0)$, and then introduce the
mono-parametric family of functions $\delta\varphi_{\epsilon}$ as a
perturbation of the function $\varphi$ inside the set $U$,
\[
\delta\varphi_{\epsilon} :=\min\bigl\{ (\phi_{+}-\phi_{\tiD}), 
\max\{(\phi_{-}-\phi_{\tiD}),\varphi+\epsilon\zeta\} \,\bigr\},
\]
where $\epsilon$ is a positive, otherwise arbitrary real number. The
construction above implies that $\delta\varphi_{\epsilon}\in U$, and
also that the functional $J_{\tiL}$ is G\^ateaux differentiable at the
function $\varphi$ along the function $\delta\varphi_{\epsilon}
-\varphi$, where the latter statement follows from
\begin{align*}
DJ_{\tiL}(\varphi)(\delta\varphi_{\epsilon} -\varphi) &= 
\lim_{t\to 0^{+}} \frac{1}{t}\,
\bigl[ J_{\tiL}\bigr(\varphi +t \,[\delta\varphi_{\epsilon} -\varphi]\bigr) 
- J_{\tiL}(\varphi) \bigr]\\
&= \lim_{t\to 0^{+}} \frac{1}{t}\,
\bigl[ J_{\tiL}\bigr(t \,\delta\varphi_{\epsilon} + (1-t)\,\varphi]\bigr) 
- J_{\tiL}(\varphi) \bigr],
\end{align*}
which is well-defined because the set $U$ is convex. Therefore, we can
choose the particular direction $\psi = (\delta\varphi_{\epsilon}
-\varphi)$ in Eq.~(\ref{VM-ineq1}), which implies
\begin{equation}
\label{VM-ineq2}
0 \leqs DJ_{\tiL}(\varphi)(\delta\varphi_{\epsilon} -\varphi).
\end{equation}
It is now be convenient to use the equivalent expression
$\delta\varphi_{\epsilon} = (\varphi +\epsilon\zeta) +\zeta_{\epsilon}
-\zeta^{\epsilon}$, where we have introduced the cut-off functions
\[
\zeta^{\epsilon} := \max\bigl\{
0, (\varphi+\epsilon\zeta) -(\phi_{+}-\phi_{\tiD})\bigr\},
\qquad
\zeta_{\epsilon} := -\min\bigl\{
0, (\varphi+\epsilon\zeta) -(\phi_{-}-\phi_{\tiD})\bigr\}.
\]
Notice that $\zeta^{\epsilon}$, $\zeta_{\epsilon}$ are non-negative
continuous functions belonging to $W^{1,2}_{\tiD}$. The inequality
in~(\ref{VM-ineq2}) implies
\begin{equation}
\label{VM-ineq3}
DJ_{\tiL}(\varphi)(\zeta) \geqs \frac{1}{\epsilon}\, \bigl[
DJ_{\tiL}(\varphi)(\zeta^{\epsilon}) 
- DJ_{\tiL}(\varphi)(\zeta_{\epsilon}) \bigr].
\end{equation}
We now show that each term on the right hand side in the inequality
above approaches zero as $\epsilon$ approaches zero. The first term on
the right hand side in Eq.~(\ref{VM-ineq3}) satisfies the following
inequalities,
\begin{align}
\nonumber
DJ_{\tiL}(\varphi)(\zeta^{\epsilon}) &= 
  DJ_{\tiL}(\varphi)(\zeta^{\epsilon})
- DJ_{\tiL}(\phi_{+}-\phi_{\tiD})(\zeta^{\epsilon})
+ DJ_{\tiL}(\phi_{+}-\phi_{\tiD})(\zeta^{\epsilon})\\
\nonumber
&\geqs \bigl[ 
DJ_{\tiL}(\varphi) 
- DJ_{\tiL}(\phi_{+}-\phi_{\tiD})\bigr](\zeta^{\epsilon})\\
\label{VM-ineq4}
&= A_{\tiL}(\phi-\phi_{+})(\zeta^{\epsilon})
+ \bigl[ f_{w}(\phi) - f_{w}(\phi_{+})\bigr](\zeta^{\epsilon}),
\end{align}
where the property that $\phi_{+}$ is a super-solution of
Eq.~(\ref{HC-LYs1}) was used to obtain the second line in the
inequality above. We analyze the last inequality, term by term. We
will need the subset $\cM^{\epsilon}\subset \cM$ defined as follows
\[
\cM^{\epsilon}:= \{x\in \cM : \zeta^{\epsilon}>0\}.
\]
This definition implies that $\mbox{meas}(\cM^{\epsilon})\to 0$ as
$\epsilon\to 0$. Then, the term involving the operator $A_{\tiL}$
satisfies the following inequalities,
\begin{align*}
A_{\tiL}(\phi-\phi_{+})(\zeta^{\epsilon}) &=
\bigl(\nabla[\phi-\phi_{+}], \nabla\zeta^{\epsilon}\bigr)
+ \bigl( K\Tr_{\tiN}[\phi-\phi_{+}],
\Tr_{\tiN}\zeta^{\epsilon}\bigr)_{\tiN}\\
&= \bigl(\nabla[\phi-\phi_{+}], 
\nabla[\phi-\phi_{+}+\epsilon\zeta]\bigr)_{\cM^{\epsilon}}\\
&\quad + \bigl( K\Tr_{\tiN}[\phi-\phi_{+}],
\Tr_{\tiN}[\phi-\phi_{+}+\epsilon\zeta]
\bigr)_{\partial\cM_{\tiN}^{\epsilon}}\\
&\geqs \epsilon \,\bigl(\nabla[\phi-\phi_{+}], 
\nabla\zeta\bigr)_{\cM^{\epsilon}}
+ \epsilon\,\bigl( K\Tr_{\tiN}[\phi-\phi_{+}],
\Tr_{\tiN}\zeta\bigr)_{\partial\cM_{\tiN}^{\epsilon}},
\end{align*}
where the subscripts $\cM^{\epsilon}$ and
$\partial\cM_{\tiN}^{\epsilon}$ on the inner products mean the $L^2$
inner product on these domains, and where we used that
$\zeta^{\epsilon}= (\phi-\phi_{+}) +\epsilon\zeta$ on
$\cM^{\epsilon}$. In order to analyze the term with the functional
$f_{w}$, it is convenient to introduce a representation based on
Gelfand triple structure, $W^{1,2}_{\tiD}\subset
L^2\equiv\bigl[L^2\bigr]^{*}\subset W^{-1,2}_{\tiD}$,
\[
f_{w}(\tilde\phi)(\un\varphi) = 
\lim_{n\to\infty}(f_{wn}(\tilde\phi),\un\varphi),
\qquad \forall\, \un\varphi\in W^{1,2}_{\tiD},
\]
with the functions $f_{wn}\in L^2$ and $\tilde\phi\in U$. Using this
representation it is not difficult to establish the following
inequalities
\begin{align*}
\bigl[ f_{w}(\phi) - f_{w}(\phi_{+})\bigr](\zeta^{\epsilon}) &=
\lim_{n\to\infty} \bigl( \bigl[
f_{wn}(\phi) -f_{wn}(\phi_{+})\bigr],
\zeta^{\epsilon} \bigr)_{\cM^{\epsilon}}\\
&\geqs- \lim_{n\to\infty} \bigl( 
|f_{wn}(\phi) -f_{wn}(\phi_{+})|,
\zeta^{\epsilon} \bigr)_{\cM^{\epsilon}}\\
&\geqs -\lim_{n\to\infty} \bigl( 
|f_{wn}(\phi) -f_{wn}(\phi_{+})|,
(\phi -\phi_{+}) \bigr)_{\cM^{\epsilon}}\\
&\quad- \epsilon \lim_{n\to\infty} \bigl( 
|f_{wn}(\phi) -f_{wn}(\phi_{+})|,\zeta \bigr)_{\cM^{\epsilon}}\\
&\geqs - \epsilon \lim_{n\to\infty} \bigl( 
|f_{wn}(\phi) -f_{wn}(\phi_{+})|,\zeta \bigr)_{\cM^{\epsilon}}.
\end{align*}
Combining the inequalities obtained for the operator $A_{\tiL}$ and
the functional $f_{w}$, and using them with Eq.~(\ref{VM-ineq4}), we
obtain
\begin{align*}
DJ_{\tiL}(\varphi)(\zeta^{\epsilon}) &\geqs \epsilon \Bigl[
\bigl(\nabla[\phi-\phi_{+}], 
\nabla\zeta\bigr)_{\cM^{\epsilon}}
+ \bigl( K\Tr_{\tiN}[\phi-\phi_{+}],
\Tr_{\tiN}\zeta\bigr)_{\partial\cM_{\tiN}^{\epsilon}}\\
&\qquad - \lim_{n\to\infty} \bigl( 
|f_{wn}(\phi) -f_{wn}(\phi_{+})|,\zeta \bigr)_{\cM^{\epsilon}}
\Bigr]. 
\end{align*}
An analogous calculation can be performed on the second term on the
right hand side in Eq.~(\ref{VM-ineq3}), and the result is
\begin{align*}
DJ_{\tiL}(\varphi)(\zeta_{\epsilon}) &\leqs -\epsilon \Bigl[
\bigl(\nabla[\phi-\phi_{-}], 
\nabla\zeta\bigr)_{\cM_{\epsilon}}
+ \bigl( K\Tr_{\tiN}[\phi-\phi_{-}],
\Tr_{\tiN}\zeta\bigr)_{\partial\cM_{\tiN\epsilon}}\\
&\qquad + \lim_{n\to\infty} \bigl( 
|f_{wn}(\phi) -f_{wn}(\phi_{-})|,\zeta \bigr)_{\cM_{\epsilon}}
\Bigr],
\end{align*}
where now we have used the property that the function $\phi_{-}$ is a
sub-solution of Eq.~(\ref{HC-LYs1}), and have introduced the analogous
subset $\cM_{\epsilon}\subset \cM$ defined as
\[
\cM_{\epsilon}:= \{x\in \cM : \zeta_{\epsilon}>0\}.
\]
These last two inequalities used in Eq.~(\ref{VM-ineq3}) imply that
\[
DJ_{\tiL}(\phi)(\zeta) \geqs o(\epsilon),
\quad \mbox{~~as~~}\epsilon \to 0,
\]
since both $\mbox{meas}(\cM^{\epsilon})$ and
$\mbox{meas}(\cM_{\epsilon})$ approach zero as $\epsilon$ approaches
zero. The same calculation must hold for $-\zeta$, and therefore we
conclude that $DJ_{\tiL}(\phi)(\zeta)=0$ for all $\zeta\in
C^{\infty}_{\tiD}(\cM,0)$. This space is dense in the space
$W^{1,2}_{\tiD}$, so we conclude that
\[
DJ_{\tiL}(\phi) =0,
\]
which establishes the Theorem.\qed

\subsection{Results using barrier methods}
\label{S:WF-Existence-LB}

We now use the barrier method to show that there exist weak solutions
to the Dirichlet-Robin boundary value formulation for the Hamiltonian
constraint equation. The barriers found in \Sec\ref{S:WF-LB} are used
to modify the original Eq.~(\ref{HC-LYs1}) into an equation with a
monotone decreasing source. This allows us to construct an iteration
which converges to a fixed point of a particular mapping, which is
constructed so that the fixed point also solves the Hamiltonian
constraint equation. The modification of the original Hamiltonian
constraint equation is called here a shift of the equation, and
imposes a restriction on the regularity of the equation coefficients.
This shift is not needed in the variational method, which is a
reason that variational methods are able to produce results with
weaker regularity, requiring fewer assumptions on the data.

The barrier method and the variational method are both constructive,
in the sense that they provide an algorithm to construct the solution,
which could be of interest in numerical relativity.  In the latter
method, one can build algorithms based on gradient descent that can
guarantee progress (descent) at each iteration. In fact, the most
effective numerical algorithms for the constraints tend to be a
combination of these two ideas: global inexact-Newton methods are
basically highly-tuned fixed-point iterations that maximize their
contraction rate, and which are ``globalized'' by enforcing descent in
an associated energy functional~\cite{mH01a}.

We therefore include the barrier technique here to give the most
complete picture of what can be shown using both techniques on compact
manifolds with boundary.

\begin{theorem}{\bf (Hamiltonian constraint)}
\label{T:HC-E}
Consider the weak formulation for the Hamiltonian constraint given in
\Sec\ref{S:WF-HC}. Assume that the following conditions hold:
\begin{enumerate}[(i)]
\item \label{T:HC-Ei}
The coefficients functionals $a_{\tau}^{*}$, $a_{\rho}^{*}$ and
$a_{w}^{*}$ given in Eq.~(\ref{HC-coeff}) have the form
\[
a_{\tau}^{*}(\un\varphi)= (a_{\tau},\un\varphi),\quad
a_{w}^{*}(\un\varphi)= (a_{w},\un\varphi),\quad
a_{\rho}^{*}(\un\varphi)= (a_{\rho},\un\varphi),\quad
\forall \, \un\varphi\in W^{1,2}_{\tiD},
\]
where the functions $a_{\tau}$, $a_{\rho}$ and $a_{w}$ belong to
$L^{p/2}_{+}$ with $p = 3$. Fix a positive extension of the Dirichlet
boundary data, as discussed in \Sec\ref{S:WF}, and the Robin boundary
data as follows
\[
\phi_{\tiD}\in L^{\infty}\cap W^{1,2},\quad
\hat\phi_{\tiN}^{*}\in W^{-\frac{1}{2},2}(\partial\cM_{\tiN},0).
\]

\item \label{T:HC-Eiii}
In the case that the Ricci scalar $R\geqs 0$, then assume that either
$a_{\rho}^{\tivee} >0$ or that $a_{\sigma}^{\tivee} >\sigma_0$, where
the positive constant $\sigma_0$ is defined in
Eq.~(\ref{WF-LB-def-s0}); In the case that the Ricci scalar $R<0$,
then assume that $a_{\tiR}^{\tivee} >0$, $a_{\rho}^{\tivee}\geqs 0$ and
$a_{\sigma}^{\tivee} \geqs 0$;

\item
\label{T:HC-Eiv}
Assume that the constants $a_{\tau}^{\tivee}$, $K^{\tivee}$,
$\hat\phi_{\tiN}^{\tivee}$, and $\phi_{\tiD}^{\tivee}$ defined in
\Sec\ref{S:WF-LB} are all positive.
\end{enumerate}
Then, there exists a function $\phi\in [\phi_{-},\phi_{w+}] \cap
A^{1,2}\subset W^{1,2}$ which is a solution of Eq.~(\ref{HC-LYs1}), where
$\phi_{w+}$ is the super-solution found in Lemma~\ref{L:HC-LSp}, and
$\phi_{-}$ is the sub-solution given in Lemma~\ref{L:HC-GSb-R-} for
the case $R<0$, and given in Lemma~\ref{L:HC-GSb-r+} for the case
$R\geqs 0$. 
\end{theorem}

\Remark 
The proof of Theorem~\ref{T:HC-E}
begins by shifting the equation~(\ref{HC-LYs1}) in an
appropriate way without changing its solutions. Then it is shown that
a solution of the shifted equation exists iff there exists a fixed
point of a certain map. It is then established that this map is
compact, and thanks to the shifting it is also monotone
increasing. These properties establish the existence of a fixed
point. The last step in the proof is to show that this fixed point is
a solution of the original Eq.~(\ref{HC-LYs1}). 

\Remark
We note that the weak boundary value problem for the Hamiltonian and
momentum constraint equations introduced in \Sec\ref{S:WF} is
well-defined for source functions satisfying Eq.~(\ref{WF-coeff}),
that is, $a_{\tau}$, $a_{\rho}$, and $a_{w}$ belong to
$L^{6/5}$. However, Theorem~\ref{T:HC-E} above requires that these
coefficients belong to $L^{3/2}$. This extra regularity is needed to
shift the Hamiltonian constraint equation. It is not clear if there
exists a different shifting procedure that also works for for
coefficients in $L^p$ with $6/5 \leqs p < 3/2$. This issue is also
present in \cite{dM05}, where rough solutions are found in $W^{k,2}$
for $k>3/2$; here we are basically asking for higher Lebesgue index
$p$ instead the higher Sobolev index $k$ in
\cite{dM05} to make the shift possible.

\Proof {\it (Theorem~\ref{T:HC-E}.)~}
The assumption~{\it(\ref{T:HC-Eiv})} is required in
Lemma~\ref{L:HC-LSp} for the existence of the (local) super-solution
$\phi_{w+}$ given in Eq.~(\ref{HC-LSp}). The
assumptions~{\it(\ref{T:HC-Eiii})-(\ref{T:HC-Eiv})} are required to
find a sub-solution: In the case of Ricci scalar $R< 0$ the global
sub-solution is given in Lemma~\ref{L:HC-GSb-R-}; In the case that the
Ricci scalar $R\geqs 0$ the global sub-solution is given in
Lemma~\ref{L:HC-GSb-r+} in the case $a_{\rho}^{\tivee} >0$, and is
given in Lemma~\ref{L:HC-GSb-s+} in the case $a_{\sigma}^{\tivee}
>\sigma_0$. Summarizing, in these cases there exists at least {\em
local} sub- and super-solutions for the Hamiltonian constraint
Eq.~(\ref{HC-LYs1}), which is sufficient for our needs here.

The condition $K^{\tivee} >0$ in assumption~{\it(\ref{T:HC-Eiv})}
implies that the constant $\hat\ttk$ in Eq.~(\ref{HC-cttk}) is
positive in both cases where $\partial\cM_{\tiD} =\emptyset$ and
$\partial\cM_{\tiD}\neq\emptyset$; so, the operator $A_{\tiL}$ is
invertible.

We now use the sub- and super-solutions $\phi_{-}$, $\phi_{w+}$ to
restrict the domain of the functionals $f_{w\tiF}$ and $f_{w}$ to the
set $[\phi_{-} ,\phi_{w+} ]\subset L^{2}$. The interval itself and the
functionals $f_{w\tiF}$, $f_{w}$ are well-defined, due to the property
$0<\phi_{-}\leqs\phi_{w+}$. Let $\alpha\in L^{p/2}$ be the function
given by
\begin{equation}
\label{T:HC-E-def-alpha}
\alpha := 5 \,\phi_{w+}^4 a_{\tau} + |a_{\tiR}| 
+ 3\,\frac{\phi_{w+}^2}{\phi_{-}^6}\,a_{\rho} 
+ 7\,\frac{\phi_{w+}^6}{\phi_{-}^{14}}\,a_{w}, 
\end{equation}
and introduce a function $s\in L^{p/2}$ such that $(s-\alpha)\in
L^{p/2}_{+}$. Then, define the shifted operators
\begin{align}
\label{HC-def-As}
A_{\tiL}^s & :W^{1,2} \to W^{-1,2}_{\tiD},&
A^s_{\tiL}\phi(\un\varphi) 
&:= A_{\tiL}\phi(\un\varphi) +  (s\phi,\un\varphi),\\
\label{HC-def-fs}
f_{w}^s & :[\phi_{-},\phi_{w+}] \subset L^2 \to W^{-1,2}_{\tiD},&
f^s_{w}(\phi)(\un\varphi) 
&:= f_{w}(\phi)(\un\varphi) - (s\phi,\un\varphi).
\end{align}
First, note that the operator $A_{\tiL}^s$ is well-defined, since for
all $s\in L^{3/2}$, and $\phi$, $\un\varphi\in W^{1,2}$ the
generalized H\"older inequality ($p=1$, $p_1=3/2$, $p_2=p_3=6$, in the
notation given at the end of \Sec\ref{S:NC}) implies
\[
(s\phi,\un\varphi)
\leqs \|s\|_{3/2} \, \|\phi\|_{6} \, \|\un\varphi\|_{6}
\leqs c^2\, \|s\|_{3/2} \,\|\phi\|_{1,2}\,\|\un\varphi\|_{1,2},
\]
where we used the fact that the imbedding $W^{1,2}\to L^6$ is
continuous with imbedding constant $c>0$. The functional $f_{w}^s$ is
also well-defined because the shift term can be bounded as follows
\[
(s\phi,\un\varphi)
\leqs \|s\|_{6/5} \, \|\phi\|_{\infty} \, \|\un\varphi\|_{6}
\leqs c\, \|s\|_{3/2} \,\|\phi\|_{\infty}\,\|\un\varphi\|_{1,2}.
\]
In fact, the first inequality shows that the shift on the functional
$f_{w}$ is well-defined for the shift function $s\in
L^{6/5}$. However, we have just seen that the shift on the operator
$A_{\tiL}$ is well-defined only for the shift function $s\in
L^{3/2}\subset L^{6/5}$.

This operator $A_{\tiL}^s$ is invertible since $A_{\tiL}$ is
invertible (due to the hypothesis $K^{\tivee}>0$) and since the
function $s$ is non-negative (see for example~\cite{Gilbarg-Trudinger}
for a proof). This shifted operator $A_{\tiL}^s$ satisfies the maximum
principle, a result shown in Lemma~\ref{L:Example-MP} in the Appendix.
Therefore, Lemma~\ref{L:Linverse} in that Appendix shows that
$(A_{\tiL}^s)^{-1}$ is a monotone increasing operator.

Second, note that the function $s$ satisfies $(s-\alpha)\in
L^{p/2}_{+}$, which implies that the operator $f_{w}^s$ is monotone
decreasing. The latter means that given functions $\phi_2$, $\phi_1\in
[\phi_{-},\phi_{w+}]$ with $\phi_2-\phi_1\in L^{\infty}_{+}$, the
functional $f_{w}^s$ satisfies $-[f_{w}^s(\phi_2) -
f_{w}^s(\phi_1)]\in W^{-1,2}_{\tiD +}$. The proof of this property is
the following: given such functions $\phi_2$ and $\phi_1$, compute
\begin{align}
\nonumber
\bigl( f_{w}^s(\phi_2) - f_{w}^s(\phi_1) \bigr)(\un\varphi) 
&= \bigl( f_{w}(\phi_2) -f_{w}(\phi_1)\bigr)(\un\varphi)
- (s [\phi_2-\phi_1],\un\varphi)\\
\nonumber
&= \bigl( [f_{w\tiF}(\phi_2) -f_{w\tiF}(\phi_1)],\un\varphi \bigr)
- (s [\phi_2-\phi_1],\un\varphi)\\
\nonumber
&= \bigl( a_{\tau} 
\bigl[ (\phi_2)^5 - (\phi_1)^5 \bigr],\un\varphi \bigr)
+ \bigl( a_{\tiR} [\phi_2 - \phi_1],\un\varphi \bigr)\\
\nonumber
&\quad - (s [\phi_2-\phi_1],\un\varphi) - \bigl( a_{\rho} 
\bigl[ (\phi_2)^{-3} - (\phi_1)^{-3}\bigr],\un\varphi \bigr)\\
&\quad - \bigl( a_{w} 
\bigl[ (\phi_2)^{-7} - (\phi_1)^{-7}\bigr],\un\varphi \bigr).
\label{T:HC-E-1}
\end{align}
Now, the conditions $0<\phi_1\leqs\phi_2$ and $\phi_1$, $\phi_2\in
[\phi_{-},\phi_{w+}]$ imply the following inequalities,
\begin{align}
\nonumber
(\phi_2)^5-(\phi_1)^5 &= 
\Bigl(\sum_{j=0}^{4} (\phi_2)^j (\phi_1)^{4-j} \Bigr) 
(\phi_2- \phi_1) \\
\label{HC-E-exp1}
&\leqs 5\, (\phi_{w+})^4 \, (\phi_2- \phi_1),\\
\nonumber
-\bigl[ (\phi_2)^{-3} -(\phi_1)^{-3} \bigr] &= 
\frac{1}{(\phi_2\phi_1)^3}\,
\Bigl(\sum_{j=0}^{2} (\phi_2)^j (\phi_1)^{2-j} \Bigr) 
(\phi_2- \phi_1) \\
\label{HC-E-exp2}
&\leqs 3\, \frac{(\phi_{w+})^2}{(\phi_{-})^6} \, (\phi_2- \phi_1),\\
\nonumber
-\bigl[ (\phi_2)^{-7} -(\phi_1)^{-7} \bigr] &= 
\frac{1}{(\phi_2\phi_1)^7}\,
\Bigl(\sum_{j=0}^{6} (\phi_2)^j (\phi_1)^{6-j} \Bigr) 
(\phi_2 - \phi_1) \\
\label{HC-E-exp3}
&\leqs 7\, \frac{(\phi_{w+})^6}{(\phi_{-})^{14}} \, (\phi_2- \phi_1).
\end{align}
These inequalities and Eq.~(\ref{T:HC-E-1}) imply
\[
\bigl( f_{w}^s(\phi_2) - f_{w}^s(\phi_1) \bigr)(\un\varphi) 
\leqs \bigl( [\alpha -s] (\phi_2-\phi_1),\un\varphi \bigr),
\]
where $\alpha$ is given in Eq.~(\ref{T:HC-E-def-alpha}). The choice
$s\in L^{p/2}$ and $s\geqs \alpha$ implies that
\[
\bigl( f_{w}^s(\phi_2) - f_{w}^s(\phi_1) \bigr)(\un\varphi) 
\leqs 0\qquad
\forall \,\un\varphi\in W^{1,2}_{\tiD +},
\]
which establishes that $f_{w}^s$ is monotone decreasing.

Having introduced the shifted operators $A_{\tiL}^s$ and $f_{w}^s$, we
now remark that a function $\phi\in [\phi_{-},\phi_{w+}]\cap
W^{1,2}$ is solution of $A_{\tiL}\phi + f_{w}(\phi) = 0$ iff $\phi$ is
solution of $A_{\tiL}^s\phi + f^s_{w}(\phi) = 0$. So far we have
the following structure:
\begin{gather*}
f_{w}^s : [\phi_{-},\phi_{w+}]\subset L^2 
\to W^{-1,2}_{\tiD},\\
(A_{\tiL}^s)^{-1}: W^{-1,2}_{\tiD} \to A^{1,2} \subset W^{1,2},\\
I: W^{1,2} \to L^2,
\end{gather*}
where $I$ is the identity imbedding, which is a compact
map. Therefore, the operator
\begin{equation}
\label{T:HC-E-def-Tw}
T_{w}^s : [\phi_{-},\phi_{w+}] \subset L^2 \to L^2,
\qquad
T^s_{\biw}(\phi) := -I\, (A_{w}^s)^{-1} f_{w}^s(\phi),
\end{equation}
is well-defined. Both the operator $(A_{\tiL}^s)^{-1}$ and the
functional $-f_{w}^s$ are monotone increasing, therefore the operator
$T_{w}^s$ is also monotone increasing, a result that is proven in
Lemma~\ref{L:T-decreasing}. Furthermore, this operator $T_{w}^s$ is
compact, because it is a composition of continuous maps and the
compact imbedding $I :W^{1,2}\to L^2$ (for example
see~\cite{DunfordSchwartz-I} page~486, Theorem~4, and also see the
imbedding Theorems in~\cite{Adams75} chapter VI). We established that
the functions $\phi_{-}$ and $\phi_{w+}$ are sub- and super-solutions
of Eq.~(\ref{HC-LYs1}), respectively. Therefore,
Lemma~\ref{L:T-sub-super} in the Appendix shows that these functions
$\phi_{-}$ and $\phi_{w+}$ satisfy the inequalities in the order given
by $L^2_{+}$,
\[
\phi_{-} \leqs T_{w}^s(\phi_{-}),\qquad 
\phi_{w+} \geqs T_{w}^s(\phi_{w+}).
\]
Since the order cone in $L^2$ is normal, all the hypotheses in
Theorem~\ref{T:FPI} in the Appendix are satisfied. Thus, there exists
$\phi\in [\phi_{-},\phi_{w+}]\subset L^2$ a fixed point of $T_{w}^s$.

We now show that the fixed point $\phi$ satisfies that $\phi\in
[\phi_{-},\phi_{w+}]\cap W^{1,2}$. This result is a consequence of
$T_{w}^s$ being bounded in $W^{1,2}$. Indeed, given any function
$\varphi\in [\phi_{-} ,\phi_{w+}]$ we have that
\begin{equation}
\label{T:HC-E-Tbound}
\|T_{w}^s(\varphi)\|_{1,2} 
= \|-(A_{\tiL}^s)^{-1}f_{w}^s(\varphi)\|_{1,2}
\leqs c_{\tiL} \, \|f_{w}^s(\varphi)\|_{-1,2}.
\end{equation}
Recalling the definition of the functional $f_{w}^s$, that is,
\[
f_{w}^s(\varphi)(\un\varphi) =
(a_{\tau}\varphi^5,\un\varphi) 
+(a_{\tiR}\varphi,\un\varphi) 
-(a_{\rho}\varphi^{-3},\un\varphi)
-(a_{w}\varphi^{-7},\un\varphi)
-\hat\phi_{\tiN}^{*}(\Tr_{\tiN}\un\varphi)
-(s\varphi,\un\varphi),
\]
we have the following inequalities,
\begin{align*}
|f_{w}^s(\varphi)(\un\varphi)| &\leqs
(a_{\tau},\un\varphi) \,\phi_{w+}^5
+(|a_{\tiR}|\varphi,\un\varphi) 
+(a_{\rho},\un\varphi)\, \phi_{-}^{-3}
+(a_{w},\un\varphi)\, \phi_{-}^{-7}\\
&\quad + |\hat\phi_{\tiN}^{*}(\Tr_{\tiN}\un\varphi)|
+ (s ,\un\varphi)\, \phi_{w+} \\
&\leqs
\Bigl[ a_{\tau}^{\tiwedge}\,\phi_{w+}^5
+ a_{\tiR}^{\tiwedge} \,\phi_{w+} 
+ a_{\rho}^{\tiwedge} \, \phi_{-}^{-3}
+ a_{w}^{\tiwedge} \, \phi_{-}^{-7}\\
&\quad + c_t \|\hat\phi_{\tiN}^{*}\|_{-\frac{1}{2},2,\tiN} 
+ s^{\tiwedge}\, \phi_{w+} \Bigr] \,\|\un\varphi\|_{1,2},
\end{align*}
where $s^{\tiwedge}$ is defined in an analogous way as
$a_{\tau}^{\tiwedge}$ in Eq.~(\ref{WF-LB-tR-up}), and $c_t$ is a
positive constant such that $\|\Tr_{\tiN}\un\varphi\|_{\frac{1}{2}
,2,\tiN}\leqs c_t \|\un\varphi\|_{1,2}$ for all $\un\varphi\in
W^{1,2}_{\tiD}$. Therefore, introducing the constant
\[
c_{w} := \bigl[ a_{\tau}^{\tiwedge}\,\phi_{w+}^5
+ (a_{\tiR}^{\tiwedge}+ s^{\tiwedge}) \,\phi_{w+} 
+ a_{\rho}^{\tiwedge} \, \phi_{-}^{-3}
+ a_{w}^{\tiwedge} \, \phi_{-}^{-7}
+ c_t \|\hat\phi_{\tiN}^{*}\|_{-\frac{1}{2},2,\tiN} \bigr],
\]
we have the inequality
\[
\sup_{0\neq\un\varphi\in W^{1,2}_{\tiD}}
\frac{|f_{w}^s(\varphi)(\un\varphi)|}
{~\|\un\varphi\|_{1,2}} 
\leqs c_{w},
\]
which yields to the desired inequality
\[
\|T_{w}^s(\varphi)\|_{1,2} \leqs k_{w}
\qquad \forall \,\varphi\in [\phi_{-},\phi_{w+}],
\]
with $k_{w} = c_{\tiL}c_{w}$. Therefore, the fixed point point
$\phi\in [\phi_{-},\phi_{w+}]$ satisfies
\[
\|\phi\|_{1,2} = \|T_{w}^s(\phi)\|_{1,2} \leqs k_{w},
\]
which establishes the property $\phi\in [\phi_{-},\phi_{w+}]\cap
W^{1,2}$. Therefore, we can apply the operator $A_{\tiL}^s$ on both
sides of the equation $\phi=T_{w}^s(\phi)$, and then the fixed point
function $\phi$ satisfies both the shifted and the non-shifted
Hamiltonian constraint equations. The latter establishes that the
function $\phi$ is a solution of the Hamiltonian constraint
Eq.~(\ref{HC-LYs1}).\qed

\subsection{Regularity of solutions}
\label{S:HC-Reg}

The following result states that when the regularity of the boundary
data agrees with the equation coefficients regularity, the solution
obtained by barrier methods is actually more regular than is stated in
Theorem~\ref{T:HC-E}.  Note that the Proposition~\ref{P:HC-E-Reg} below
does not apply in the case of the solutions found by variational
methods. In this latter case the coefficients in the functional $f_w$
belong to $W^{-1,2}_{\tiD}$, so the bootstrap argument mentioned below
does not apply.
\begin{proposition}{\bf (Regularity)}
\label{P:HC-E-Reg}
Assume the hypotheses in Theorem~\ref{T:HC-E}, and in addition assume
that the boundary $\partial\cM$ is $C^{2}$. Assume that the boundary
data satisfy $\hat\phi_{\tiN}^{*}(\Tr_{\tiN}\un\varphi) =
(\hat\phi_{\tiN},\Tr_{\tiN}\un\varphi)_{\tiN}$ for all $\un\varphi\in
W^{1,2}_{\tiD}$ and the following condition holds
\[
\phi_{\tiD}\in W^{2,(p/2)},\quad
\hat\phi_{\tiN}\in W^{\frac{1}{(p/2)'},(p/2)}(\partial\cM_{\tiN}),
\quad p= 3.
\]
Then, the function $\phi\in [\phi_{-},\phi_{w+}]\cap W^{1,2}$
solution of the Hamiltonian constraint Eq.~(\ref{HC-LYs1}) belongs to
the space $W^{2,(p/2)}$.
\end{proposition}

\Outline {\it (Proposition~\ref{P:HC-E-Reg}.)~}
A proof can be based on linear elliptic estimates (see for example
Theorem~9.11 in~\cite{Gilbarg-Trudinger}) and a standard bootstrap
argument.
\qed

\section{Coupled system}
\label{S:CS}

Here we combine the results for the individual constraints derived
earlier to establish a new non-CMC result for the coupled system. In
\Sec\ref{S:CS-Ex} we use the global barriers found in
\Sec\ref{S:WF-LB} to establish existence of non-CMC solutions to the
coupled constraints through fixed-point iteration and compactness
arguments directly, rather than by using the Contraction Mapping
Theorem as was done in the original work of Isenberg and Moncrief
in~\cite{jIvM96}.

It is interesting to note that for the main result on the non-CMC
coupled system in \cite{jIvM96}, the near-CMC condition on the trace
of the extrinsic curvature is actually used twice: once to obtain the
global super-solution, and a second distinct time to construct a
contraction for using the Contraction Mapping Theorem to get existence
and uniqueness.  Here, a weak version of the near-CMC condition must 
also be employed in \Sec\ref{S:WF-LB} to drive a global super-solution 
for the Hamiltonian constraint in our weaker setting.
However, by using a compactness argument for the coupled system 
in \Sec\ref{S:CS-Ex} rather than the Contraction Mapping 
Theorem, we avoid the second use of the near-CMC condition.
If a global super-solution can be constructed without the near-CMC 
assumption, then our compactness argument would give existence of 
solutions to the coupled system in the fully general ``far-from-CMC''
case.  What our proof technique gives up is uniqueness of solutions 
to the coupled system, which comes for free with existence when the 
contraction argument is used as in \cite{jIvM96}.

\subsection{Existence of weak solutions}
\label{S:CS-Ex}

This section is dedicated to establishing existence of solutions to
the weak Dirichlet-Robin boundary value problem
(\ref{WF-LYs1})-(\ref{WF-LYm1}).  In the case that $R<0$ there is no
condition on the matter fields other than $\rho\geqs 0$, but in the
case that the Ricci scalar $R\geqs 0$ it is required that either
$a_{\rho}^{\tivee} >0$ or $a_{\sigma}^{\tivee} >\sigma_0$, with the
positive constant $\sigma_0$ defined in Eq.~(\ref{WF-LB-def-s0}). The
equation coefficients are required to have stronger regularity than
those previously required in Secs.~\ref{S:WF-HC} and
~\ref{S:WF-Existence-LB}. Although our problem formulation is
different (compact domains with boundary), our results can be viewed
as extending the result in~\cite{jIvM96} to weaker solution spaces,
and extending their result for $R=-1$ to scalar curvature having any
sign. The work presented can similarly be viewed as extending the CMC
results on rough solutions in~\cite{dM05} to the non-CMC case, for
compact domains with boundary.  As remarked earlier, the ``near-CMC''
assumption required for the Contraction Mapping Argument
in~\cite{jIvM96} is not required for the compactness argument below.

\begin{theorem}
\label{T:CS-Ex}{\bf(Non-CMC)}
Consider the weak formulation for the Hamiltonian and momentum
constraints defined in \Sec\ref{S:WF}. Assume the background metric
$h\in C^2(\overline\cM,2)$ and that the following conditions hold:
\begin{enumerate}[(i)]
\item
\label{T:CS-Exi}
Fix a number $p>3$ and denote by $q:= 6p/(3+p)$. Fix source and
boundary functions
\begin{gather*}
\tau\in L^{q}, \quad 
(\nabla\tau)^{*}\in \tbW^{-1,q}_{\tiID},\quad 
\sigma\in L^{q}(\cM,2),\quad
\rho \in L^{q/2}_{+},\quad 
\tbj\in \tbL^{q/2},\\
\phi_{\tiD}\in W^{1,p},\quad 
\hat\phi_{\tiN}^{*}\in W^{-\frac{1}{p},p}(\partial\cM_{\tiN},0),
\quad \tbw_{\tiID}\in \tbW^{1,q},\quad 
\hat\tbw_{\tiIN}^{*}\in W^{-\frac{1}{q},q}(\partial\cM_{\tiIN},1),
\end{gather*}
with $(\rho^2-\tbj\cdot\tbj\,) \in \mbox{int}(L^{q/4}_{+})$ in the
case $\tbj\neq 0$;
\item
\label{T:CS-Exii}
In the case that the Ricci scalar $R$ of the background metric is
non-negative, then assume that either the constant $a_{\rho}^{\tivee}
>0$ or the constant $a_{\sigma}^{\tivee} >\sigma_0$, with the positive
constant $\sigma_0$ defined in Eq.~(\ref{WF-LB-def-s0}); In the case
that the Ricci scalar $R$ is negative, then assume that
$a_{\tiR}^{\tivee} >0$, $a_{\rho}^{\tivee}\geqs 0$ and
$a_{\sigma}^{\tivee}\geqs 0$;
\item
\label{T:CS-Exiii}
Assume that the function $a_{\tau}$ and the constant $\ttK_1$ defined
in Eq.~(\ref{CS-def-K1}) satisfy that $a_{\tau}^{\tivee} - \ttK_1 >0$;
also assume that the constants $K^{\tivee}$,
$\hat\phi_{\tiN}^{\tivee}$, and $\phi_{\tiD}^{\tivee}$ are all
positive.
\end{enumerate}
Then, there exists a solution 
\[
\phi\in [\phi_{-},\phi_{+}]\cap A^{1,p} \subset W^{1,p},\qquad
\tbw \in \tbA^{1,q} \subset \tbW^{1,q},\qquad
p > 3, \quad
q=\frac{6p}{3+p},
\]
of the weak Dirichlet-Robin boundary value problem for the
Hamiltonian and momentum constraint
Eqs.~(\ref{WF-LYs1})-(\ref{WF-LYm1}), where $\phi_{+}$ is the
super-solution found in Lemma~\ref{L:CS-GSp}, and $\phi_{-}$ is the
sub-solution given in Lemma~\ref{L:HC-GSb-R-} for the case $R<0$, and
given in Lemma~\ref{L:HC-GSb-r+} for the case $R\geqs 0$.
\end{theorem}

\Proof {\it (Theorem~\ref{T:CS-Ex}.)~}
Notice that the definition of the numbers $p$ and $q$ satisfies that
$3 < q < p$. The assumption $\tau\in L^{q}$ indicates that $a_{\tau}
=\tau^2/12\in L^{q/2}$, which implies that the linear functional
$a_{\tau}^{*}$ given by as $a_{\tau}^{*}(\un\varphi)=
(a_{\tau},\un\varphi)$ for all $\un\varphi\in W^{1,p'}_{\tiD}$ is a
well-defined element $a_{\tau}^{*}\in W^{-1,p}_{\tiD}$. The proof of
the latter statement is based in the H\"older inequality, which implies
\[
|a_{\tau}^{*}(\un\varphi)| \leqs \|a_{\tau}\|_{(q/2)} \, 
\|\un\varphi\|_{(q/2)'};
\]
since $(q/2)=3p/(3+p)$, the relations $\frac{1}{(q/2)}
+\frac{1}{(q/2)'} =1$ and $\frac{1}{p}+\frac{1}{p'}=1$ imply that
$(q/2)' = 3p'/[3-p']$. Now the coefficient $p >3$ implies that $p'<
3/2$, so we conclude that the imbedding $W^{1,p'} \subset L^{(q/2)'}$
is continuous (see~\cite{Gilbarg-Trudinger}, Corollary~7.11 in
\Sec7.7), and then there exists a positive constant $c_s$ such that
\[
|a_{\tau}^{*}(\un\varphi)| \leqs c_s\,\|a_{\tau}\|_{(q/2)} \, 
\|\un\varphi\|_{1,p'},
\]
which establishes that $a_{\tau}^{*}\in W^{-1,p}_{\tiD}$. The
coefficient functions $a_{\rho} =\kappa\rho/4$ and $a_{\sigma}
=\sigma^2/8$ belong to $L^{q/2}$, and the same holds for the
coefficient $a_{\cL w} = (\cL\biw)^2/8$ whenever the vector
$\biw\in\biW^{1,q}$. A similar argument as above shows that the
functionals
\[
a_{\rho}^{*}(\un\varphi):= (a_{\rho}, \un\varphi),\quad
a_{\sigma}^{*}(\un\varphi):= (a_{\sigma}, \un\varphi),\quad
a_{\cL w}^{*}(\un\varphi):= (a_{\cL w}, \un\varphi),
\qquad \forall \, \un\varphi\in W^{1,p'}_{\tiD}
\]
are well-defined elements in $W^{-1,p}_{\tiD}$. The choice of the
Dirichlet and Robin boundary data and the Gelfand triple structure
reviewed in the Appendix imply that the functional $f$ defined in
Eq.~(\ref{WF-def-f}) is a well-defined map
\[
f: [\phi_1,\phi_2]\subset L^{\infty}\times \biW^{1,q}
\to W^{-1,p}_{\tiD}.
\]
The assumption that the function $\bij\in L^{q/2}$ implies that
$\bij\in L^{r/2}$ with $(r/2)= 3q/(3+q)$, since for $q > 3$ holds that
$3 < r < q$. Hence, the functional $\bib_j^{*}(\un\bomega):=
(\kappa\bij,\un\bomega)$ for all $\un\bomega\in\biW^{1,q'}_{\tiID}$ is
a well-defined element $\bib_j^{*}\in\biW^{-1,q}_{\tiID}$. The proof
is again based in the H\"older inequality
\[
|\bib_{j}^{*}(\un\bomega)|\leqs \|\kappa\bij\|_{(r/2)} \, 
\|\un\bomega\|_{(r/2)'},
\]
The condition $q > 3$ implies the inequality $q' < 3/2$ and the
relations $\frac{1}{q}+\frac{1}{q'} =1$ and $\frac{1}{(r/2)} +
\frac{1}{(r/2)'}=1$ imply that $(r/2)' = 3q'/[3-q']$. From the latter
relation and the inequality $q'< 3/2$ we conclude that the imbedding
$W^{1,q'}\subset L^{(r/2)'}$ is continuous, so there exists a positive
constant $c_s$ such that
\[
|\bib_j^{*}(\un\bomega)| \leqs c_s\,\|\bij\|_{(r/2)} \, 
\|\un\bomega\|_{1,q'},
\]
which establishes that $\bib_{j}^{*}\in W^{-1,q}_{\tiID}$. The
assumption $(\nabla\tau)^{*}\in \biW^{-1,q}_{\tiID}$ implies that
$\bib_{\tau}(\un\bomega):=(\frac{2}{3}\nabla\tau,\un\bomega)$ for all
$\un\bomega\in\biW^{1,q'}_{\tiID}$ is a well-defined element
$\bib_{\tau}^{*}\in\biW^{-1,q}_{\tiID}$. The choice of the Dirichlet
and Robin boundary data and the Gelfand triple structure reviewed in
the Appendix imply that the functional $\bif$ defined in
Eq.~(\ref{WF-def-If}) is a well-defined map
\[
\bif: [\phi_1,\phi_2]\subset L^{\infty}\to \biW^{-1,q}_{\tiID}. 
\]

The regularity assumptions on the equation coefficients and the
assumptions~{\it(\ref{T:CS-Exii})-(\ref{T:CS-Exiii})} are sufficient
conditions to establish the existence of a global sub-solution
$\phi_{-}$ for the Hamiltonian constraint Eq.~(\ref{WF-LYs1}). In the
case that the $R<0$ this result is proved in Lemma~\ref{L:HC-GSb-R-},
and in the case that $R\geqs 0$ this result is proved either in
Lemma~\ref{L:HC-GSb-r+} in the case $a_{\rho}^{\tivee} >0$, or in
Lemma~\ref{L:HC-GSb-s+} in the case $a_{\sigma}^{\tivee} >\sigma_0$.
The regularity assumptions on the equation coefficients and the
assumptions~{\it(\ref{T:CS-Exiii})} are the sufficient conditions to
establish the existence of a global super-solution $\phi_{+}$ for the
Hamiltonian and momentum constraint
Eqs.~(\ref{WF-LYs1})-(\ref{WF-LYm1}), which was established in
Lemma~\ref{L:CS-GSp}.

We now use these global sub- and super-solutions of
Eqs.~(\ref{WF-LYs1})-(\ref{WF-LYm1}) to define the domain
$[\phi_{-},\phi_{+}]\subset L^{\infty}$ of the operators $f$ and
$\bif$ given in Eqs.~(\ref{WF-def-f})-(\ref{WF-def-If}). The
inequality $K^{\tivee}>0$ in assumption~{\it(\ref{T:CS-Exiii})}
implies that the operator $A_{\tiIL} :W^{1,2}\to W^{-1,2}_{\tiD}$
defined in Eq.~(\ref{WF-def-AL}) is invertible, which was established
in Theorem~\ref{T:w-MC}. The regularity result in
Proposition~\ref{P:HC-E-Reg} implies that the operator $A_{\tiL}
:W^{1,p}\to W^{-1,p}_{\tiD}$ with the same action as defined in
Eq.~(\ref{WF-def-AL}) is also invertible. Regarding the Hamiltonian
constraint equation, we introduce the same shifting done in the proof
of Theorem~\ref{T:HC-E}, that is, fix a function $s\in L^{q/2}$, given
by
\[
s := 5 \,\phi_{+}^4 a_{\tau} + |a_{\tiR}| 
+ 3\,\frac{\phi_{+}^2}{\phi_{-}^6}\,a_{\rho} 
+ 7\,\frac{\phi_{+}^6}{\phi_{-}^{14}}\,a_{w}, 
\]
and introduce the shifting operators
\begin{align*}
A_{\tiL}^s & :W^{1,p} \to W^{-1,p}_{\tiD},&
A^s_{\tiL}\phi(\un\varphi) 
&:= A_{\tiL}\phi(\un\varphi) +  (s\phi,\un\varphi),\\
f^s & :[\phi_{w-},\phi_{w+}] \subset L^{\infty} \times \biW^{1,q} 
\to W^{-1,p}_{\tiD},&
f^s(\phi,\biw)(\un\varphi) 
&:= f(\phi,\biw)(\un\varphi) - (s\phi,\un\varphi).
\end{align*}
Then, we have the following structure,
\[
\begin{gathered}
f^s : [\phi_{-},\phi_{+}]\subset L^{\infty} \times \biW^{1,q} 
\to W^{-1,p}_{\tiD},\\
(A_{\tiL}^s)^{-1} : W^{-1,p}_{\tiD} \to A^{1,p} \subset W^{1,p},\\
I : W^{1,p} \to L^{\infty},
\end{gathered}
\qquad
\begin{gathered}
\bif : [\phi_{-},\phi_{+}]\subset L^{\infty} \to \biW^{-1,q}_{\tiID},\\
(A_{\tiIL})^{-1} : \biW^{-1,q}_{\tiID} \to \biA^{1,q},
\end{gathered}
\]
where the map $I :W^{1,p}\to L^{\infty}$ is the identity imbedding,
which is compact for $p > 3$. Therefore, the following operators
are well-defined,
\begin{align*}
&S : [\phi_{-},\phi_{+}]\subset L^{\infty} \to \biW^{1,q},& 
&S(\phi) := -(A_{\tiIL})^{-1}\bif (\phi),\\
&T_{\biw}^s: [\phi_{-},\phi_{+}]\subset L^{\infty} \to L^{\infty},&
&T_{\biw}^s(\phi) := -I(A_{\tiL}^s)^{-1}f(\phi,\biw).
\end{align*}
Since we can choose $\phi_{w+}$ in Theorem~\ref{T:HC-E} to be the
constant $\phi_{+}$ found in Lemma~\ref{L:CS-GSp}, then this
Theorem~\ref{T:HC-E} and the regularity results in
Proposition~\ref{P:HC-E-Reg} imply that exists a fixed point
$\varphi\in [\phi_{-},\phi_{+}]\cap W^{1,p}$ of the iteration
\[
\varphi_{k+1} := T_{\biw}^s(\varphi_k),\qquad
\varphi_{0} = \phi_{-}.
\]
Therefore, the sequence $\{\phi^{n},\biw^{n}\}$ given by
$\phi^0=\phi_{-}$, $\biw^0=\biw_{\tiID}$, and
\[
\biw^{n} = S \bigl(\phi^{n-1}\bigr),\qquad
\phi^{n} = T_{\biw^{n}}^s \bigl( \phi^{n} \bigr),
\qquad n\in \N,
\]
is well-defined, where $\phi^{n}\in [\phi_{-},\phi_{+}]\cap
W^{1,p}$ is a fixed point of the operator $T_{\biw^{n}}$, with
$n\in\N$. By definition and by Theorem~\ref{T:HC-E}, each element in
this sequence satisfies the equations
\[
A_{\tiL}\phi^{n} + f\bigl(\phi^{n},\biw^{n}\bigr) =0,\qquad
A_{\tiIL}\biw^{n} + \bif \bigl(\phi^{n-1}\bigr)=0.
\]
We will show that the sequence $\{\phi^n ,\biw^n\}\subset
[\phi_{-},\phi_{+}]\cap W^{1,p}\times\biW^{1,q}$ is bounded. The
proof is as follows. First, the elliptic estimates for the momentum
constraint given in \Sec\ref{S:SMC}, which imply, that there exists
positive constants $\tilde\ttK_1$, $\tilde\ttK_2$ such that
\[
\|\biw^n\|_{1,q} \leqs \hat\ttK_1 \, \phi_{+}^6 
+ \hat\ttK_2,\qquad \forall \, n\in\N.
\]
Second, a calculation similar to the one performed after
Eq.~(\ref{T:HC-E-Tbound}) changing norms in $W^{-1,2}$ with norms in
$W^{-1,p}$ for $p>3$ implies that
\[
\|T_{w^n}^s(\phi^n)\|_{\infty} \leqs
c_0\,\|T_{w^n}^s(\phi^n)\|_{1,p} \leqs k_{w^n},
\]
\begin{align*}
k_{w^n} &:= c_0c_{\tiL}\,\bigl[ 
\tilde a_{\tau}^{\tiwedge}\,\phi_{+}^5
+ (\tilde a_{\tiR}^{\tiwedge}+ \tilde s_n^{\tiwedge}) \,\phi_{+} 
+ \tilde a_{\rho}^{\tiwedge} \, \phi_{-}^{-3}\\
&\quad + \tilde a_{w^n}^{\tiwedge} \, \phi_{-}^{-7}
+ c_0 \|\hat\phi_{\tiN}^{*}\|_{-\frac{1}{p},p,\tiN} \bigr],
\end{align*}
where we have introduced the constants $\tilde a_{\tau}^{\tiwedge}$,
$\tilde a_{\tiR}^{\tiwedge}$, $\tilde a_{\rho}^{\tiwedge}$, and
$\tilde a_{w^n}^{\tiwedge}$, which are defined in a similar way as in
Eqs.~(\ref{WF-LB-tR-up})-(\ref{WF-LB-rw-up}) changing the norms in
$W^{1,2}_{\tiD}$ by norms in $W^{1,p}_{\tiD}$. We have also
introduced the number $\tilde s^{\tiwedge}$ given by
\[
\tilde s^{\tiwedge}_n := 5 \,\phi_{+}^4 \tilde a_{\tau}^{\tiwedge}
+ a_{\tiR}^{\tiwedge} + 3\,\frac{\phi_{+}^2}{\phi_{-}^6}\,
\tilde a_{\rho}^{\tiwedge} + 7\,\frac{\phi_{+}^6}{\phi_{-}^{14}}\,
\tilde a_{w^n}^{\tiwedge}. 
\]
The dependence on $n$ in $k_{w^n}$ is due to the term $\tilde
a_{w^n}^{\tiwedge}$. However, Eq.~(\ref{CS-aw-bound}) implies that
$a_{w^n}^{\tiwedge}$ can be bounded for all $n\in\N$, and we obtain
the inequality
\[
\|T_{w^n}^s(\phi^n)\|_{\infty} \leqs k_0,
\]
\begin{align*}
k_0 &:= c_0 c_{\tiL}\,\bigl[ 
\tilde a_{\tau}^{\tiwedge}\,\phi_{+}^5
+ (\tilde a_{\tiR}^{\tiwedge} 
+ \tilde s^{\tiwedge}_{\max}) \,\phi_{+} 
+ \tilde a_{\rho}^{\tiwedge} \, \phi_{-}^{-3}\\
&\quad+ (\tilde \ttK_1 \phi_{+}^{12}+ \tilde\ttK_2)\,\phi_{-}^{-7}
+ c_0 \|\hat\phi_{\tiN}^{*}\|_{-\frac{1}{p},p,\tiN} \bigr],\\
\tilde s^{\tiwedge}_{\max} &:= 
5 \,\phi_{+}^4 \tilde a_{\tau}^{\tiwedge}
+\tilde a_{\tiR}^{\tiwedge} + 3\,\frac{\phi_{+}^2}{\phi_{-}^6}\,
\tilde a_{\rho}^{\tiwedge} + 7\,\frac{\phi_{+}^6}{\phi_{-}^{14}}\,
 (\tilde \ttK_1 \phi_{+}^{12} + \tilde\ttK_2).
\end{align*}
This establishes that the sequence $\{\phi^n ,\biw^n\}$ is bounded in
$W^{1,p}\times\biW^{1,q}$. The latter space is a reflexive Banach
space, so the sequence $\{\phi^n ,\biw^n\}$ has a weakly convergent
subsequence, that is, there exist elements $\phi\in W^{1,p}$ and
$\biw\in\biW^{1,q}$ such that
\[
\phi^{n_j} \wto \phi \in W^{1,p},\qquad
\biw^{n_j} \wto \biw \in \biW^{1,q}.
\]
The imbeddings $W^{1,p}\to L^{\infty}$ and
$\biW^{1,q}\to\biL^{\infty}$ are compact since $3< q < p$, which
implies that
\[
\phi^{n_j} \to \phi \mbox{~~in~~} L^{\infty},\qquad
\biw^{n_j} \to \biw \mbox{~~in~~} \biL^{\infty},
\]
that is, the convergence is strong in these spaces. We now first note
that $\phi\in [\phi_{-},\phi_{+}]$, since the sequence $\phi^{n_j}\to
\phi$ in the supremum norm and the interval $[\phi_{-},\phi_{+}]$ is a
closed set in $L^{\infty}$. Second, we now show that indeed
\[
\biw^{n_j} \to \biw \mbox{~~in~~} \biW^{1,q},
\]
that is, the sequence $\{\biw^{n_j}\}$ converges strongly to $\biw$ in
$\biW^{1,q}$. The proof is to show that $\{\biw^{n_j}\}$ is a Cauchy
sequence in $\biW^{1,q}$. This is shown by the following calculation,
where we rename $\{\phi^{n_j}, \biw^{n_j}\}$ simply as $\{\phi^n,
\biw^n\}$. Then, we obtain,
\begin{align*}
\|\biw^n-\biw^m\|_{1,q} &\leqs c_1 \,
\|\bib_{\tau}^{*}\|_{-1,q} \, 
\Bigl\|\bigl[ (\phi^n)^6-(\phi^m)^6\bigr]\Bigr\|_{\infty}\\
&= c_1\,\|\bib_{\tau}^{*}\|_{-1,q}\,
\Bigl\|\Bigl[ \sum_{j=0}^5 (\phi^n)^j(\phi^m)^{5-j}\Bigr]\, 
\bigl(\phi^n-\phi^m \bigr)\Bigr\|_{\infty}\\
&\leqs  c_1\,\|\bib_{\tau}^{*}\|_{-1,q}\, \Bigl[
\sum_{j=0}^5 \|\phi^n\|_{\infty}^j \,\|\phi^m\|_{\infty}^{5-j} 
\Bigr] \, \bigl\|\phi^n-\phi^m\bigr\|_{\infty},
\end{align*}
which leads to
\[
\|\biw^n-\biw^m\|_{1,q} \leqs 6c_1  \,\phi_{+}^6 \,
\|\bib_{\tau}^{*}\|_{-1,q}\, \bigl\|\phi^n-\phi^m\bigr\|_{\infty}.
\]
Since $\phi^n$ is Cauchy in $L^{\infty}$, we have established that
$\biw^n$ is a Cauchy sequence in $\biW^{1,q}$.

The final step in the proof is to verify that $\phi$ and $\biw$
satisfy the constraint equations~(\ref{WF-LYs1})-(\ref{WF-LYm1}).
Since $\phi\in [\phi_{-},\phi_{+}]$, the function $T_{w}^s(\phi) \in
W^{1,p}\subset L^{\infty}$ is well-defined. What we have to show is
that
\begin{equation}
\label{CS-sol}
T_{w}^s(\phi) = \phi, \qquad S(\phi) =\biw.
\end{equation}
The first equation in (\ref{CS-sol}) can be written conveniently as follows:
\[
T_{w}^s(\phi)-\phi = \Bigl( T_{w}^s(\phi) - T_{w}^s(\phi^{n}) \Bigr)
+ \Bigl( T_{w}^s(\phi^{n}) - T_{w^n}^s(\phi^{n}) \Bigr)
+(\phi^{n} -\phi),
\]
therefore,
\begin{equation}
\label{CS-sol-1}
\|T_{w}^s(\phi) -\phi\|_{\infty} \leqs 
\| T_{w}^s(\phi) - T_{w}^s(\phi^{n}) \|_{\infty}
+ \| T_{w}^s(\phi^{n}) - T_{w^n}^s(\phi^{n}) \|_{\infty}
+\|\phi^{n} -\phi\|_{\infty}.
\end{equation}
The first term on the right hand side above satisfies the inequalities
\begin{align*}
\|T_{w}^s(\phi)-T_{w}^s(\phi^n)\|_{\infty} 
&\leqs c_0\,\|T_{w}^s(\phi)-T_{w}^s(\phi^n)\|_{1,p}\\ 
&= c_0\,\|(A_{\tiL}^s)^{-1}
\bigl( f^s(\phi,\biw) - f^s(\phi^n,\biw) \bigr)\|_{1,p}\\
&\leqs c_0c_{\tiL}\, \|f^s(\phi,\biw) - f^s(\phi^n,\biw)\|_{-1,p},
\end{align*}
and the last line above can be bounded as follows,
\begin{align*}
\|f^s(\phi,\biw) - f^s(\phi^n,\biw)\|_{-1,p} &\leqs 
\|a_{\tau}^{*}\|_{-1,p}\, \|\phi^5-(\phi^n)^5\|_{\infty}\\
&\quad + \bigl( \|a_{\tiR}^{*}\|_{-1,p} + \|s^{*}\|_{-1,p} \bigr)
\,\|\phi-\phi^n\|_{\infty} \\
&\quad + \|a_{\rho}^{*}\|_{-1,p} \,
\|\phi^{-3} - (\phi^n)^{-3}\|_{\infty}\\
&\quad + \|a_{w}^{*}\|_{-1,p}\, 
\|\phi^{-7} - (\phi^n)^{-7}\|_{\infty}.
\end{align*}
Notice that $\|a_{\tau}^{*}\|_{-1,p} = \tilde a_{\tau}^{\tiwedge}$,
and the same holds for $a_{\tiR}^{*}$, $s^{*}$, $a_{\rho}^{*}$ and
$a_{w}^{*}$. These relations together with
Eqs.~(\ref{HC-E-exp1})-(\ref{HC-E-exp3}) imply that
\begin{multline} 
\|f^s(\phi,\biw) - f^s(\phi^n,\biw)\|_{-1,p} \\
\begin{aligned}
&\leqs \Bigl[ 5\phi_{+}^4 \tilde a_{\tau}^{\tiwedge} 
+ (\tilde a_{\tiR}^{\tiwedge}+ \tilde s^{\tiwedge})
+ 3\frac{\phi_{+}^2}{\phi_{-}^6} \, \tilde a_{\rho}^{\tiwedge}
+ 7 \frac{\phi_{+}^6}{\phi_{-}^{14}}\, \tilde a_{w}^{\tiwedge} 
\Bigr]\, \|\phi - \phi^n\|_{\infty}\\
&= \bigl( \tilde\alpha^{\tiwedge} + \tilde s^{\tiwedge}\bigr)\,
\|\phi-\phi^n\|_{\infty}.
\end{aligned}
\end{multline}
Theorem~\ref{T:HC-E} requires $s\geqs \alpha$, with $\alpha$ given in
Eq.~(\ref{T:HC-E-def-alpha}). Choosing $s=\alpha$ we obtain,
\[
\|f(\phi,\biw) 
- f(\phi^n,\biw)\|_{-1,p} \leqs 2 \tilde\alpha^{\tiwedge}\,
\|\phi-\phi^n\|_{\infty},
\]
with 
\[
\tilde \alpha^{\tiwedge} := 5 \,\phi_{+}^4 \tilde a_{\tau}^{\tiwedge}
+ \tilde a_{\tiR}^{\tiwedge}
+ 3\,\frac{\phi_{+}^2}{\phi_{-}^6}\,\tilde a_{\rho}^{\tiwedge}
+ 7\,\frac{\phi_{+}^6}{\phi_{-}^{14}}\,\tilde a_{w}^{\tiwedge}.
\]
The number $\tilde\alpha^{\tiwedge}$ can be bounded independently of
$\biw$ since we have the inequality $\tilde
a_{w}^{\tiwedge}\leqs\tilde\ttK_1\,\phi_{+}^{12} +\tilde\ttK_2$, which
is obtained in a similar way as the inequality in
Eq.~(\ref{CS-aw-bound}), just changing the norms used in that result
to the appropriate norms needed here. Therefore, the following bound
holds,
\[
\|T_{w}^s(\phi) - T_{w}^s(\phi^n)\|_{\infty} 
\leqs k_{T} \, \|\phi -\phi^n\|_{\infty},
\]
with the constant $k_{T}$ given by
\begin{equation}
\label{CS-def-kT}
k_{T} := 2 c_0c_{\tiL} \,\Bigl[
5 \,\phi_{+}^4 \tilde a_{\tau}^{\tiwedge}
+ \tilde a_{\tiR}^{\tiwedge}
+ 3\,\frac{\phi_{+}^2}{\phi_{-}^6}\,\tilde a_{\rho}^{\tiwedge}
+ 7\,\frac{\phi_{+}^6}{\phi_{-}^{14}}\,
\bigl( \tilde \ttK_1\,\phi_{+}^{12} + \tilde \ttK_2 \bigr) \Bigr].
\end{equation}
(Although we do not need to exploit this fact here, note that from the
definition of $k_{T}$ it can be seen that there always exists source
functions and boundary data small enough such that $0\leqs k_{T} <1$,
a condition that implies that this map $T_{w}^s$ is a
$k_{T}$-contraction, as is it defined in~\cite{Zeidler-I} page 17.)

The second term on the right hand side in Eq.~(\ref{CS-sol-1})
satisfies the following bounds
\begin{align*}
 \| T_{w}^s(\phi^{n}) - T_{w^n}^s(\phi^{n}) \|_{\infty} 
&\leqs c_0\,\| T_{w}^s(\phi^{n}) - T_{w^n}^s(\phi^{n}) \|_{1,p}\\ 
&\leqs c_0\,\bigl\| (A_{\tiL}^s)^{-1} \bigl[ f^s(\phi^{n},\biw) 
- f^s(\phi^{n},\biw^n) \bigr] \bigr\|_{1,p}\\
&\leqs c_0c_{\tiL} \, \| f^s(\phi^{n},\biw)
- f^s(\phi^{n},\biw^n) \|_{-1,p}\\
&\leqs c_0c_{\tiL} \, \| a_{\biw}^{*} - a_{\biw^n}^{*}\|_{-1,p}\, 
\|(\phi^{n})^{-7} \|_{\infty}\\
&\leqs c_0c_{\tiL} \, \phi_{-}^{-7} \,
\| a_{\biw}^{*} - a_{\biw^n}^{*}\|_{-1,p}.
\end{align*}
Recall that the functional $a_{w}^{*}$ can be expressed as
$a_{w}^{*}(\un\varphi) = (a_{w},\un\varphi)$ for all $\un\varphi\in
W^{1,p'}_{\tiD}$, with $a_{w} = (\sigma+\cL \biw)^2/8$. We then
conclude that $a_{w}\in L^{q/2}$, with $q=6p/(3+p)$, which implies that
for $p > 3$ we have the inequality $3< q < p$. Hence, there exists a
positive constant $c_s$ such that $\|a_{w}^{*}\|_{-1,p}\leqs
c_s\|a_{w}\|_{(q/2)}$. So we have the inequality
\[
\| T_{w}^s(\phi^{n}) - T_{w^n}^s(\phi^{n}) \|_{\infty} \leqs 
c_0c_{\tiL}c_s \, \phi_{-}^{-7} \,
\| a_{\biw} - a_{\biw^n}\|_{(q/2)}.
\]
Now, the function $a_{w}-a_{w^n}$ can be written as
\begin{align*}
a_{\biw} - a_{\biw^{n}} &= \frac{1}{8}\, \bigl[
(\sigma+\cL\biw)^2 - (\sigma+\cL\biw^{n})^2\bigr]\\
&= \frac{1}{8}\, \bigl[ 2\sigma+\cL(\biw+\biw^{n})\bigr]\,
 \cL(\biw - \biw^{n}),
\end{align*}
with each factor in $L^{q}$, so a simple case of the generalized
H\"older inequality (see the last part of \Sec\ref{S:NC}) implies that
\[
\| a_{w} - a_{w^n}\|_{(q/2)} \leqs
\frac{1}{8}\, \|2\sigma+\cL(\biw+\biw^{n})\|_{q} \, 
\|\cL(\biw - \biw^n)\|_{q}.
\]
Then, we have the further inequalities
\begin{align*}
\| a_{w} - a_{w^n}\|_{(q/2)} &\leqs
\frac{1}{8}\,\bigl(2 \|\sigma\|_{q} 
+ \|\cL(\biw+\biw^{n})\|_{q} \bigr) \, 
\|\cL(\biw - \biw^n)\|_{q}\\
&\leqs \frac{c_{\cL}}{8}\,\bigl(2 \|\sigma\|_{q} 
+ c_{\cL} \, \|\biw+\biw^{n}\|_{1,q} \bigr) \, 
\|\biw - \biw^n\|_{1,q}\\
&\leqs \frac{c_{\cL}}{4}\,\bigl[ \|\sigma\|_{q} 
+ c_{\cL} \, (\tilde\ttK_1 \,\phi_{+}^{6} + \tilde\ttK_2) \bigr] \, 
\|\biw - \biw^n\|_{1,q}.
\end{align*}
Denote $k_2 = c_0c_{\tiL}c_s
c_{\cL}\phi_{-}^{-7}\bigl[\|\sigma\|_{q} + c_{\cL} \,
(\tilde\ttK_1 \,\phi_{+}^{6} +\tilde\ttK_2)\bigr]/4$, then
\[
 \| T_{w^n}^s(\phi^{n}) - T_{\biw}^s(\phi^{n}) \|_{\infty}
\leqs k_2 \, \|\biw^n-\biw\|_{1,q}.
\]
Therefore, the inequality in Eq.~(\ref{CS-sol-1}) implies
\[
\|T_{w}^s(\phi)-\phi\|_{\infty} \leqs
k_{T}\, \|\phi-\phi^n\|_{\infty} 
+ k_2 \, \|\biw -\biw^n\|_{1,q}
+ \|\phi^n - \phi\|_{\infty},
\]
and all the terms in the right hand side approaches zero when $n$
approaches infinity, so we conclude that $T_w^s(\phi)=\phi$. 

We now show that
the second equation in~(\ref{CS-sol}) also holds using the 
following argument.
We begin with the convenient representation
\[
S(\phi)-\biw = \bigl[ S(\phi) - S(\phi^{n}) \bigr]
+(\biw^{n+1} -\biw).
\]
This gives
\begin{equation}
\label{CS-sol-2}
\|S(\phi)-\biw \|_{1,q} \leqs 
\| S(\phi) - S(\phi^{n}) \|_{1,q}
+ \|\biw^{n+1} -\biw \|_{1,q}.
\end{equation}
The first term on the right hand side can be bounded as follows,
\begin{align*}
\| S(\phi) - S(\phi^{n}) \|_{1,q} 
&=\bigl\|-(A_{\tiIL})^{-1}\bigl(\bif(\phi)
-\bif(\phi^n)\bigr)\bigr\|_{1,q}\\
&\leqs c_{\tiIL}\, \|\bif(\phi)-\bif(\phi^n)\bigr\|_{-1,q}\\
&\leqs c_{\tiIL} \, \bigl\|\bib_{\tau}^{*}\|_{-1,q}\,
\|\phi^6-(\phi^n)^6\|_{\infty},
\end{align*}
where to get the last line we used the product property elements in
$L^{\infty}$ and elements in $W^{-1,q}_{\tiID}$, which is discussed
in the Gelfand triple part of the Appendix. Recalling the identity
\[
\phi^6-(\phi^n)^6 = (\phi-\phi^n) \sum_{j=0}^5 (\phi)^j(\phi^n)^{5-j},
\]
and that $\phi$, and $\phi^n\in [\phi_{-},\phi_{+}]$, one finds
\[
\| S(\phi) - S(\phi^{n}) \|_{1,q} \leqs
k_3 \, \| \phi-\phi^{n}\|_{\infty},
\]
with $k_3 = 6c_{\tiIL}\,\|\bib_{\tau}^{*}\|_{-1,q}\,\phi_{+}^5$. Finally,
inequality~(\ref{CS-sol-2}) and the inequalities above imply
\[
\|S(\phi)-\biw \|_{1,q} \leqs  k_3 \, \| \phi-\phi^{n}\|_{\infty} 
+  \|\biw^{n+1} -\biw \|_{1,q}.
\]
The right hand side in equation above approaches zero as $n$
approaches infinity. Therefore we conclude that $S(\phi)=\biw$. This
result establishes the Theorem.\qed

\subsection{Regularity of solutions}
\label{S:CS-Reg}

A bootstrap type argument shows that the regularity of weak
solutions is actually related to the minimum regularity of the
equation coefficients and of the boundary data.

\begin{proposition}{\bf(Non-CMC Regularity)}
\label{P:CS-Reg}
Assume the hypotheses in Theorem~\ref{T:CS-Ex}, assume that the
boundary set $\partial\cM$ is $C^2$, and recall the parameters
$q=6p/(3+p)$ and $p > 3$. If the extension $\phi_{\tiD}$ of the
Dirichlet boundary data and the Robin data $\hat\phi_{\tiN}$ for the
Hamiltonian constraint equation (\ref{WF-LYs}) satisfy
\[
\phi_{\tiD} \in W^{2,(q/2)},\qquad
\hat\phi_{\tiN} \in 
W^{\frac{1}{(q/2)'},(q/2)}(\partial\cM_{\tiN},0),
\]
then the solution $\phi\in [\phi_{-},\phi_{+}]\cap W^{1,p}$ and
$\tbw\in\tbW^{1,q}$ of the weak Dirichlet-Robin boundary value
formulation for the Hamiltonian and momentum constraint
Eqs.~(\ref{WF-LYs1})-(\ref{WF-LYm1}) found in Theorem~\ref{T:CS-Ex}
also satisfies $\phi\in [\phi_{-},\phi_{+}]\cap W^{2,(q/2)}$ and
$\tbw\in\tbW^{1,q}$.
\end{proposition}

\Outline {\it (Proposition~\ref{P:CS-Reg}.)~}
The proof can again be based on linear elliptic estimates (see
Theorem~9.11 in~\cite{Gilbarg-Trudinger}, and~\cite{Ciarlet97},
Vol. II, page 296) and a standard bootstrap argument.
\qed

\section{Summary}
\label{S:summary}

In this article, we considered the conformal decomposition of
Einstein's constraint equations introduced by Lichnerowicz and York,
on a compact manifold with boundary. We began by developing some basic
technical results for the momentum constraint operator, and then
established existence and uniqueness of $W^{1,2}$-solutions to the
momentum constraint (with conformal factor as fixed data) using
variational methods.  Among the technical results we established were
generalized Korn inequalities for the conformal Killing operator on a
compact manifold with boundary, Lemma~\ref{L:GKIbc}, which does not
appear to be in the literature.  An alternative invertibility argument
for the divergence of the conformal Killing operator is given using
Riesz-Schauder theory, which yielded similar results in the case
where the Dirichlet part of the boundary is non-empty. In both cases,
the assumptions on the data were quite weak so that standard
techniques cannot be used to establish additional regularity.

We then considered the Hamiltonian constraint (with momentum vector as
fixed data); using order cones in Banach spaces, we derived weak sub-
and super-solutions to the Hamiltonian constraint.  These can be viewed 
as non-trivial generalizations of the barriers constructed previously 
in the literature to a setting with much weaker assumptions on the data. 
We also establish some related {\em a~priori} $L^{\infty}$-bounds
on any $W^{1,2}$-solution to the Hamiltonian constraint 
(Theorem~\ref{T:ape}).
Although such results are standard for semi-linear scalar
problems with monotone nonlinearities (for example, see~\cite{jJ85}),
our results hold for a class of non-monotone nonlinearities that
includes the Hamiltonian constraint nonlinearity and appear to be new.
The generalized sub- and super-solutions are subsequently used
together with variational methods to establish existence (and
uniqueness when scalar curvature $R\geqs 0$) of solutions to the
Hamiltonian constraint in $L^{\infty} \cap W^{1,2}$.  Our arguments
allowed the scalar curvature $R$ to have any sign; the case of
non-negative $R$ required the additional assumption that either the
matter energy density or the trace-free, divergence-free part of the
extrinsic curvature be positive. Again, we made very weak assumptions
on the data so that standard techniques cannot be used to establish
additional regularity of the solutions.  Due to the lack of
G\^ateaux-differentiability of the nonlinearity in the space
$W^{1,2}$, the connection between the energy used in the variational
argument and the Hamiltonian constraint as its Euler condition for
stationarity was non-trivial, and was established through several
Lemmas. Although our problem formulation is slightly different
(bounded domains with matter), the final result for weak solutions of
the Hamiltonian constraint could be viewed as lowering the regularity
of the recent result of Maxwell~\cite{dM05} on ``rough'' CMC solutions
in $W^{k,2}$ for $k>3/2$ down to $L^{\infty} \cap W^{1,2}$.  We also
gave an alternative non-variational argument using the more standard
barrier methods, which requires more assumptions on the data, and
yields essentially the Maxwell result for our problem formulation.

We then combined the weak solution results for the individual
Hamiltonian and momentum constraints to establish an existence result
for the coupled system in the case of nonconstant mean curvature,
through fixed-point iteration and compactness arguments rather than
through the Contraction Mapping Theorem as used in the original 1996 work
of Isenberg and Moncrief.  This result requires more regularity than
that needed for the results established for the individual
constraints, with solutions for the conformal factor in $W^{1,p}$ for
$p>3$, and momentum vector in $W^{1,q}$ for $q=6p/(3+p)$, but still
extends the existing theory for the system in two ways. First,
although our problem formulation is somewhat different (bounded
domains with matter), the results could be viewed as extending the
1996 result of Isenberg and Moncrief on nonconstant mean curvature
with Ricci scalar $R=-1$, to weaker solution spaces, and to cases
where the Ricci scalar $R$ can have sign.  Second, again although the
problem formulation is different, the result could be viewed as
extending the recent rough solution work of Maxwell from the CMC case
to the non-CMC case.

It is interesting to note that for the main result on the non-CMC
coupled system in \cite{jIvM96}, the near-CMC condition on the trace
of the extrinsic curvature is actually used twice: once to obtain the
global super-solution, and a second distinct time to construct a
contraction for using the Contraction Mapping Theorem to get existence
and uniqueness.  By using a compactness argument directly rather than
the Contraction Mapping Theorem, we avoid the second use of the
near-CMC condition.  If a global super-solution can be constructed
without the near-CMC assumption, then our compactness argument would
give existence of solutions to the coupled system in the fully general
``far-from-CMC'' case.  What our proof technique gives up is
uniqueness of solutions to the coupled system, which comes for free
with existence when the contraction argument used as in \cite{jIvM96}.

The variational approach used for the Hamiltonian constraint in the
article should allow for the treatment of the case where the
coefficient of the leading nonlinear term $\phi^5$ becomes slightly
negative, by using the Mountain Pass approach as in the recent
work of Hebey, Pacard, and Pollack.\footnote{
E.~Hebey, F.~Pacard, and D.~Pollack.
{\em A variational analysis of {E}instein-scalar field {L}ichnerowicz
equations on compact {R}iemannian manifolds}.
Available as gr-qc/0702031v1, 2007.
}
The variational approach presented here also for the momentum
constraint might make possible the combined variational treatment of
systems involving the Hamiltonian and/or momentum constraints as part
of a large variational system.  Finally, if the existing non-constant
sub- and super-solutions in the literature for the Hamiltonian
constraint can be extended to our less regular setting, it would allow
for weakening some of the assumptions on the signs of the coefficients
used for our results above.\footnote{
J.~Isenberg.  {\em Private communication}, 2007.
}

Although our presentation was for 3-manifolds, the results in the
paper remain valid for higher spatial dimensions with minor adjustments,
and the techniques we employed should extend to other cases such as closed
and (fully or partially) open manifolds through use of techniques such
as weighted Sobolev spaces.

\section{Acknowledgements}
The authors thanks Jim Isenberg for several helpful insights and
comments on the man\-u\-script. MH thanks Robert Bartnik, Jim
Isenberg, Vince Moncrief, and Niall O'Murchadha for many useful
discussions about this problem over several years.  MH thanks David
Bernstein for germinating a deep interest in this and related problems
in mathematical physics.  MH thanks Kip Thorne, Lee Lindblom, and Herb
Keller for hospitality, support, and enthusiasm for this work over a
number of years.  GN thanks Sergio Dain for useful discussions
regarding generalizations of Korn's inequalities to the conformal
Killing operator.  MH and GN thank Gantumur Tsogtgerel for a number of
helpful comments on the manuscript.
GN thanks the UCSD Mathematics Department for their hospitality.

MH was supported in part by NSF Awards~0715145, 0411723,
and 0511766, and DOE Awards DE-FG02-05ER25707 and DE-FG02-04ER25620. 
JK was supported in part by a UCSD Academic Enrichment
Fellowship and a UCSD/CalIT2 Summer Research Fellowship.
GN was supported in part by NSF Awards~0715145 and 0411723.

\appendix
\section{Some tools from nonlinear functional analysis}
\label{S:APP}

\subsection{Gelfand triples}
\label{S:GT}

A readable reference for Gelfand triples is \Sec17.1
in~\cite{Wloka87}. See also \Sec23.4 in~\cite{Zeidler-IIA}, where they
are called evolution triples. The vector spaces $(X,H,X^{*})$ form a
{\bf Gelfand triple} iff the space $H$ is a Hilbert space, $X$ is a
reflexive Banach space with dual space $X^{*}$, and there exists a
continuous imbedding $I:X\to H$ such that $I(X)$ is dense in $H$. It
can be shown that: If $I:X\to H$ is continuous and $I(X)$ dense in
$H$, then the dual map $I^{*}:H^{*}\to X^{*}$ is continuous; in
addition, since $X$ is reflexive, $I^{*}(H^{*})$ is dense in
$X^{*}$. For the proof see~\cite{Zeidler-IIA}, page~417. Denote by
$R:H^{*}\to H$ the {\bf Riesz map} defined as follows: given an
element $h^{*}\in H^{*}$ the element $Rh^{*}\in H$ is given by
$h^{*}(\un h)=(Rh^{*},\un h)_{\tiH}$ for all $\un h\in H$. It can be
shown that this map is a bijection, therefore it is invertible, and
its inverse satisfies that for all $h\in H$ the element $R^{-1}h\in
H^{*}$ and the following equation holds, $R^{-1}h(\un h) = (h,\un
h)_{\tiH}$ for all $\un h\in H$.  A Gelfand triple is usually denoted
as
\[
X\stackrel{I}{\longrightarrow} 
H\equiv H^{*} 
\stackrel{I^{*}}{\longrightarrow}X^{*}.
\]
Gelfand triples are useful to study weak formulations of elliptic
PDE. An example of a Gelfand triple is given by the Sobolev spaces
$(W^{1,p},L^2,W^{-1,p'})$, with $p'=p/(p-1)$.

The property of a Gelfand triple we are most interested in is that
$I^{*}(H^{*})$ is dense in $X^{*}$. This property together with the
existence of the Riesz map imply that for all $x^{*}\in X^{*}$ there
exists a sequence $\{h_n\}\subset H$ such that the elements
$x_n^{*}\in X^{*}$, defined as $x_n^{*}(\un x) := (h_n, I\un
x)_{\tiH}$ for all $\un x\in X^{*}$, satisfy
\begin{equation}
\label{GT-eq1}
x^{*} = \lim_{n\to\infty} x_n^{*},\qquad
\mbox{in~} X^{*}.
\end{equation}
The elements $x_n^{*}$ can be written in terms of the Riesz map $R$
and the imbedding $I$ as follows, $x_n^{*} = I^{*}R^{-1}h_n$. The
definition of the map $I^{*}$ implies that for all $h^{*}\in H^{*}$
holds
\[
I^{*}h^{*}\in X^{*},\qquad
I^{*}h^{*}(\un x) = h^{*}(I\un x)\qquad
\forall\, \un x\in X.
\]
Therefore, Eq.~(\ref{GT-eq1}) can be expressed as follows: for all
$x^{*}\in X^{*}$ there exists a sequence $\{h_n\}\subset H$ such that
\[
x^{*}(\un x) = \lim_{n\to\infty}(h_n,I\un x)_{\tiH}\qquad
\forall\, \un x\in X.
\]
\begin{definition}{\bf(Product)}
\label{D:GT-prod}
Let $X\stackrel{I}{\longrightarrow} H\equiv
H^{*}\stackrel{I^{*}}{\longrightarrow}X^{*}$ be a Gelfand triple, and
let $V$ be a vector space such that there exists an imbedding
$\cI:V\to H$. Furthermore, assume that for all $v\in V$ and $h\in H$
there exists a map $v,h\mapsto (\cI v)h \in H$, where the element
$(\cI v)h$ in $H$ satisfies that for every $v\in V$ there exists a
positive constant $c_v$ such that
\[
\bigl( (\cI v) h, \un h\bigr)_{\tiH} \leqs c_v |(h,\un h)_{\tiH}|
\qquad \forall\, h,\un h \in H.
\]
Then, given any $x^{*}\in X^{*}$ define the map $v,x^{*}\mapsto
(vx)^{*}\in X^{*}$ as follows:
\[
(vx)^{*}(\un x) := \lim_{n\to \infty} 
x^{*}_n\bigl((\cI v)(I\un x)\bigr)\qquad
\forall \un x\in X.
\]
\end{definition}

An example of the situation above is the following: Let the Gelfand
triple be given by the Sobolev spaces of scalar valued functions
$(W^{1,p},L^2,W^{-1,p'})$, let the subspace $V=L^{\infty}$, and let
the map $v,h\mapsto vh$ be defined as pointwise multiplication a.e. in
the domains of $v$ and $h$. Then, all the properties in
Def.~\ref{D:GT-prod} are satisfied. Indeed, the name ``product'' for
the definition above originates in this example. This product given in
Def.~\ref{D:GT-prod} satisfies the following property:
\[
\|(vx)^{*}\|_{\tiX^{*}} \leqs c_v \, \|x^{*}\|_{\tiX^{*}}
\qquad \forall\, x^{*}\in X^{*} \mbox{~and~}\forall\, v\in V.
\]
The proof is the following calculation: 
\[
\|(vx)^{*}\|_{\tiX^{*}} =  \lim_{n\to\infty}\|(vx_n)^{*}\|_{\tiX^{*}}
=\lim_{n\to\infty} \sup_{0\neq \un x\in X}
\frac{|\bigl( (\cI v)h_{n}, (I\un x)\bigr)_{\tiH}|}{~\|\un x\|_{\tiX}};
\]
noticing that $|\bigl (\cI v)h_n,(I\un x)\bigr)_{\tiH}| \leqs c_v
|(h_n,(I\un x))_{\tiH}|$ holds for all $\un x\in X$, then it also holds
for the supremum in $\un x$, which leads us to the following
inequalities,
\[
\|(vx)^{*}\|_{\tiX^{*}} \leqs c_v\, \lim_{n\to\infty}
 \sup_{0\neq \un x\in X}
\frac{|\bigl( h_{n},(I\un x)\bigr)_{\tiH}|}{~\|\un x\|_{\tiX}} 
= c_v\, \lim_{n\to\infty}\|x_n^{*}\|_{\tiX^{*}}
= c_v\,\|x^{*}\|_{\tiX^{*}}.
\]

We now show that the map $v,x^{*}\mapsto (vx)^{*}$ is well-defined in
the sense that it is independent of the sequence $x^{*}_n$ that
approximates $x^{*}$. The proof is the following: Let
$\{x^{*}_{h_n}\}$ and $\{\tilde x^{*}_{\tilde h_n}\}$ be sequences in
$X^{*}$ such that as $n\to\infty$ holds
\[
x^{*}_{h_n} \to x^{*},\quad
\tilde x^{*}_{\tilde h_n} \to x^{*}\quad \mbox{in} \quad X^{*}.
\]
Introduce the sequences $\{(vx)^{*}_{h_n}\}$ and $\{(v\tilde
x_{\tilde h_n})^{*}\}$ in $X^{*}$ such that 
\[
(vx_{h_n})^{*} \to (vx)^{*},\quad
(v\tilde x_{\tilde h_n})^{*} \to (\widetilde{vx})^{*}\quad 
\mbox{in} \quad X^{*}.
\]
Then, the following argument shows that
\begin{align}
\nonumber
\|(vx)^{*} - (\widetilde{vx})^{*}\|_{\tiX^{*}} &\leqs
  \|(vx)^{*} - (vx_{h_n})^{*}\|_{\tiX^{*}}
+ \|(vx_{h_n})^{*} - (v\tilde x_{\tilde h_n})^{*}\|_{\tiX^{*}}\\
\nonumber
&\quad 
+ \|(v \tilde x_{\tilde h_n})^{*} - (\widetilde{vx})^{*}\|_{\tiX^{*}}\\
\nonumber
&\leqs \epsilon
+ c_v\, \|x_{h_n}^{*} - \tilde x_{\tilde h_n}^{*}\|_{\tiX^{*}}
+\epsilon\\
\nonumber
&\leqs 2\epsilon + c_v \bigl(
\|x_{h_n}^{*} - x^{*}\|_{\tiX^{*}}
+\|x^{*} - \tilde x_{\tilde h_n}^{*}\|_{\tiX^{*}} \bigr)\\
\label{A:GT-ineq1}
&\leqs 2(1+c_v)\epsilon,
\end{align}
where the second inequality comes from the following one,
\[
\|(vx_{h_n})^{*} - (v\tilde x_{\tilde h_n})^{*}\|_{\tiX^{*}}
= \sup_{0\neq \un x\in X} 
\frac{|\bigl(v[h_n-\tilde h_n],I\un x\bigr)_{\tiH}|}{~\|\un x\|_{\tiX}}
\leqs c_v\, \|x_{h_n}^{*} - \tilde x_{\tilde h_n}^{*}\|_{\tiX^{*}}.
\]
Since $\epsilon\to 0$ as $n\to\infty$, we then conclude from
Eq.~(\ref{A:GT-ineq1}) that $(vx)^{*} = (\widetilde{vx})^{*}$.

\subsection{G\aa rding inequality and Riesz-Schauder theory}
\label{S:A-LM-F}

We recall now (without giving a proof) the well-known result of Riesz
and Schauder. Standard references for this result are
in~\cite{Gilbarg-Trudinger} page~76, in~\cite{Wloka87} page~166, and
in~\cite{Zeidler-I} page 372.
\begin{theorem}{\bf (Riesz-Schauder)}
\label{T:Riesz-Schauder}
Let $X$ be a Banach space, $K:X\to X$ be a linear and compact map, and
$\cI_{\tiX} :X\to X$ be the identity map. Then, the following
statements hold:
\begin{enumerate}[(i)]
\item
$\dim N_{(\cI_{\tiX}-K)} = \dim N_{(\cI_{\tiX}^{*}-K^{*})}$;

\item
Given $f\in X$ the equation
\begin{equation}
\label{eq-K}
(\cI_{\tiX} - K) x = f
\end{equation}
has a solution iff $x_n^{*}(f)=0$ for all $x_n^{*}\in
N_{(\cI_{\tiX}^{*}-K^{*})}$;

\item
If $\dim N_{(\cI_{\tiX}^{*}-K^{*})} =0$, then the condition
$x_n^{*}(f)=0$ is trivially satisfied for all elements $f\in X$, hence
for every $f\in X$ there exist a unique element $x\in X$ solution of
Eq.~(\ref{eq-K}). Furthermore, the operator $(\cI_{\tiX}-K)^{-1}$,
whose existence is asserted here, is linear and bounded.

\item
If $\dim N_{(\cI_{\tiX}^{*}-K^{*})}> 0$ and the condition
$x_n^{*}(f)=0$ is satisfied, then the solutions $x$ of
Eq. (\ref{eq-K}) are not unique, and given any solution $x$ then $\hat
x = x +x_n$ is also a solution, with $x_n \in N_{(\cI_{\tiX}-K)}$;
\end{enumerate}
\end{theorem}

We now use the Riesz-Schauder Theorem above to show whether a linear
equation involving a bounded bilinear form satisfying G\aa rding's
inequality has solutions. Let $(X,H,X^{*})$ be a Gelfand triple, as it
is defined in the previous subsection of this Appendix. Introduce a
bilinear form $a:X\times X\to\R$ and consider the following problem:
Given an element $f^{*}\in X^{*}$ find an element $x\in X$ solution of
the equation
\begin{equation}
\label{w-eq}
a(x,\un x) = f^{*}(\un x) \qquad
\forall \, \un x\in X.
\end{equation}
It is convenient to reformulate this problem in terms of operators
instead of bilinear forms. Introduce the operator $A: X\to X^{*}$,
with action $Ax(\un x) := a(x,\un x)$ for all $x,\un x\in X$. Then,
the problem above has the following form: Given an element $f^{*}\in
X^{*}$ find an element $x\in X$ solution of the equation
\begin{equation}
\label{w-eq1}
A x = f^{*}.
\end{equation}
It is also convenient to introduce the Banach adjoint operator
$A^{*} :X\to X^{*}$ defined as $A^{*}x(\un x) := A\un x(x)$ for all
$x,\un x\in X$. We are identifying $X$ with its double dual space
$X^{**}$.  Let $N_{A}$, and $N_{A^{*}}$ be the null spaces of the
operators $A$ and $A^{*}$, respectively.
\begin{theorem}
\label{T:GI-FT}
Let $(X,H,X^{*})$ be a Gelfand triple, and in addition assume that the
imbedding $I :X\to H$ is compact. Let $A :X\to X^{*}$ be a linear,
bounded operator satisfying G\aa rding's inequality, that is, there
exist positive constants $k_0$ and $K_0$ such that
\[
\|Ax\|_{\tiX^{*}} \leqs K_0 \,\|x\|_{\tiX},\qquad
k_0 \,\| x\|^2_{\tiX} \leqs \|I x\|^2_{\tiH} + Ax(x),\qquad
\forall \,x \in X.
\]
Then, $\dim N_A = \dim N_{A^{*}}$ and Eq.~(\ref{w-eq1}) has a solution
$x\in X$ iff $f^{*}(\tilde x_n) =0$ for all $\tilde x_n \in
N_{A^{*}}$.  Furthermore, the following statements hold:
\begin{enumerate}[(i)]
\item \label{T:GI-FT-i}
If $\dim N_{A^{*}} = 0$, then there exists a unique $x\in X$ solution
of Eq.~(\ref{w-eq1}) for all $f^{*}\in X^{*}$; Furthermore, there
exists a positive constant $c_0$ such that the following estimate
holds,
\begin{equation}
\label{GI-FT-est}
\|x\|_{\tiX} \leqs c_0\,\|Ax\|_{\tiX^{*}}
\qquad \forall \, x\in X;
\end{equation}
\item \label{T:GI-FT-ii}
If $\dim N_{A^{*}} >0$ and the condition $f^{*}(\tilde x_n) =0$ for
all $\tilde x_n \in N_{A^{*}}$ holds, then the solution $x$ is not
unique, since $x' := x + x_n$ is also a solution, with $x_n\in
N_A$.
\end{enumerate}
\end{theorem}

\Proof {\it (Theorem~\ref{T:GI-FT}.)~}
Given the operator $A$, introduce the operator $A_{\tiX}: X\to X^{*}$
with action
\[
A_{\tiX}x(\un x) := Ax(\un x) + (Ix,I\un x)_{\tiH}.
\]
The assumptions that the operator $A$ is bounded and satisfies G\aa
rding's inequality imply that the operator $A_{\tiX}$ is bounded and
coercive, respectively, hence, invertible. Notice that $A_{\tiX}$ can
be written in terms of operators as follows: $A_{\tiX} = A + J$, where
$J :X\to X^{*}$ is given by $J:= I^{*} R^{-1}I$, since
\[
Jx(\un x) = I^{*}R^{-1}Ix(\un x)= R^{-1}Ix(I\un x)= (Ix,I\un x)_{\tiH}.
\]
The Eq.~(\ref{w-eq1}) can be re-expressed as follows:
\[
Ax = f^{*}\LRI (A_{\tiX} - J)x = f^{*}
\LRI (\cI_{\tiX} -  A_{\tiX}^{-1} J) x = A_{\tiX}^{-1} f^{*},
\]
where $\cI_{\tiX}:X\to X$ is the identity map. Introduce the notation
$f_{\tiX} := A_{\tiX}^{-1}f^{*}\in X$ and the operator $K :X\to X$
given by $K:= A_{\tiX}^{-1}J$. So, $x$ is solution of
Eq.~(\ref{w-eq1}) iff it solves the equation
\begin{equation}
\label{K-eq}
(\cI_{\tiX} - K) x = f_{\tiX}.
\end{equation}
Since the imbedding $I:X\to H$ is compact, and the remaining maps that
define $K$ are continuous, we conclude that $K$ is compact (for
example see~\cite{DunfordSchwartz-I} page~486, Theorem~4). Then, the
operator $\cI_{\tiX}-K$ is a Fredholm operator of index zero, and
Theorem~\ref{T:Riesz-Schauder} implies that $\dim N_{\cI_{\tiX}-K} =
\dim N_{\cI_{\tiX}^{*}-K^{*}}$. By construction we have that
$N_{\cI_{\tiX}-K} = N_A$. One can also show that $x_n^{*}\in
N_{\cI_{\tiX}^{*}-K^{*}}$ iff $\tilde x_n:= (A_{\tiX}^{-1})^{*}
x^{*}_n\in N_{A^{*}}$. Due to $A_{\tiX}$ is a bijection, this shows
that
\[
\dim N_A = \dim N_{A^{*}}.
\]
Theorem~\ref{T:Riesz-Schauder} implies that Eq.~(\ref{K-eq}) has
solution iff $x_n^{*}(f_{\tiX})=0$ for all $x_{n}^{*}\in
N_{\cI_{\tiX}^{*} - K^{*}}$. This condition can be rewritten as
follows:
\[
0 = x_n^{*} (f_{\tiX})= x_n^{*}(A_{\tiX}^{-1}f^{*})
=f^{*}\bigl((A_{\tiX}^{-1})^{*}x_n^{*}\bigr)
= f^{*}(\tilde x_n)\qquad \forall \, \tilde x_n \in N_{A^{*}},
\]
which is the condition appearing in Theorem~\ref{T:GI-FT}. In the case
that $\dim N_A = 0$, then $\dim N_{A^{*}}=0$, and so the condition
$x_n^{*}(f_{\tiX})=0$ is trivially satisfied for all $f_{\tiX}\in X$.
Therefore, Theorem~\ref{T:Riesz-Schauder} implies that for every
element $f^{*}\in X^{*}$ there always exists a unique solution $x\in
X$ of Eq.~(\ref{w-eq1}). This statement defines the operator $A^{-1} :
X^{*} \to X$, and Theorem~\ref{T:Riesz-Schauder} asserts that this
operator is linear and bounded, the latter property implies that there
exists a positive constant $c_0$ such that
\[
\|x\|_{\tiX} \leqs c_0\, \|Ax\|_{\tiX^{*}}
\qquad \forall\, x\in X.
\]
This establishes part~{\it(\ref{T:GI-FT-i})} in Theorem~\ref{T:GI-FT}. In the
case that $\dim N_A > 0$ and the condition $x_n^{*}(f_{\tiX})=0$ is
satisfied, then Theorem~\ref{T:Riesz-Schauder} says that a solution
$x\in X$ exists, and $x' = x + x_n$ is also a solution, where $x_n \in
N_A$. This establish part~{\it(\ref{T:GI-FT-ii})} in
Theorem~\ref{T:GI-FT}.\qed

\subsection{Variational methods}
\label{S:VM}

These notes follow the main ideas in Chapter~4 of Part Two
in~\cite{Jost-LiJost98}, and \Sec1, \Sec2 in Chapter~1
in~\cite{Struwe96}. An introduction into this subject is \Sec7.1
in~\cite{McOwen96}. The main result of this Section is
Theorem~\ref{T:VM-M}. We could not find in the literature this result
precisely in this form, needed for the Hamiltonian constraint problem,
so for completeness we included the proof of the Theorem.

Given a Banach space $X$, a subset $U\subset X$ is called {\bf closed
under weak convergence} (\closedw \!) iff for all sequence
$\{x_n\}\subset U$ such that $x_n\wto x_0$ in X holds that $x_0\in
U$. Every \closedw set in a Banach space is closed, but the converse
statement is not true. A particular class of closed sets that are also
\closedw are closed and convex sets. Given a vector space $V$, a
subset $U\subset V$ is called {\bf convex} iff for all $\tilde x$,
$\hat x\in U$ the elements $[t\tilde x +(1-t)\hat x]\in U$ for
$t\in[0,1]$. The proof of the above statement is based in a result by
Mazur (see Theorem~2.2.4 on page 142 in~\cite{Jost-LiJost98}) that
says: In a Banach space, for every sequence $\{x_n\}$ such that
$x_n\wto x_0$ there exists a sequence $\{y_m\}$ such that $y_m\to
x_0$, where the element $y_m$ are constructed as a convex combinations
of the $x_n$, that is,
\begin{equation}
\label{VM-cc}
y_m := \sum_{n=1}^m \lambda_nx_n,\qquad
\Bigl(\lambda_n\geqs 0,\quad
\sum_{n=1}^m\lambda_n =1\Bigr).
\end{equation}
Using this result is not difficult to show that every closed and
convex set $U$ in a Banach space $X$ is also \closedw, as the
following argument shows: given $\{x_n\}\subset U$ such that $x_n\wto
x_0$ in $X$, use Mazur's idea to construct the sequence $\{y_m\}$ as a
convex combination of the $x_n$ such that $y_m\to x_0$ in
$X$. However, $U$ is convex, so $\{y_m\}\subset U$, and it is also
closed, so $x_0\in U$. This establishes that $U$ is also \closedw.

Let $X$ be a Banach space, and introduce the functional $J:
X\to\overline\R$. The symbol $\overline\R$ means that there might
exist points $x_0\in U$ such that there exists a sequence $x_n\to x_0$
with $\lim_{n\to\infty}J(x_n) =\infty$ or equal $-\infty$. If such
points exist, then $J$ is an unbounded operator with domain $D_J$
strictly included in $X$. This idea is summarized with the notation
$\overline\R:=\R\cup\{+\infty\}\cup\{-\infty\}$. The notation $J:
X\to\overline\R$ is more convenient than the notation $J: D_J\subset
X\to\R$ in cases where the continuity or the differentiability of the
functional is not important in the situation under study. An example
is the problem of finding the local or global minimum of a functional
using direct methods, which do not include computing the Euler
equations for the functional. Given any subset $U\subset X$ of a
Banach space $ X$, the functional $J:U\subset X\to\overline\R$ is
called {\bf proper} on $U$ iff for all $\{x_n\}\subset U$ such that
$\|x_n\|_{\tiX}\to\infty$ holds $J(x_n)\to +\infty$. A particular case
of proper functionals are coercive functionals, where $J$ is called
{\bf coercive} on $U$ iff there exist positive constants $c_0$, $c_1$,
such that for all $x\in U$ holds $J(x)\geqs c_0\,\|x\|_{\tiX}^2
-c_1$. The functional $J$ is {\bf bounded below} on $U$ iff there
exists $\alpha_0\in\R$ such that $J(x)\geqs\alpha_0$ for all $x\in U$.
All coercive functionals are bounded below by $-c_1$.

The functional $J: X\to\overline\R$ is {\bf lower semi-continuous}
(lsc) at the element $x_0\in X$ iff for all sequence $\{x_n\}\subset
X$ with $x_n\to x_0$ in $X$ holds $J(x_0)\leqs\liminf_{n\to\infty}
J(x_n)$. Any continuous functional is lsc. The functional $J:
X\to\overline\R$ is called {\bf lower semi-continuous under weak
convergence} (\lscw\!) at the element $x_0\in X$ iff for all sequence
$\{x_n\}\subset X$ with $x_n\wto x_0$ in $X$ holds
\[
J(x_0) \leqs \liminf_{n\to\infty} J(x_n).
\]
Given any set $U\subset X$ a functional $J:U\subset X\to\overline\R$
is lsc (respectively \lscw\!) on $U$ if it is lsc (respectively
\lscw\!)  on all points in $U$. Not every lsc functional is \lscw. The
later property is a stronger condition on the functional than the
former property, due to the set of all sequences that converge weakly
in a Banach space is bigger than the set of all sequences that
converge strongly. An example of a \lscw functional, mentioned
in~\cite{Struwe96}, is the norm in an arbitrary Banach space, as the
following argument shows: Let $X$ be a Banach space, $\{x_n\}\subset
X$ be any sequence such that $x_n\wto x_0$, then there always exists
an element $x_{x_0}^{*}\in X^{*}$ such that $x^{*}_{x_0}(x_0) =
\|x^{*}_{x_0}\|_{\tiX^{*}}\,\|x_0\|_{\tiX}$; then we obtain
\[
\|x^{*}_{x_0}\|_{\tiX^{*}}\,\|x_0\|_{\tiX} =x^{*}_{x_0}(x_0) 
=\liminf_{n\to\infty}x^{*}_{x_0}(x_n) 
\leqs \|x^{*}_{x_0}\|_{\tiX^{*}}\,\liminf_{n\to\infty} \|x_n\|_{\tiX},
\]
and this implies that $\|x_0\|_{\tiX}\leqs\liminf_{n\to\infty}\|
x_n\|_{\tiX}$, establishing our assertion. The well-known case of the
norm in a Hilbert space $H$ being \lscw follows from the previous
argument choosing $x^{*}_{x_0}(\un x) = (x_0,\un x)_{\tiH}$, where
$\un x$ is any element in $H$.

\begin{theorem}{\bf (Existence of a minimizer)}
\label{T:VM-M}
Let $X$ be a reflexive Banach space, and $U\subset X$ be a \closedw
subset. Let $J:U\subset X\to\R$ be a proper, bounded below, and \lscw
functional. Then, there exists an element $x_0\in U$ minimizer of the
functional $J$ in the set $U$, that is,
\[
J(x_0) = \inf_{x\in U} J(x).
\]
\end{theorem}

\Proof {\it (Theorem~\ref{T:VM-M}.)~}
The functional $J$ is bounded below in $U$, therefore there exists a
positive constant $\alpha_0$ such that for all $x\in U$ holds
$J(x)\geqs\alpha_0$. Then, there exists a minimizing sequence, that
is, a sequence $\{x_n\}\subset U$ such that $J(x_n)\to\alpha_0$ as
$n\to\infty$. The functional $J$ is proper on $U$, therefore the
sequence $\{x_n\}$ is bounded. The Banach space $X$ is reflexive,
which implies that there exists $x_0\in X$ such that $x_n\wto x_0$ as
$n\to\infty$. The set $U$ is \closedw, therefore $x_0\in U$. Finally,
the functional $J$ is \lscw and the sequence $\{x_n\}$ is a minimizing
sequence, which imply that
\[
J(x_0) \leqs \liminf_{n\to\infty} J(x_n) = \alpha_0 
\leqs \inf_{x\in U}J(x).
\]
The definition of infimum implies $\di J(x_0) = \inf_{x\in U}J(x)$,
which establishes the Theorem.\qed

Let $U\subset X$ be a convex set in a Banach space $X$. A functional
$J:U\subset X\to\overline\R$ is called {\bf convex} iff for all $x$,
$\un x\in U$ and $t\in [0,1]$ holds
\[
J(t x+(1-t)\un x) \leqs tJ(x) + (1-t) J(\un x).
\]
A convex functional $J$ is called {\bf strictly convex} iff for all
$x$, $\un x\in U$, with $x\neq \un x$, and $t\in (0,1)$ holds
\[
J(t x+(1-t)\un x) < tJ(x) + (1-t) J(\un x).
\]
Besides these main Theorems above, the following Lemma is also needed
in the proof of Theorem~\ref{T:VM-EM}.
\begin{lemma}
\label{L:VM-cf}
Let $X$ be a Banach space, $U\subset X$ be a closed, convex set, and
$J:U\subset X\to\overline\R$ be a convex and lsc functional. Then, the
functional $J$ is \lscw.
\end{lemma}

\Proof {\it (Lemma~\ref{L:VM-cf}.)~}
Let $\{x_n\}\subset U$ be any sequence such that $x_n\wto x_0$ in $X$,
and denote $\alpha_0:=\liminf_{n\to\infty}J(x_n)$. Earlier in this
Section it was shown, using a Mazur's sequence, that a closed and
convex set in a Banach space is also \closedw, therefore $x_0\in
U$. Using once again Mazur's result, let $\{y_m\}\subset U$ be a
convex combination of the elements $x_n$ such that $y_m\to x_0$ in
$U$. Furthermore, let the convex combination of the elements $x_n$
start at $n=N$ for some number $N\in\N$ instead of $n=1$, that is,
\[
y_m=\sum_{n=N}^m\lambda_nx_n,\qquad
\Bigl(\lambda_n\geqs 0,\quad
\sum_{n=N}^m\lambda_n =1\Bigr).
\]
The functional $J$ is convex, therefore,
\begin{equation}
\label{VM-cf1}
J(y_m)\leqs \sum_{n=N}^m \lambda_n J(x_n),
\end{equation}
By definition of the constant $\alpha_0$ and after selecting a
subsequence if necessary, given any positive number $\epsilon$ there
exists a number $N(\epsilon)\in\N$ such that
\[
J(x_n) < \alpha_0 + \epsilon,\qquad
\forall\,n \geqs N(\epsilon).
\]
Then, Eq.~(\ref{VM-cf1}) implies
\[
J(y_m) < \Bigl(\sum_{n=N(\epsilon)}^m \lambda_n\Bigr) 
\,(\alpha_0 + \epsilon) = \alpha_0 + \epsilon.
\]
so by choosing $\epsilon$ small enough we conclude that
$\liminf_{m\to\infty} J(y_m)\leqs\alpha_0$, or alternatively,
\[
\liminf_{m\to\infty} J(y_m)\leqs\liminf_{n\to\infty}J(x_n).
\]
Finally, recalling that the sequence $y_m\to x_0$ and that the
functional $J$ is lsc, we have that
$J(x_0)\leqs\liminf_{m\to\infty}J(y_m)$, which together with equation
above says,
\[
J(x_0)\leqs\liminf_{n\to\infty}J(x_n).
\]
This equation establishes the Lemma.\qed

For completeness we now state the following result, which establishes
the existence and uniqueness of minimizers for certain type of convex
functionals.
\begin{theorem}{\bf (Minimizers of convex functionals)}
\label{T:VM-mcf}
Let $X$ be a Banach space and $U\subset X$ be a closed, convex
set. Let $J:U\subset X\to\overline\R$ be a convex, lsc, and coercive
functional. Then, there exists an element $x\in U$ minimizer of the
functional $J$ in the set $U$. Furthermore, if the functional $J$ is
strictly convex, then the min\-i\-miz\-er $x$ is unique.
\end{theorem}

\Proof {\it (Theorem~\ref{T:VM-mcf}.)~}
We know that a closed, convex set $U$ in a reflexive Banach space $X$
is a \closedw set. And a convex and lsc functional $J$ on a \closedw
set $U$ is \lscw, result proved in Lemma~\ref{L:VM-cf}. Since the
functional $J$ is also coercive, then $J$ is proper and bounded
below. Therefore, Theorem~\ref{T:VM-M} implies that the functional $J$
has a minimizer $x\in U$. 

Assume now that the functional $J$ is strictly convex, and assume that
there exist two minimizers $x_1$, $x_2\in U$ of the functional $J$,
that is,
\[
J_0 := J(x_1) =J(x_2) = \inf_{x\in X} J(x).
\] 
We will now construct a contradiction.
Assume that the minimizers are different, $x_1\neq
x_2$, and introduce the elements $x_t := t x_1 + (1-t) x_2$, for $t\in
(0,1)$. The strict convexity of the functional $J$ implies
\[
J(x_t) = J\bigl(tx_1+(1-t)x_2\bigr) 
< t J(x_1) + (1-t) J(x_2) = t J_0 + (1-t) J_0 
= J_0.
\]
We then conclude that $J(x_t) < J(x_1)=J(x_2)$, contradicting the
assumption that the elements $x_1$, $x_2$ are minimizers of
$J$. Therefore the minimizer must be unique. This establishes the
Theorem.\qed

We finish this Section with a calculation that is useful to verify
whether a G\^ateaux differentiable functional is convex or strictly
convex. Let $B_{\epsilon}(x_0)\subset X$ be an open ball of radius
$\epsilon$ centered at the element $x_0\in X$. If a convex functional
$J:U\subset X\to X$ is G\^ateaux differentiable in the convex set $U$
with G\^ateaux derivative $DJ$, then, there exists a positive and
small enough number $\epsilon$ such that the following inequality
holds
\begin{equation}
\label{VM-conv-gat1}
DJ(x)v \leqs J(x+v) - J(x)\qquad
\forall \, x\in \mbox{int}(U), 
\quad \forall \,v \in B_{\epsilon}(0).
\end{equation}
The proof is the following calculation: Fix a positive number
$\epsilon$, then given both $x\in \mbox{int}(U)$ and $v\in
B_{\epsilon}(0)$ there exists a small enough number $\epsilon$ such
that $x+v\in U$. The convexity of the set $U$ and of the functional
$J$ imply that
\[
J\bigl(t (x+v) + (1-t)x\bigr) \leqs t J(x+v) + (1-t) J(x),
\]
which in turn implies
\[
\frac{1}{t}\bigl[ J(x+tv)-J(x)\bigr] \leqs J(x+v) - J(x).
\]
Then, Eq.~(\ref{VM-conv-gat1}) follows by taking the limit $t\to
0^{+}$ in the inequality above. A similar proof establishes the
following result: If the functional $J$ is a strictly convex and
G\^ateaux differentiable in a convex set $U$, then there exists a
positive and small enough number $\epsilon$ such that the following
inequality holds
\begin{equation}
\label{VM-conv-gat2}
DJ(x)v < J(x+v) - J(x)\qquad
\forall \, x\in \mbox{int}(U), 
\quad \forall \,v \in B_{\epsilon}(0), \quad v\neq 0.
\end{equation}

\subsection{Ordered Banach spaces}
\label{S:OBS}

\begin{figure}[ht]
\begin{center}
\includegraphics[height=3cm,width=4cm]{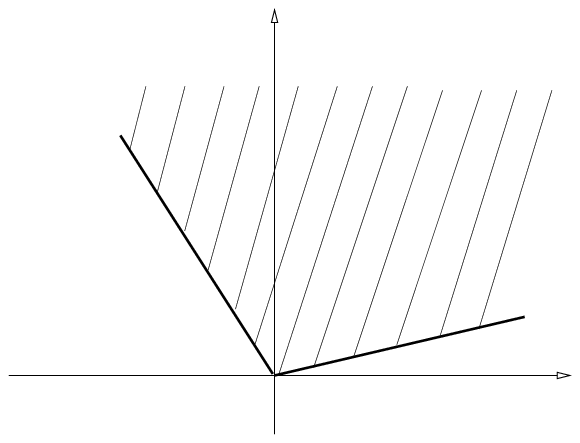}
\includegraphics[height=3cm,width=4cm]{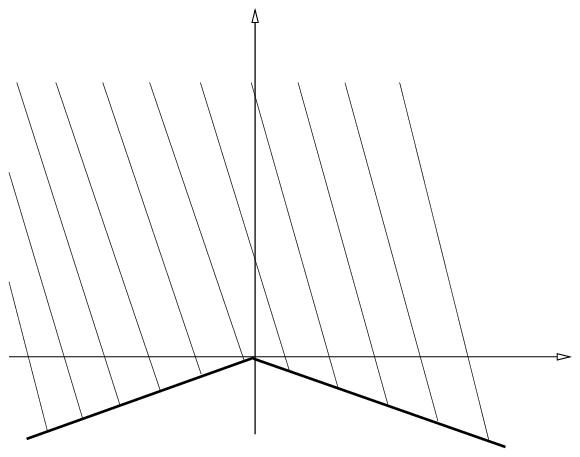}
\includegraphics[height=3cm,width=4cm]{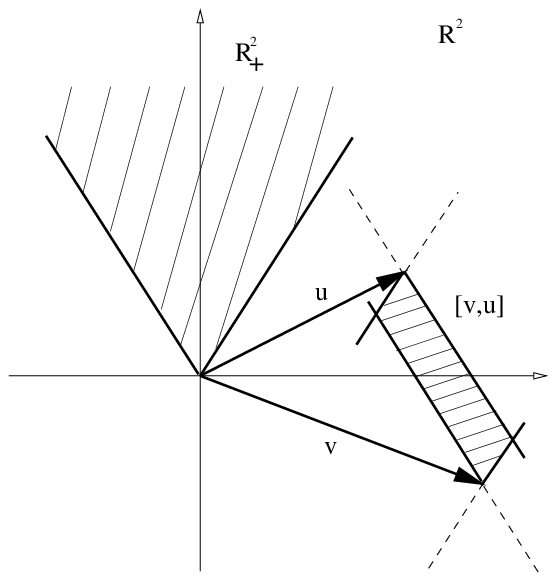}
\end{center}
\caption{The shaded regions in the first picture represents an
order cone, while the second picture represents a cone that is not an
order cone. The shaded region between $u$ and $v$ in the third
picture represents the closed interval $[v,u]$, constructed with the 
order cone $\R^2_{+}$, which is also represented by a shaded region.}
\label{F:cones}
\end{figure}

These notes follow the main ideas and definitions given Chapter~7.1,
page~275, in \cite{Zeidler-I}, while some examples were taken
from~\cite{hA76} and~\cite{Du06}. Let $X$ be a Banach space, $\R_{+}$
be the non-negative real numbers. A subset $C\subset X$ is a {\bf
cone} iff given any $x\in C$ and $a\in\R_{+}$ the element $ax\in C$. A
subset $X_{+}\subset X$ is an {\bf order cone} iff the following
properties hold:
\begin{enumerate}[{\it(i)}]
\item
The set $X_{+}$ is non-empty, closed, and $X_{+}\neq \{0\}$;
\item
Given any $a$, $b\in\R_{+}$ and $x$, $\un x\in X_{+}$ then 
$ax +b\un x \in X_{+}$;
\item 
If $x\in X_{+}$ and $-x\in X_{+}$, then $x=0$.
\end{enumerate}
The second property above says that every order cone is in fact a
cone, and that the set $X_{+}$ is convex. The space $X=\R^2$ is a
convenient Banach space to picture non-trivial examples of cones and
order cones, as can be seen in Fig.~\ref{F:cones}. A pair $X$,
$X_{+}$ is called an {\bf ordered Banach space} iff $X$ is a Banach
space and $X_{+}\subset X$ is an order cone. The reason for this name
is that the order cone $X_{+}$ defines several relations on elements
in $X$, called order relations, as follows:
\[
\begin{gathered}
u \geqs v \mbox{~~iff~~} u-v \in X_{+},\\
u \gg v \mbox{~~iff~~} u-v \in \mbox{int}(X_{+}),
\end{gathered}
\qquad
\begin{gathered}
u > v \mbox{~~iff~~} u\geqs v \mbox{~~and~~} u\neq v,\\
u \ngeqs v \mbox{~~iff~~} u\geqs v \mbox{~is false};
\end{gathered}
\]
finally it is also used the notation $u\leqs v$, $u< v$, and $u\ll v$
to mean $v\geqs u$, $v>u$, $v\gg u$, respectively. A simple example of
an ordered Banach space is $\R$ with the usual order. Another example
can be constructed when this order on $\R$ is transported into
$C^0(\overline \cM,0)$, the set of scalar-valued functions on a set
$\cM\subset\R^n$, with $n\geqs 1$. An order on $C^0(\overline\cM,0)$
is the following: the functions $u$, $v\in C^0(\overline\cM,0)$
satisfy $u\geqs v$ iff $u(x) \geqs v(x)$ for all $x\in \cM$. The
following Lemmas summarize the main properties of order relations
in Banach spaces.
\begin{lemma}
\label{L:o-1}
Let $X$, $X_{+}$ be an ordered Banach space. Then, for all elements
$u$, $v$, $w \in X$, hold: (i) $u \geqs u$; (ii) If $u\geqs v$ and $v
\geqs u$, then $u=v$; (iii) If $u \geqs v$ and $v \geqs w$, then
$u\geqs w$.
\end{lemma}

\Proof {\it (Lemma~\ref{L:o-1}.)~}
The property that $u-u=0\in X_{+}$ implies that $u\geqs u$. If
$u\geqs v$ and $v\geqs u$ then $u-v \in X_{+}$ and $-(u-v)\in X_{+}$,
therefore $u-v =0$. Finally, if $u\geqs v$ and $v\geqs w$, then $u-v
\in X_{+}$ and $v-w\in X_{+}$, which means that $u-w = (u-v) + (v-w)
\in X _{+}$.\qed

Furthermore, the order relation is compatible with the vector space
structure and with the limits of sequences.
\begin{lemma}
\label{L:o-2}
Let $X$, $X_{+}$ be an ordered Banach space. Then, for all $u$, $\hat
u$, $v$, $\hat v$, $w \in X$, and $a$, $b\in \R$, hold
\begin{enumerate}[(i)]
\item \label{L:o-2i}
If $u\geqs v$ and $a\geqs b \geqs 0$, then $au \geqs bv$;
\item \label{L:o-2ii}
If $u\geqs v$ and $\hat u \geqs \hat v$, then $u+\hat u \geqs v+\hat v$;
\item \label{L:o-2iii}
If $u_n \geqs v_n$ for all $n\in \N$, 
then $\lim_{n\to\infty}u_n \geqs \lim_{n\to\infty}v_n$.
\end{enumerate}
\end{lemma}

\Proof {\it (Lemma~\ref{L:o-2}.)~}
The first two properties are straightforward to prove, and we do not
do it here. The third property holds because the order cone is a
closed set. Indeed, $u_n \geqs v_n$ means that $u_n-v_n \in X_{+}$ for
all $n\in\N$, and then $\lim_{n\to\infty}(u_n-v_n)\in X_{+}$ because
$X_{+}$ is closed, then Property {\it(\ref{L:o-2iii})} follows.\qed

The remaining order relations have some other interesting properties.
\begin{lemma}
\label{L:o-3}
Let $X$, $X_{+}$ be an ordered Banach space. Then, for all $u$, $v$,
$w \in X$, and $a\in \R$, hold: (i) If $u\gg v$ and $v\gg w$, then $u
\gg w$; (ii) If $u\gg v$ and $v\geqs w$, then $u \gg w$; (iii) If
$u\geqs v$ and $v\gg w$, then $u \gg w$; (iv) If $u \gg v$ and $a > 0$,
then $au \gg av$.
\end{lemma}
The Proof of Lemma~\ref{L:o-3} is similar to the previous Lemma, and
is not reproduced here. Given an ordered Banach space $X$, $X_{+}$,
and two elements $u\geqs v$, introduce the intervals
\[
[v,u] := \{w\in X : v\leqs w \leqs u \},\qquad
(v,u) := \{w\in X : v \ll w \ll u \}.
\]
Analogously, introduce the intervals $[v,u)$ and $(v,u]$. See
Fig.~\ref{F:cones} for an example in $X=\R^2$.
Useful order cones for solving PDE are those that define an order
structure in the Banach space which is related with the norm and the
notion of boundness. These type of order cones are called normal. More
precisely, an order cone $X_{+}$ in a Banach space $X$ is called {\bf
normal order cone} iff there exists $0<a\in\R$ such that for all $u$,
$v\in X$ with $0 \leqs v\leqs u$ holds $\|v\| \leqs a\, \|u\|$.
\begin{lemma}
\label{L:normal}
If $X$, $X_{+}$ is an ordered Banach space with normal order cone
$X_{+}$, then every closed interval in $X$ is bounded.
\end{lemma}

\Proof {\it (Lemma~\ref{L:normal}.)~}
Let $w\in [v,u]$, then $v\leqs w\leqs u$, and so $0\leqs w-v \leqs
u-v$. Since the cone $X_{+}$ is normal, this implies that there exists
$a>0$ such that $\|w-v\|\leqs a\,\|u-v\|$. Then, the inequalities
$\|w\|\leqs\|w-v\| +\|v\|\leqs a\,\|u-v\| +\|v\|$, which hold for all
$w\in [v,u]$, establish the Lemma.\qed

Not every order cone is normal. For example, consider the Sobolev
spaces $W^{k,p}$ of scalar-valued functions on an $n$-dimensional,
compact manifold $\cM$, with Lipschitz continuous boundary, where $k$
is a non-negative integer, and $p>1$ is a real number. An order cone
in $W^{k,p}$ is defined translating the order on the real numbers,
almost everywhere in $\cM$, that is,
\[
W^{k,p}_{+} :=\{ u\in W^{k,p} : u \geqs 0 \mbox{ a.e. in } \cM\}.
\]
In the case $k=0$, that is, we have $W^{0,p} = L^p$, the order cone
above is a normal cone \cite{hA76,Zeidler-I}. However, in the case
$k\geqs 1$ the cone above cannot be normal, since on the one hand, the
cone definition involves information only of the values of $u(x)$ and
not of its derivatives; on the other hand, the norm in $W^{k,p}$
contains information of both the values of $u(x)$ and its derivatives.
Since there is no boundary conditions on $\partial\cM$ in the
definition of $W^{k,p}$, there is no way to relate the values of a
function in $\cM$ with the values of its derivatives. (In other words,
there is no Poincar\'e inequality for elements in $W^{k,p}$, with
$k\geqs 1$.)

An order cone $X_{+}\subset X$ is {\bf generating} iff $\Span (X_{+})
= X$. An order cone $X_{+}\subset X$ is called {\bf total} iff
$\Span(X_{+})$ is dense in $X$. Total order cones are important
because the order structure associated with them can be translated
from the space $X$ into its dual space $X^{*}$.
\begin{lemma}
\label{L:dual}
Let $X$, $X_{+}$ be an ordered Banach space. If $X_{+}$ is a total
order cone, then an order cone in $X^{*}$ is given by the set
$X^{*}_{+} \subset X^{*}$ defined as
\[
X^{*}_{+} := \{ u^{*} \in X^{*} : u^{*}(v) \geqs 0 \quad
\forall \, v\in X_{+} \}.
\]
\end{lemma}

\Proof {\it (Lemma~\ref{L:dual}.)~}
We check the three properties in the definition of the order cone. The
first property is satisfied because $X_{+}$ is an order cone, so there
exists $v\neq 0$ in $X_{+}$, and then there exists $u^{*}\neq 0$ in
$X^{*}$ such that $u^{*}(v)=1 \geqs 0$, so $X^{*}_{+}$ is
non-empty. Trivially, $0\in X^{*}_{+}$. Finally, $X^{*}_{+}$ is closed
because the order relation $\geqs$ for real numbers is used in its
definition. The second property of an order cone is satisfied, because
given any $u^{*}$, $v^{*}\in X^{*}_{+}$ and any non-negative $a$,
$b\in\R$, then for all $\un u\in X_{+}$ holds
\[
(au^{*}+bv^{*})(\un u) = 
au^{*}(\un u) + bv^{*}(\un u) \geqs 0
\]
since each term is non-negative. This implies that $(au^{*}+bv^{*})\in
X^{*}_{+}$. The third property is satisfied because the order cone
$X_{+}$ is total. Suppose that the element $u^{*}\in X^{*}_{+}$ and
$-u^{*}\in X^{*}_{+}$, then for all $\un u\in X_{+}$ holds that
$u^{*}(\un u) \geqs 0$ and $-u^{*}(\un u) \geqs 0$, which implies that
$u^{*}(\un u) =0$ for all $\un u\in X_{+}$. Therefore, $u^{*}\in
X_{+}^{\bperp}\subset X^{*}$, where the super-script $\bperp$ in
$X^{\bperp}_{+}$ means the Banach annihilator of the set $X_{+}$,
which is a subset of the space $X^{*}$. Therefore, we conclude that
$u^{*}\in\bigl[\Span (X_{+})\bigr]^{\bperp}$. Since the order cone is
total, $\overline{\Span (X_{+})} = X$, that implies $\bigl[\Span
(X_{+})\bigr]^{\bperp} = \{0\}$, so $u^{*}=0$. This establishes the
Lemma.\qed

An order cone $X_{+}$ in a Banach space $X$ is called a {\bf solid
cone} iff $X_{+}$ has non-empty interior. The following result asserts
that solid order are generating. We remark that the converse is not
true. In the examples below we present function spaces frequently used
in solving PDE with order cones having empty interior which are indeed
generating.
\begin{lemma}
\label{L:int}
Let $X$, $X_{+}$ be an order Banach space. If $X_{+}$ is a solid cone,
then $X_{+}$ is generating.
\end{lemma}

\Proof {\it (Lemma~\ref{L:int}.)~}
The cone $X_{+}$ has a non-empty interior, so there exists
$x_0\in\mbox{int}(X_{+})$ and $x_0\neq 0$. This means that given any
$x\in X$ there exists $0<a\in\R$ small enough such that both $x_{+} :=
x_0 + a x$ and $x_{-} := x_0 - ax$ belong to $\mbox{int}(X_{+})$. But
then, $x = (x_{+} - x_{-})/(2a)$, so $x\in\Span(X_{+})$. This
establishes the Lemma. \qed

Here is a list of examples of several order cones used in function
spaces. All these examples use order cones obtained from the usual
order in $\R$. In particular, they refer to scalar-valued functions on
an $n$- dimensional, compact manifold $\cM$ with Lipschitz boundary.
\begin{itemize}
\item 
Introduce on $C^k$ the cone
$
C^k_{+} := \{u\in C^k : u(x) \geqs 0 \quad \forall x \in \cM\}.
$
This is an order cone for all non-negative integer $k$. The cone is a
normal cone in the particular case $k=0$. The cone is solid for all
$k\geqs 0$, therefore it is a generating cone.

\item 
Introduce on $L^{\infty}$ the cone
$L^{\infty}_{+} := \{u\in L^{\infty} : u \geqs 0 
\quad \mbox{a.e. in } \cM\}$.
This is a normal, order cone. It is a solid cone, therefore is
generating.

\item 
Introduce on $W^{k,\infty}$ the cone
$W^{k,\infty}_{+} := \{u\in W^{k,\infty} : u \geqs 0 
\quad \mbox{a.e. in } \cM\}$.
This is an order cone. It is not normal for $k\geqs 1$. The cone is
solid, therefore it is generating.

\item 
Introduce on $L^p$ the cone
$L^p_{+} := \{u\in L^p : u \geqs 0 \quad \mbox{a.e. in } \cM\}$.
This is a normal, order cone every real numbers $p\geqs 1$. The cone
is not solid, however it is a generating cone.

\item 
Introduce on $W^{k,p}$ the cone
$W^{k,p}_{+} := \{u\in W^{k,p} : u \geqs 0 \quad \mbox{a.e. in } \cM\}$.
This is an order cone every real numbers $p\geqs 1$. The cone is not
normal for $k\geqs 1$. The cone is not solid for $kp \leqs n$, and it
is solid for $kp > n$. In both cases, the cone is generating.
\end{itemize}

\subsection{Maximum principles}
\label{S:OBS-MP}

We have not seen in the literature an approach to maximum principles
on ordered Banach spaces in the generality we present it in this
Section.  Let $X$, $X_{+}$ and $Y$, $Y_{+}$ be ordered Banach
spaces. An operator $A : D_A \subset X\to Y$ satisfies the {\bf
maximum principle} iff for every $u$, $v\in D_A$ such that $Au-Av\in
Y_{+}$ holds that $u-v\in X_{+}$. In the particular case that the
operator $A$ is linear, then it satisfies the maximum principle iff
for all $u\in X$ such that $Au\in Y_{+}$ holds that $u\in X_{+}$. The
main example is the Laplace operator acting on scalar-valued functions
defined on different domains. It is shown later on in this Appendix
that the inverse of an operator that satisfies the maximum principle
is monotone increasing. The following result gives a simple sufficient
condition for an operator to satisfy the maximum principle. This
result is useful on weak formulations of PDE.
\begin{lemma}
\label{L:OBS-MP-suff}
Let $X$, $X_{+}$ be an ordered Banach space, and $A: X\to X^{*}$ be a
linear and coercive map. Assume that $X_{+}$ is a generating order
cone, and that for all $u\in X$ such that $Au\in X_{+}^{*}$ there
exists a decomposition $u= u^{+} - u^{-}$ with $u^{+}$, $u^{-}\in
X_{+}$ that also satisfies $Au^{+}(u^{-}) =0$. Then, the operator $A$
satisfies the maximum principle.
\end{lemma}

\Proof {\it (Lemma~\ref{L:OBS-MP-suff}.)~}
Since the order cone $X_{+}$ is generating, the space $X^{*}$ is also
an ordered Banach space. Denote its order cone by $X^{*}_{+}$.  The
assumption that the order cone $X_{+}$ is generating also implies that
for any element $u\in X$ there exists a decomposition $u = u^{+} -
u^{-}$ with $u^{+}$, $u^{-}\in X_{+}$. By hypothesis, there exists at
least one decomposition with the extra property that
$Au^{+}(u^{-})=0$.  Now, by definition of the order in the space
$X^{*}$ we have that
\[
Au\in X_{+}^{*} \LRI Au(\un u)\geqs 0 \quad \forall \, \un u \in X_{+}.
\]
Pick as test function $\un u = u^{-}$. Then,
\[
0 \leqs Au(u^{-}) = A(u^{+} - u^{-})(u^{-}) 
=Au^{+}(u^{-}) - Au^{-}(u^{-}) = -Au^{-}(u^{-}),
\]
where the last equality comes from the condition
$Au^{+}(u^{-})=0$. Therefore, we have
\[
Au^{-}(u^{-}) \leqs 0 \RI u^{-}=0,
\]
because $A$ is coercive. So we showed that $u=u^{+} \in X_{+}$. This
establish the Lemma.\qed

An example is the weak form of the Laplace operator on scalar
functions in the homogeneous Dirichlet problem on a compact manifold
$\cM$ with Lipschitz boundary. Consider the case $X = W^{1,2}_0$, with
$Y = X^{*} = W^{-1,2}$, and $X_{+}=W^{1,2}_{+}$, while $Y_{+} =
W^{-1,2}_{+}$. The Laplace operator in this case is given by $A:X \to
X^{*}$ with action $Au(v):=(\nabla u,\nabla v)$.  It is not difficult
to check that this operator satisfies the hypothesis in
Lemma~\ref{L:OBS-MP-suff}. Therefore, this operator satisfies the
maximum principle, that is, $Au\in W^{-1,2}_{+}$ implies $u\in
W^{1,2}_{+}$, that is, $u\geqs 0$ a.e. in the manifold $\cM$.  This
result is in agreement with Theorem~8.1 in
\cite{Gilbarg-Trudinger}, where it is stated that: ``If $Au
\geqs 0$, then $\inf_{\cM}u\geqs -\inf_{\partial\cM}u^{-}$.''
Here we introduced the cut-off function $u^{-} := - \min(u,0) \geqs
0$. Recalling that in our case the domain of $A$ contains only
functions that vanish at the boundary, then $\inf_{\cM} u \geqs 0$,
that is, $u\geqs 0$ in $\cM$.

The following example is again the Laplace operator that appears in
equations when they are written in weak form, but this time using more
complicated operator domains due to more complicated boundary
conditions in the PDE equation. Let $(\cM,h)$ be a 3-dimensional
Riemannian manifold, where $\cM$ is a smooth, compact manifold with a
Lipschitz boundary $\partial\cM$, and $h\in C^{2}(\overline \cM,2)$ is
a positive definite metric.  Assume that the boundary set can be
decomposed as follows, $\partial\cM
=\partial\cM_{\tiD}\cup\partial\cM_{\tiN}$ and
$\overline{\partial\cM}_{\tiD}\cap\overline{\partial\cM}_{\tiN}
=\emptyset$. Recall the definition of the Sobolev spaces
\[
W^{1,2}_{\tiD} := \{u\in W^{1,2}(\cM,\R) : \Tr_{\tiD} u=0\},\qquad
W^{-1,2}_{\tiD} := \bigl[ W^{1,2}_{\tiD} \bigr]^{*}.
\]
Then, define the operator 
\begin{equation}
\label{OBS-def-AL}
A_{\tiL}^s: W^{1,2}_{\tiD}\to W^{-1,2}_{\tiD},\qquad
A_{\tiL}^s\phi(\un\phi) := a_{\tiL}(\phi,\un\phi)+ (s\phi,\un\phi);
\end{equation}
where $a_{\tiL}$ is the bilinear form
\[
a_{\tiL} : W^{1,2}_{\tiD} \times W^{1,2}_{\tiD} \to \R, \qquad
a_{\tiL}(\phi,\un\phi) := (\nabla\phi,\nabla\un\phi) 
+ (K\, \Tr_{\tiN} \phi,\Tr_{\tiN}\un\phi)_{\tiN},
\]
and the Robin coefficient $K\in
L^{\infty}(\partial\cM_{\tiN},0)$ satisfies the bounds
\begin{equation}
\label{OBS-ttk-cond}
\hat\ttk\, \|\Tr_{\tiN}\phi\|^2_{\tiN} \leqs 
(K \Tr_{\tiN}\phi, \Tr_{\tiN}\phi)_{\tiN},
\qquad \forall \phi\in W^{1,2},
\end{equation}
with $\hat\ttk$ a positive constant. Assume that the function
$s\in L^{3/2}_{+}$, so the second term in the definition of the
operator $A_{\tiL}^s$ is well defined. 
\begin{lemma}
\label{L:Example-MP}
The operator $A_{\tiL}^s$ defined in Eq.~(\ref{OBS-def-AL}) satisfies
the maximum principle.
\end{lemma}

\Proof {\it (Lemma~\ref{L:Example-MP}.)~}
We now verify all the hypothesis in Lemma~\ref{L:OBS-MP-suff}. The
cone $W^{1,2}_{\tiD +} = W^{1,2}_{+}\cap W^{1,2}_{\tiD}$ is generating
in $W^{1,2}_{\tiD}$ therefore, $W^{-1,2}_{\tiD}$ is also an ordered
space. The constant $\hat\ttk$ is positive and the function $s$ is
non-negative, which implies that the operator $A_{\tiL}^s$ is
coercive. Using the usual decomposition of a function $u$ into
$u^{+}(x)=\max_{\cM}(u(x),0)$ and $u^{-}(x)=-\min_{\cM}(u(x),0)$, then
it is not difficult to show that $A_{\tiL}^su^{+}(u^{-})=0$, because the
two parts of the decomposition of $u$ are defined on non-intersecting
parts of $\cM$. Therefore, Lemma~\ref{L:Example-MP} follows from
Lemma~\ref{L:OBS-MP-suff}.\qed

\subsection{Monotone operators}
\label{S:OBS-MO}

Let $X$, $X_{+}$ and $Y$, $Y_{+}$ be two ordered Banach spaces. An
operator $F:X\to Y$ is {\bf monotone increasing} iff for all $x$, $\un
x\in X$ such that $x-\un x \in X_{+}$ holds that $F(x) -F(\un x)\in
Y_{+}$. An operator $F:X\to Y$ is {\bf monotone decreasing} iff for
all $x$, $\un x\in X$ such that $x-\un x \in X_{+}$ holds that
$-\bigl[F(x) -F(\un x)\bigr]\in Y_{+}$. The following result is a
useful relation between linear, invertible operators that satisfy the
maximum principle and monotone increasing operators.
\begin{lemma}
\label{L:Linverse}
Let $X$, $X_{+}$ and $Y$, $Y_{+}$ be two ordered Banach spaces.  Let
$A:X\to Y$ be a linear, invertible operator satisfying the maximum
principle. Then, the inverse operator $A^{-1}:Y\to X$ is monotone
increasing.
\end{lemma}

\Proof {\it (Lemma~\ref{L:Linverse}.)~}
Let $y$, $\un y\in Y$ be such that $y-\un y\in Y_{+}$. Then,
\[
A\bigl(A^{-1}(y-\un y)\bigr)\in Y_{+} \RI
A^{-1}(y-\un y)\in X_{+} \LRI
A^{-1}y -A^{-1}\un y \in X_{+}.
\]
This establishes that the operator $A^{-1}$ is monotone increasing.
\qed

We are interested in a class of nonlinear problems where the principal
part involves a linear operator $A:X \to Y$ satisfying the maximum
principle, and the non-principal part involves a nonlinear operator
$F:X\to Y$ which has monotonicity properties; problems of this type
can be written as follows: Find an element $x\in X$ solution of the
equation
\begin{equation}
\label{eqn:affine}
Ax + F(x)=0.
\end{equation}
We now establish some results for this class of problems.

\begin{lemma}
\label{L:T-decreasing}
Let $X$, $X_{+}$ and $Y$, $Y_{+}$ be two ordered Banach spaces. Let
$A:X\to Y$ be a linear, invertible operator satisfying the maximum
principle. Let $F:X\to Y$ be a monotone decreasing (increasing)
operator. Then, the operator $T:X\to X$ given by $T:= -A^{-1}F$ is
monotone increasing (decreasing).
\end{lemma}

\Proof {\it (Lemma~\ref{L:T-decreasing}.)~}
Assume first that the operator $F$ is monotone decreasing.  So, given
any $x$, $\un x\in X$ such that $x-\un x\in X_{+}$, the following
inequalities hold,
\begin{align*}
x-\un x \in X_{+} &\RI -\bigl[F(x) - F(\un x)\bigr] \in Y_{+},\\
&\LRI  A\bigl(-A^{-1}\bigl[ F(x) - F(\un x) \bigr]\bigr)\in Y_{+},\\
&\RI -A^{-1}\bigl[ F(x) - F(\un x) \bigr]\in X_{+},\\
&\LRI -\bigl[ A^{-1}F(x) -A^{-1}F(\un x) \bigr]\in X_{+},\\
&\LRI T(x) - T(\un x) \in X_{+},
\end{align*}
which establishes that the operator $T$ is monotone increasing.  In
the case that the operator $F$ is monotone increasing, then the first
line in the proof above changes into $x -\un x\in X_{+}$ implies that
$F(x) - F(\un x)\in Y_{+}$, and then all the remaining inequalities in
the proof above are reverted. This establishes the Lemma.\qed

The next result translates the inequalities that satisfy sub- and
super-so\-lu\-tions to the equation $Ax+F(x)=0$, into inequalities for
the operator $T=-A^{-1}F$.
\begin{lemma}
\label{L:T-sub-super}
Assume the hypothesis in Lemma~\ref{L:T-decreasing}. 

If there exists an element $x_{+}\in X$ such that $Ax_{+}+F(x_{+})\in
Y_{+}$, then this element satisfies that
$x_{+}-T(x_{+})\in X_{+}$.

If there exists an element $x_{-}\in X$ such that $-\big[
Ax_{-}+F(x_{-})\bigr]\in Y_{+}$, then this element satisfies that
$-\bigl[x_{-}-T(x_{-})\bigr]\in X_{+}$.
\end{lemma}

\Proof {\it (Lemma~\ref{L:T-sub-super}.)~}
The first statement in the Lemma can be shown as follows,
\begin{align*}
Ax_{+}+F(x_{+}) \in Y_{+} &\Leftrightarrow
A\bigl( x_{+} +A^{-1}F(x_{+})\bigr) \in Y_{+}\\
&\Rightarrow  x_{+} +A^{-1}F(x_{+}) \in X_{+},
\end{align*}
which then establishes that $x_{+}-T(x_{+})\in X_{+}$.  In a similar
way, the second statement in the Lemma can be shown as follows,
\begin{align*}
-\bigl[Ax_{-}+F(x_{-})\bigr] \in Y_{+} &\Leftrightarrow
A\bigl(- x_{-} -A^{-1}F(x_{-})\bigr) \in Y_{+}\\
&\Rightarrow  -x_{-} -A^{-1}F(x_{-}) \in X_{+},
\end{align*}
which then establishes that $-\bigl[x_{-}-T(x_{-})\bigr]\in X_{+}$.
This establishes the Lemma.\qed

The last result can be found as Theorem~7.A in \cite{Zeidler-I}, page
283, and Corollary~7.18 on page 284. We reproduce it here for
completeness, without the proof.
\begin{theorem}{\bf (Fixed point for increasing operators)}
\label{T:FPI}
Let $X$ be an ordered Banach space, with a normal order cone $X_{+}$.
Let $T:[x_{-}, x_{+}]\subset X\to X$ be a monotone increasing, compact
map. If $-\bigl[x_{-} - T(x_{-})\bigr]\in X_{+}$ and $x_{+}-
T(x_{+})\in X_{+}$, then the iterations
\begin{align*}
x_{n+1} &:= T(x_n),\qquad x_0 = x_{-},\\
\hat x_{n+1} &:= T(\hat x_n),\qquad \hat x_0 = x_{+},
\end{align*}
converge to $x$ and $\hat x\in [x_{-},x_{+}]$, respectively, and the
following estimate holds,
\begin{equation}
\label{eqn:bounds}
x_{-} \leqs x_n \leqs x 
\leqs \hat x \leqs \hat x_n \leqs x_{+},
\qquad \forall n=\N.
\end{equation}
\end{theorem}

For nonlinear problems of the form~(\ref{eqn:affine}), one can use
Theorem~\ref{T:FPI} for monotone nonlinearities to conclude the
following.

\begin{corollary}
{\bf (Semi-linear equations with sub-/super-solutions)}
\label{C:equiv2}
Let $X$, $X_{+}$ and $Y$, $Y_{+}$ be two ordered Banach spaces where
$X_{+}$ is a normal order cone. Let $A: X\to Y$ be a linear,
invertible operator satisfying the maximum principle. Let $x_{+}$,
$x_{-}\in X$ be elements such that $(x_{+}-x_{-})\in X_{+}$, and then
assume that the operator $F:[x_{-}, x_{+}]\subset X\to Y$ is monotone
decreasing and compact. If the elements $x_{-}$ and $x_{+}$ satisfy
the relations
\begin{equation}
\label{ss-sol2}
-\bigl[ Ax_{-} + F(x_{-})\bigr] \in Y_{+},\qquad
Ax_{+} + F(x_{+}) \in Y_{+},
\end{equation}
then there exists a solution $x\in [x_{-},x_{+}]\subset X$ of the
equation $Ax + F(x)=0$.
\end{corollary}

\Proof {\it (Corollary~\ref{C:equiv2}.)~}
The operator $A$ is invertible, then rewrite the equation $Ax+F(x)=0$
as a fixed-point equation,
\begin{equation}
\label{eqn:fixed-point2}
x = -A^{-1}F(x) =: T(x).
\end{equation}
By Lemma \ref{L:T-decreasing}, we know that the map $T :X\to X$ is
monotone increasing. Moreover, this operator $T$ it is compact, since
is the composition of the continuous mapping $-A^{-1}$ and the compact
map $F$. The elements $x_{-}$ and $x_{+}$ satisfy Eq.~(\ref{ss-sol2}),
therefore, by Lemma~\ref{L:T-sub-super}, they are also sub- and
super-solutions for the fixed-point equation involving the map $T$. It
follows from Theorem \ref{T:FPI} that there exists an element $x\in X$
solution to the fixed-point equation~(\ref{eqn:fixed-point2}), and
this solution satisfies the bounds $x_{-}\leqs x\leqs x_{+}$.\qed

\subsection{{\em A~priori} estimates in ordered Banach spaces}
\label{S:OBS-APRIORI}

Many problems of the form in Eq.~(\ref{eqn:affine}) do not have
monotone nonlinearities. However, in the case that there exist sub-
and super-solutions to Eq.~(\ref{eqn:affine}) it is possible to
introduce a ``shift'' into the equation. This shift transforms a
problem that does not have a monotone nonlinearity into one that does,
without destroying the maximum principle property required of the
linear part.  However, the disadvantage of the shift technique is that
it requires additional regularity in the equation coefficients than
the regularity needed for the original equation to be well-defined.
On the other hand, it is possible to construct arguments leading to
{\em a~priori} order cone estimates on any possible solution (whether
or not it exists) with very weak assumptions on the nonlinearity.
Although such results are standard for semi-linear scalar problems
with monotone nonlinearities (for example, see~\cite{jJ85}), our
result below holds for a class of semi-linear problems with
non-monotone nonlinearities and appears to be new.  Problems with
monotone nonlinearities fit into this class, but it also includes a
much larger set of nonlinearities.  (See the second assumption
in~\ref{ape-i} in Lemma~\ref{L:ape} below.)

The following result (Lemma~\ref{L:ape} below) gives sufficient
conditions for establishing {\em a~priori} order cone estimates on
solutions to certain PDE-like operator equations in ordered Banach
spaces.  These order estimates can be translated into norm estimates
in the case that the order cone is normal.  (See Corollary~\ref{C:ape}
following Lemma~\ref{L:ape} below.)  Note that the bounds established
in Lemma~\ref{L:ape} below are not necessarily sub- and
super-solutions; establishing the bounds by first showing they are
sub- and super-solutions and then using Corollary~\ref{C:equiv2} would
require a monotone nonlinearity, or use of the shifting technique
requiring additional regularity assumptions.

\begin{lemma}{\bf({\em A~priori} order estimates)}
\label{L:ape}
Let $Y$, $Y_{+}$ be an ordered Banach space with a generating 
order cone $Y_{+}$. Let $F :Y\to Y^{*}$ be a continuous map.  Let $A:
Y\to Y^{*}$ be a linear, continuous operator with $\dim N_{A}\geqs
1$. Assume that there exists a subspace $X\subset Y$, with an induced
order cone $X_{+}=X\cap Y_{+}$, such that $A_{\tiX} :X\to X^{*}$, the
restriction of the operator $A$ to the space $X$, is coercive. Let
$u\in Y$ be a solution of the equation $Au + F(u)=0$. 
\begin{enumerate}[(i)]
\item
\label{ape-i}
If there exists an element $y_{\tiwedge}\in N_{A}$ such that $(u -
y_{\tiwedge})^{+}\in X_{+}$, and for all $y\in Y$ such that $(y-
y_{\tiwedge})\in Y_{+}$ holds that $F(y)\bigl(
(y-y_{\tiwedge})^{+}\bigr)\geqs 0$; Then, the solution $u\in Y$
satisfies $y_{\tiwedge} -u\in Y_{+}$.
\item
\label{ape-ii}
If there exists an element $y_{\tivee}\in N_{A}$ such that 
$(u - y_{\tivee})^{-}\in X_{+}$, and
for all $y\in Y$ such that $-(y- y_{\tivee})\in Y_{+}$ holds that
$F(y)\bigl((y-y_{\tivee})^{-}\bigr)\leqs 0$.
Then, the solution $u\in Y$ satisfies $u- y_{\tivee}\in Y_{+}$. 
\end{enumerate}
\end{lemma}

\Proof {\it (Lemma~\ref{L:ape}.)~}
We first show part~{\it(\ref{ape-i})}. Given the solution $u\in Y$,
introduce and element $u_{\tiD}\in Y$ be an element such that
$u-u_{\tiD}\in X$. Second, notice that the element $(u-y_{\tiwedge})$
belongs to the space $Y$, which has a generating order cone $Y_{+}$,
so we know that there exists a decomposition
\[
(u-y_{\tiwedge}) = (u-y_{\tiwedge})^{+} -(u-y_{\tiwedge})^{-},
\]
with both elements $(u-y_{\tiwedge})^{+}$, $(u-y_{\tiwedge})^{-}\in
Y_{+}$. The first assumption in~{\it(\ref{ape-i})} says that $(u -
y_{\tiwedge})^{+}\in X_{+}$ and so the element $(u -
y_{\tiwedge})^{+}$ is a valid test function for the functional
\[
\bigl[A_{\tiX} (u-u_{\tiD}) + F(u) + Au_{\tiD}\bigr] \in X^{*},
\]
so we have the following,
\begin{align*}
Au(u-y_{\tiwedge})^{+} &= A(u-y_{\tiwedge})(u-y_{\tiwedge})^{+} \\
&= A(u-y_{\tiwedge})^{+}(u-y_{\tiwedge})^{+}\\
&= A_{\tiX}(u-y_{\tiwedge})^{+}(u-y_{\tiwedge})^{+}.
\end{align*}
Therefore, we have the following inequalities,
\begin{align*}
0&= Au(u-y_{\tiwedge})^{+} 
+ F(u)\bigl((u-y_{\tiwedge})^{+}\bigl)\\
&= A_{\tiX}(u-y_{\tiwedge})^{+}(u-y_{\tiwedge})^{+}
+ F(u) \bigl((u-y_{\tiwedge})^{+}\bigl)\\
&\geqs a_0 \,\|(u-y_{\tiwedge})^{+}\|_{\tiX}^2 
+ F(u)\bigl((u-y_{\tiwedge})^{+}\bigl), \qquad a_0 >0\\
&\geqs a_0 \, \|(u-y_{\tiwedge})^{+}\|_{\tiX}^2.
\end{align*}
The last inequality implies that $(u-y_{\tiwedge})^{+}=0$, which then
says that $u-y_{\tiwedge} = -(u-y_{\tiwedge})^{-}$, and we then
conclude that $-(u-y_{\tiwedge})\in Y_{+}$.  This condition can be
written using inequalities as $u\leqs y_{\tiwedge}$. 

We now prove part~{\it(\ref{ape-ii})}. The element $(u-y_{\tivee})$
also belongs to the space $Y$, which has a generating order cone
$Y_{+}$, so we know that there exists a decomposition
\[
(u-y_{\tivee}) = (u-y_{\tivee})^{+} -(u-y_{\tivee})^{-},
\]
with both elements $(u-y_{\tivee})^{+}$, $(u-y_{\tivee})^{-}\in
Y_{+}$. The first assumption in part~{\it(\ref{ape-ii})} says that $(u
- y_{\tivee})^{-}\in X_{+}$ and so the element $(u - y_{\tivee})^{-}$
is a valid test function for the functional
\[
\bigl[A_{\tiX} (u-u_{\tiD}) + F(u) + Au_{\tiD}\bigr] \in X^{*},
\]
so we have the following,
\begin{align*}
Au(u-y_{\tivee})^{-} &= A(u-y_{\tivee})(u-y_{\tivee})^{-} \\
&= -A(u-y_{\tivee})^{-}(u-y_{\tivee})^{-}\\
&= -A_{\tiX}(u-y_{\tivee})^{-}(u-y_{\tivee})^{-}.
\end{align*}
Therefore, we have the following inequalities,
\begin{align*}
0&= Au(u-y_{\tivee})^{-} 
+ F(u)\bigl((u-y_{\tivee})^{-}\bigl)\\
&=- A_{\tiX}(u-y_{\tivee})^{-}(u-y_{\tivee})^{-}
+ F(u) \bigl((u-y_{\tivee})^{-}\bigl)\\
&\leqs - a_0 \,\|(u-y_{\tivee})^{-}\|_{\tiX}^2 
+ F(u)\bigl((u-y_{\tivee})^{-}\bigl), \qquad a_0 >0\\
&\leqs - a_0 \, \|(u-y_{\tivee})^{-}\|_{\tiX}^2.
\end{align*}
The last inequality implies that $(u-y_{\tivee})^{-}=0$, which then
says that $u-y_{\tivee} = (u-y_{\tivee})^{+}$, and we then conclude
that $(u-y_{\tivee})\in Y_{+}$. This condition can be written
using inequalities as $u\geqs y_{\tivee}$. This inequality
establishes the Lemma.\qed

\begin{corollary}
\label{C:ape}
Let $Y$, $Y_{+}$ be an ordered Banach space with a normal order cone
$Y_{+}$. Let $Z$ be a Banach space, and consider the space $W=Y\cap
Z$, with order cone $W_{+}:= Y_{+}\cap Z$. If there exist elements
$u$, $y_{\tivee}$, and $y_{\tiwedge}\in W$ such that $0\leqs
y_{\tivee}\leqs u\leqs y_{\tiwedge}$ in the order given by $W_{+}$,
then there exists a positive constant $c$ such that the following
inequalities hold
\[
c\;\|y_{\tivee}\|_{\tiZ} \leqs \|u\|_{\tiZ} 
\leqs \frac{1}{c}\;\|y_{\tiwedge}\|_{\tiZ}.
\]
\end{corollary}

\Proof {\it (Corollary~\ref{C:ape}.)~}
It follows directly from the definition of a normal order cone.\qed

As an example, Lemma~\ref{L:ape} holds with the spaces taken to be
$X=W^{1,2}_{\tiD}$, $Y=W^{1,2}$, the linear operator taken to be
$Au(v) = (\nabla u,\nabla v)$, and the nonlinear operator taken to be
a monotone operator such as $F(u)= u^5$. Lemma~\ref{L:ape} also holds
for a non-monotone nonlinear operator satisfying the assumptions for
the Lemma, such as $F(u)= u^5 - 2 u^3$.  An example of the space $Z$
where Corollary~\ref{C:ape} holds is $Z=L^{\infty}$.

\bibliographystyle{plain}
\bibliography{ref-gn}

\end{document}